\documentclass[12pt]{article}
\usepackage{epsfig}
\usepackage{graphicx}
\usepackage{bm}

\evensidemargin
0.3truein 
\newcommand{\la}{{\lambda}}
\newcommand{\ie}{{\it i.e.}}
\newcommand{\eg}{{\it e.g.}}

\newcommand{\be}{\begin{equation}}
\newcommand{\ee}{\end{equation}}
\newcommand{\beq}{\begin{equation}}
\newcommand{\eeq}{\end{equation}}
\newcommand{\bea}{\begin{eqnarray}}
\newcommand{\eea}{\end{eqnarray}}
\newcommand{\br}{\begin{eqnarray}}
\newcommand{\er}{\end{eqnarray}}
\newcommand{\ba}{\begin{array}}
\newcommand{\ea}{\end{array}}
\newcommand{\bi}{\begin{itemize}}
\newcommand{\ei}{\end{itemize}}
\newcommand{\bn}{\begin{enumerate}}
\newcommand{\en}{\end{enumerate}}
\newcommand{\bc}{\begin{center}}
\newcommand{\ec}{\end{center}}

\def\smd{\sigma_{\mu}}
\def\smu{\sigma^{\mu}}
\def\smdb{{\bar\sigma}_{\mu}}
\def\smub{{\bar\sigma}^{\mu}}

\def\bZ{{\bf Z}}
\def\bg{\bm{\gamma}}
\def\bY{{\bf Y}}
\def\bA{{\bf A}}

\def\bV{{\bf V}}
\def\bmm{{\bf m}}

\def\tl{{\tilde{L}}}
\def\tll{{\tilde{\ell}}}

\def\te{{\tilde{e^c}}}

\def\td{{\tilde{d^c}}}
\def\tq{{\tilde{Q}}}
\def\tf{{\tilde{f}}}
\def\tu{{\tilde{u^c}}}

\def\cq{{\cal Q}}
\def\tcq{{\tilde{{\cal Q}}}}
\def\ln{{\rm ln}}
\def\unity{{\hbox{1\kern-.8mm l}}}
\newcommand{\ov}{\overline}

\newcommand{\no}{\nonumber}
\newcommand{\nn}{\nonumber}

\newcommand{\lto}{\longrightarrow}

\newcommand{\BR}{{\rm BR}}
\newcommand{\CR}{{\rm CR}}

\newcommand{\ga}{\gamma}

\newcommand{\gsim}{\lower.7ex\hbox{$\;\stackrel{\textstyle>}{\sim}\;$}}
\newcommand{\lsim}{\lower.7ex\hbox{$\;\stackrel{\textstyle<}{\sim}\;$}}

\newcommand{\captions}{\sf \caption}

\begin{document}
\begin{flushright}

{DFPD-06/TH/09}\\
\today
\end{flushright}
\vspace{0.5cm}

\begin{center}

{\Large \bf \sf Phenomenology of the triplet seesaw
mechanism with  Gauge and \\
\vspace{0.3cm} Yukawa mediation of SUSY breaking } \vspace{1cm}

{ {\large\bf Filipe R. Joaquim$^{a,b,\ast}$ and Anna Rossi$^{a,\S}$}
}
\\[7mm]
{\it $^a$ Dipartimento di Fisica ``G.~Galilei'', Universit\`a di
Padova
I-35131 Padua, Italy}\\[3mm]
{\it $^b$ Istituto Nazionale di Fisica Nucleare (INFN), Sezione di
Padova, I-35131 Padua, Italy}
\\[1cm]
\vspace{-0.3cm}

{\tt  $^\ast$\,E-mail: \hspace*{-0.3cm}joaquim@pd.infn.it\\
$\!\!\!^\S$\,E-mail: \hspace*{-0.3cm}arossi@pd.infn.it}

\vspace{1cm}

{\large\bf ABSTRACT}
\renewcommand{\baselinestretch}{1.8}
\end{center}

\begin{quote}
{\large\noindent We thoroughly discuss a new supersymmetric grand
unified scenario of the triplet seesaw mechanism where the exchange
of heavy $SU(2)_W$ triplet states generates \emph{both} neutrino
masses and soft supersymmetry breaking terms. This framework,
recently proposed by us in a previous work, is highly predictive
since it contains only three free parameters connecting low-energy
neutrino parameters, lepton and quark flavour violation, sparticle
and Higgs boson spectra and electroweak symmetry breakdown. These
three parameters are the triplet mass $M_T$, the effective
supersymmetry breaking scale $B_T$ and a coupling constant
$\lambda$. We perform a complete analysis of the parameter space
taking into account the present experimental constraints and
considering different types of neutrino spectrum. A special emphasis
is given to the particular features of the sparticle and Higgs
spectra and to the model independent predictions obtained for the
processes $\mu \to e X$, $\mu\to e$ conversion in nuclei, $\tau \to
e Y$  and $\tau \to \mu Y$ $(X = \ga, ee$, $Y= \ga, e e, \mu \mu)$.
In the appendices, we present some technical aspects relevant for
our analysis.}
\end{quote}

\baselineskip=14pt

\vspace*{0.5cm}
\section{Introduction}
It has long been recognised that the realm of neutrino physics may
offer some insights on the search for physics beyond the Standard
Model (SM). The evidence of non vanishing neutrino masses and of
leptonic mixing angles, as provided by neutrino oscillations, calls
for an extension of the SM particle content. Namely, the $d=5$
operator $\bY_\nu (L H)^2 /M_L$, describing neutrino masses at the
effective level~\cite{W}, can be generated by decoupling some heavy
degrees of freedom at the scale $M_L$ where lepton number ($L$) is
broken. This is the essence of the well-celebrated seesaw mechanism
which can be realized at the tree-level by exchanging either singlet
fermions $N$~\cite{seesaw}, or $SU(2)_W$ triplets $T$ with
zero~\cite{t0} or non-zero hypercharge~\cite{ss2}, at $M_L$.
However, we should notice that, from a theoretical perspective, the
concept of non-zero neutrino masses by itself does not tell us much
about the {\it new physics} beyond $M_L$. On the other hand,
alternative signals of lepton flavour violation (LFV) (besides
neutrino oscillations) would be a clear and dramatic manifestation
of {\it new physics}, since they are strongly suppressed within the
SM by the smallness of neutrino masses. For this reason, and taking
into account the increasing sensitivity of the present and future
experiments, it is crucial to explore theoretical frameworks where
such LFV processes can be sizeable. A typical example is the minimal
supersymmetric standard model (MSSM) extended with renormalizable
interactions that give rise to the $d=5$ effective neutrino mass
operator. Hence, at least two attractive features of the
supersymmetric version of the above seesaw scenarios are worth to be
recalled:

1) supersymmetry (SUSY) alleviates the hierarchy problem of the SM,
which would be exacerbated by the presence of one more high scale,
$M_L$ (besides the Planck mass scale $M_{PL}$)~\cite{casas}

2) lepton flavour violating (LFV) processes (otherwise unobservable)
can be enhanced through  one-loop exchange of the lepton
superpartners if their masses are not too far from the electroweak
scale and do not conserve flavour~\cite{ellisnan}.

Regarding the latter aspect, most of the literature has been
focussing on the most conservative scenario of universal sfermion
masses at a scale larger than $M_L$, which is realized in either
minimal supergravity or gauge mediated supersymmetry breaking (GMSB)
models with very large SUSY breaking mediation scale. In such cases,
flavour non-conservation in the sfermion masses arises from
renormalization group (RG) effects due to flavour-violating Yukawa
couplings~\cite{lfv1,lfv2,ar} encoded, at the effective level, in
$\bY_\nu$. In this respect, it has been pointed out that, in the
seesaw realization with non-zero hypercharge triplets, the flavour
structure of the slepton mass matrix $\bmm^2_\tl$ can be univocally
determined in terms of the low-energy neutrino parameters~\cite{ar}.
In contrast, within the singlet seesaw, the determination of
$\bmm^2_\tl$ from low-energy observables requires model dependent
assumptions~\cite{DI,HMTY,rgess}.

Recently we have proposed a new supersymmetric scenario of the
triplet seesaw mechanism in which the soft SUSY breaking (SSB)
parameters of the MSSM are generated at the decoupling of the heavy
triplets. Moreover, the mass scale of such SSB terms is fixed {\it
only} by the triplet SSB bilinear term $B_T$~\cite{af1}. This
scenario turns out to be highly predictive in the sense that it
relates neutrino masses, LFV in the sfermion sector, sparticle and
Higgs boson spectra and electroweak symmetry breaking (EWSB). In
this work we aim to further elucidate and discuss a more general
version of this new framework including effects (previously
neglected in Ref.~\cite{af1}) which also entail quark flavour
violation~(QFV).

The paper is organised as follows. In Section~\ref{recalling} we
review the main features of the SUSY triplet seesaw mechanism. Its
embedding into the $SU(5)$ gauge group is described in
Section~\ref{GaugeYukawa} where the role played by the heavy
triplets as messengers of SUSY breaking is also discussed. It will
be shown that the exchange at the quantum level of the triplet
states generate all the MSSM SSB mass parameters. The results of the
related analytical evaluations are presented and analysed in
Section~\ref{SSBmass}. We proceed in Section~\ref{Flavour Structure}
with a detailed description of the flavour structure exhibited by
the SSB terms. Then, in Section~\ref{phenviab}, we bring forward the
phenomenological analysis of our framework. More specifically, we
detail the experimental constraints imposed on the parameter
space~(Section~\ref{phenconst}) and present our numerical results
(Section~\ref{paramspace}). Section~\ref{phenaspects} is devoted to
the relevant phenomenological predictions such as the sparticle and
Higgs boson mass spectra (Section~\ref{sparspec}) and the expected
size of lepton and quark flavour violation (Section~\ref{LFVQFV}).
Afterwards, we discuss the predictions for several LFV processes
($\mu \to e \ga\,, \tau \to\mu \ga\,, \tau \to e \ga\,, \mu \to e e
e\,, \tau \to \mu e e\, ,\tau\to e \mu \mu \,,\tau \to \mu \mu \mu,
\tau \to e e e$ and $\mu\to e$ conversion in nuclei) and the
peculiar correlations which arise in our scenario, together with a
complete numerical analysis (Section~\ref{numanalysis}). Our
conclusions and summary are drawn in Section~\ref{conclusions}.
Several technical aspects are collected in appendices: Appendix~A is
devoted to the generalization of the method based on the
wave-function renormalization to derive the SSB mass parameters;
Appendix~B presents our analytical calculations of the MSSM
coefficients for the LFV $\ell_j\ell_i Z$ operators; finally,
Appendix~C regards the computation of the box diagrams relevant for
the $\tau \to \mu e e$ and $\tau \to e \mu \mu$ amplitudes.

\section{Recalling the triplet seesaw mechanism}
\label{recalling}
Before starting the discussion of the main subject of our paper, let
us briefly review the key features of the supersymmetric triplet
seesaw mechanism. The requirement of a holomorphic superpotential
implies introducing the triplets as super-multiplets $T =(T^0, T^+,
T^{++}), \bar{T}=(\bar{T}^0, \bar{T}^+, \bar{T}^{++})$ in a
vector-like $SU(2)_W\times U(1)_Y$ representation, $T\sim (3,1),
\bar{T} \sim (3,-1)$~\cite{ar,hms}. The relevant superpotential
terms are:
\be\label{L-T}%
\frac{1}{\sqrt{2}}\bY^{ij}_{T} L_i T L_j + \frac{1}{\sqrt{2}}\la_1
H_1 T H_1 + \frac{1}{\sqrt{2}} \la_2 H_2 \bar{T} H_2 +  M_T T
\bar{T} + \mu H_2 H_1,
\ee%
where $i,j = e, \mu, \tau$ are family indices, $L_i$ are  the
$SU(2)_W$ lepton doublets and $H_1 (H_2)$ is the Higgs doublet with
hypercharge $Y=-1/2 (1/2)$. The matrix $\bY^{ij}_{T}$ is a  $3
\times 3$ symmetric matrix and $\la_{1,2}$ are dimensionless
unflavoured couplings. $M_T$ and $\mu$ denote the mass parameters of
the triplets and the Higgs doublets, respectively.

By decoupling the triplet states at the scale $M_T$, one obtains the
$d=5$ effective operator $\bY_\nu (L H_2)^2 /2 M_L$ where
$\bY_\nu/M_L$ is identified as follows:
\be\label{d5-T}
\frac{1}{M_L} \bY^{ij}_\nu = \frac{\la_2}{M_T} \bY^{ij}_T .
\ee
At the electroweak scale the Majorana neutrino mass matrix emerges
and is given by\footnote{The appearance of neutrino masses can be
also interpreted  in terms of non-vanishing vacuum expectation value
({\it vev}) induced by the EWSB on the scalar neutral state $T^0$,
{\it i.e.} $\langle T^0\rangle = \frac{v_2^2 \la_2}{M_T}$. In our
scenario the values of $\langle T^0\rangle$ will be smaller than
$10^{-3}~{\rm GeV}$, which is much below the upper bound of $\sim
2~{\rm GeV}$ inferred from the global fits of the electroweak
data~\cite{PDG}.}:
\be
\label{T-mass}%
{\bf m}^{ij}_\nu = \frac{v_2^2 }{M_L} \bY^{i j}_\nu=\frac{v_2^2
\la_2}{M_T} \bY^{i j}_T\,,
\ee
where $v_2= v\sin\beta =\langle H_2 \rangle ~(v=174~{\rm GeV})$. It
is worth to emphasize that the flavour structure of the matrix
$\bY_T$ is the same as that of $\bY_\nu$ and hence of the neutrino
mass $\bmm_\nu$.

Without loss of generality, we choose to work in the basis where the
charged-lepton Yukawa matrix $\bY_e$ is diagonal. Therefore, all the
information about low-energy lepton flavour violation is contained
in $\bY_\nu$ or $\bmm_\nu$: 
\be\label{mix}
\bmm_\nu = {\bf U}^\star\bmm^D_\nu {\bf
U}^\dagger\quad\;\;,\quad\;\;
\bmm^D_\nu = {\rm diag}(m_1, m_2, m_3)\,,
\ee
where $m_1, m_2, m_3$ are the neutrino mass eigenvalues and the
unitary lepton mixing matrix ${\bf U}$ can be written as
\bea\label{Um}%
{\bf U}&=& {\bf V}\cdot{\rm diag}\left({1,\rm e}^{ i\phi_1}, {\rm
e}^
{ i\phi_2}\right)\,,  \no\\
{\bf V}& = & \pmatrix{c_{12}c_{13} & s_{12}c_{13}
&s_{13}e^{-i\delta} \cr -s_{12}c_{23}-c_{12}s_{23}s_{13}e^{i\delta}
& c_{12}c_{23}-s_{12}s_{23}s_{13}e^{i\delta} &  s_{23}c_{13} \cr
s_{12}s_{23}-c_{12}c_{23}s_{13}e^{i\delta} & -c_{12}
s_{23}-s_{12}c_{23}s_{13}e^{i\delta} & c_{23}c_{13}}\,.
\eea
The mixing matrix {\bf V} is responsible for LFV and, in particular,
for neutrino oscillations. We have adopted the notation
$s_{ij}\equiv \sin\theta_{ij}$, $c_{ij}\equiv \cos\theta_{ij}$ for
the three mixing angles $\theta_{12}$, $\theta_{23}$ and
$\theta_{13}$, and denoted the ``Dirac'' and ``Majorana''
CP-violating phases by $\delta$ and $\phi_{1,2}$, respectively.

The relations (\ref{T-mass})  and (\ref{mix}) clearly show that the
high-energy structure of $\bY_T$ can be determined by the low-energy
neutrino parameters (taking also into account the RG effects on the
$d=5$ operator which, however, do not introduce unknown flavour
structures). The implications of such simple flavour structure
become dramatic when one considers LFV induced by RG effects in the
mass matrix $\bmm^2_{\tl}$ of the left-handed sleptons~\cite{ar}.
Assuming, for instance, flavour-blind SSB terms at the
grand-unification scale $M_G$ ($\bmm^2_\tl= m^2_0\unity$), the form
of the LFV entries is
\be\label{susy1}
  ( \bmm^{2 }_{\tilde{L}})_{ij} \propto
 m^2_0 (\bY^{\dagger}_T \bY_T)_{ij} {\log}\frac{M_G}{M_T} , ~~~~~~
i\neq j\,,
\ee
which, in terms of the neutrino parameters, read
\be\label{susy2}
(\bmm^{2 }_{\tilde{L}})_{ij}\propto m^2_0 \left(\frac{M_T}{\la_2
v^2_2}\right)^2 (\bmm^{\dagger}_\nu \bmm_\nu )_{ij}
{\log}\frac{M_G}{M_T}\sim m^2_0 \left(\frac{M_T}{\la_2
v^2_2}\right)^2
 \left[\bV (\bmm^{D }_\nu)^2 \bV^\dagger\right]_{ij}
{\log}\frac{M_G}{M_T}\,.
\ee
We note that the $L$-conserving combination $\bY^\dagger_T\bY_T
\propto \bV(\bmm^D_\nu) \bV^\dagger$ depends only on the neutrino
oscillations parameters, since the ``Majorana'' phases have been
absorbed. 
From here, one can derive the relative size of LFV among the
different leptonic family sectors~\cite{ar}:
\be\label{LFratio}
\frac{ ( \bmm^{2 }_{\tilde{L}})_{\tau \mu}} {( \bmm^{2
}_{\tilde{L}})_{\mu e} } \approx \frac{\left[\bV (\bmm^{D
}_\nu)^2\bV^\dagger\right]_{\tau \mu}} {\left[\bV (\bmm^{D
}_\nu)^2\bV^\dagger\right]_{\mu e}} \quad , \quad \frac{ ( \bmm^{2
}_{\tilde{L}})_{\tau e}} {( \bmm^{2 }_{\tilde{L}})_{\mu e} } \approx
\frac{\left[\bV (\bmm^{D }_\nu)^2\bV^\dagger\right]_{\tau e}}
{\left[\bV (\bmm^{D }_\nu)^2\bV^\dagger\right]_{\mu e}} .
\ee 
\emph{The above ratios depend only on the neutrino parameters which
can be measured in low-energy experiments}. This observation renders
the SUSY triplet seesaw mechanism much more predictive when compared
with the singlet one. Our scenario, therefore, constitutes a
concrete and simple realization of the so-called minimal lepton
flavour violation hypothesis which has been recently revived in the
literature~\cite{MFV1,MFV2}. In fact, relations like those shown in
Eq.~(\ref{LFratio}) cannot be obtained in the latter case due to an
ambiguity in the extraction of the high-energy neutrino parameters
from low-energy observables~\cite{DI}.

\section{Gauge and Yukawa mediated SUSY breaking scenario}
\label{GaugeYukawa}
A brief comment on the SSB pattern is now in order to motivate and
introduce the main idea of this paper. We recall that the assumption
of universality of the soft scalar masses at a high scale below $M_
{PL}$ may not be a justified one. As a matter of fact, flavour
universality at $M_{PL}$, arising from some gravity-mediated SUSY
breaking models~\cite{bim}, is likely to be spoiled by Yukawa
interactions in the energy range above
$M_G$~\cite{univers,univers2}. Therefore, the common lore of
adopting the universality condition at $M_G$ should be regarded as a
conservative approach to the issue of flavour violation (FV). In
this work, we discuss an alternative scenario which, in our opinion,
suggests a more motivated and predictive picture for the SSB pattern
at high scales. This is the consequence of the fact that the
triplet-exchange at the quantum level gives rise to the SSB
terms\footnote{In~\cite{cmrv} the authors discussed the finite
radiative contributions (arising from the decoupling of the triplet
states) relevant to the SSB trilinear couplings in connection with
the generation of the electric dipole moments.} of the MSSM,
\textit{i.e.}, the triplets play the role of SUSY breaking
messengers~\cite{af1}.

At first sight, the presence of extra $SU(2)_W$ triplet states at
intermediate energies could spoil the simple gauge coupling
unification, which can be achieved within the MSSM. However, this
can be remedied by invoking a grand unified theory (GUT) where the
triplets live in a complete GUT representation, in such a way that
gauge coupling unification can be preserved\footnote{This is not the
only possibility to maintain gauge coupling unification in the
presence of extra states at intermediate scales~\cite{gcu}.}. To
this purpose, we arrange a $SU(5)$ set-up where the $T$ ($\ov{T}$)
states fit into the 15 $(\ov{15})$ representation, $15 = S + T +Z$
with $S$, $T$ and $Z$ transforming as $S\sim (6,1,-\frac23), ~T \sim
(1,3,1),~ Z\sim (3,2,\frac16)$ under $SU(3) \times SU(2)_W \times
U(1)_Y$ (the $\ov{15}$ decomposition is obvious)\footnote{We find it
interesting that such supersymmetric $SU(5)$ with a 15,$\ov{15}$
pair may be realized in contexts based on string inspired
constructions~\cite{HS1,HS2}.}.

The SUSY breaking mechanism is parameterized by a gauge singlet
chiral supermultiplet $X=S_X+\theta \psi_X+\theta^2F_X$, whose
scalar $S_X$ and auxiliary $F_X$ components are assumed to acquire a
{\it vev} through some unspecified dynamics in the secluded sector
($\psi_X$ is one of the goldstino components). In order to prevent
the tree-level generation of SSB terms in the observable sector, the
couplings of $X$ to the MSSM fields must be forbidden. At the same
time, $X$ must couple with the 15 and $\ov{15}$ fields in order to
trigger SUSY breaking in the messenger sector. Both these
requirements can be achieved by, \eg, imposing that the $SU(5)$
model conserves the combination of baryon and lepton number $B-L=Q +
\frac45 Y$, where $Y$ are the hypercharges and
\be
Q_{10}= \frac15,\,~~~~ Q_{\bar{5}} = - \frac35,\, ~~~~ Q_{{5}_H
(\bar{5}_H )} =-\frac25 \left(\frac25\right), \,~~~~ Q_{15} =
\frac65, \,~~~~ Q_{\ov{15}}= \frac45, \,~~~~Q_{X} = -2.
\ee
The $SU(5)$ matter multiplets are understood as $\bar5  = (d^c, L)$,
$10 =(u^c,e^c,Q)$ and the Higgs doublets fit with their coloured
partners $t$ and $\bar{t}$, like ${5}_H = (t, H_2), \bar{5}_H =
(\bar{t}, H_1)$. Given this, the relevant superpotential
terms~\cite{af1}, consistent with the $SU(5)$ and $B-L$ symmetries,
are
\bea \!\!W_{SU(5)}&\!\!\!\!\!\!=\! \!\!\!\!\!\!&\!
\frac{1}{\sqrt2}(\bY_{15} \bar5~ 15 ~\bar5 +
 \la {5}_H ~\ov{15}~ {5}_H)
 + \bY_5 \bar5 ~ \bar5_H 10
+ \bY_{10} 10 ~10 ~5_H +
M_5  5_H~\bar5_H  + \xi X 15 ~ \ov{15}\,. \no \\
&& \label{su5}
\eea
%
\begin{figure}
\vglue -1.cm
\begin{center}
\includegraphics[width=12.0cm]{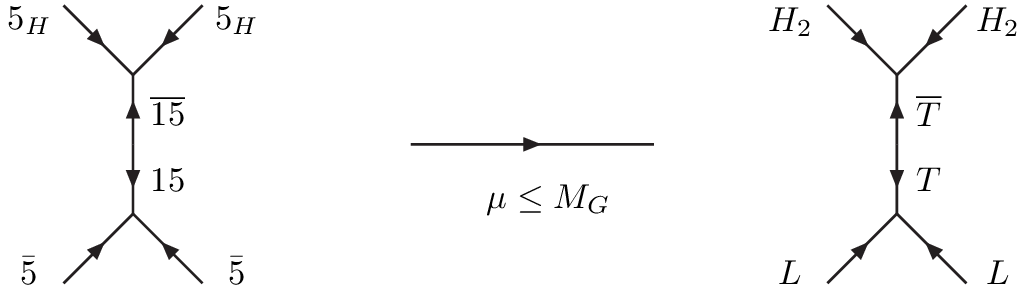}
\end{center}
\captions{\small The $d=5$ supergraph from the $15, \ov{15}$
exchange in the $SU(5)$ symmetric phase (left). Once the coloured
states $t \subset 5_H$ have been splitted from the Higgs doublet
partners $H_2$ and decoupled at the scale $M_G$, there is only the
supergraph (right) with $T, \bar{T}$ exchange generating the $L$
violating $d=5$ effective superpotential operator $(L H_2)^2$ .}
\label{f1}
\end{figure}
The form of $W_{SU(5)}$ makes explicit the fact that, thanks to the
coupling with $X$, the $15$ and $\ov{15}$ states act as {\it
messengers} of both $B-L$ and SUSY breaking to the MSSM observable
sector. Namely, while $\langle S_X\rangle$ only breaks $B-L$,
$\langle F_X\rangle$ breaks both SUSY and $B-L$. These effects are
tracked by the superpotential mass term $M_{15} 15~\ov{15}$, where
$M_{15} = \xi \langle S_X\rangle$, and the bilinear SSB term
$-B_{15} M_{15}15~\ov{15}$, with $ B_{15} M_{15}= - \xi \langle
F_X\rangle$. Once $SU(5)$ is broken to the SM group\footnote{ For
the sake of brevity, we have omitted in the $SU(5)$ invariant
superpotential (\ref{su5}) other terms, as those involving the
adjoint 24 representation responsible for the $SU(5)$ breaking.} we
find, below the GUT scale $M_G$~\cite{ar},
\bea
W& =& W_0 + W_T + W_{S,Z}  \no \\
W_0 & = & \bY_e  e^c H_1  L
+\bY_d d^c H_1  Q + \bY_u  u^c Q  H_2  + \mu H_2 H_1 \no \\
W_T&=& \frac{1}{\sqrt{2}}(\bY_{T} L T L  + \la H_2 \bar{T} H_2) +
  M_T T \bar{T} \label{WT} \no \\
 W_{S,Z} & =& \frac{1}{\sqrt{2}}\bY_S
d^c S d^c + \bY_Z  d^c  Z L + M_Z Z\bar{Z}+M_S S\bar{S} .
\label{su5b}
\eea%
Here, $W_0$ denotes  the MSSM superpotential where the standard
notation for the supermultiplets is understood. The terms relevant
for neutrino mass generation [cf.~(\ref{L-T})] are contained in
$W_T$ while the couplings and masses of the coloured fragments $S$
and $Z$ are included in $W_{S,Z}$. From the comparison of
Eqs.~(\ref{su5b}) and (\ref{L-T}), we observe that the $B-L$
invariance forbids the coupling $\la_1 H_1 T H_1$ (which is not
relevant for neutrino masses) leaving only the term proportional to
$\la\equiv\la_2$. Consequently, the number of independent real
parameters in $W_T$ is reduced to eleven: $M_T$, $\la$ and the nine
from the symmetric matrix $\bY_T$.

\begin{figure}
\begin{center}
\includegraphics[width=15.5cm]{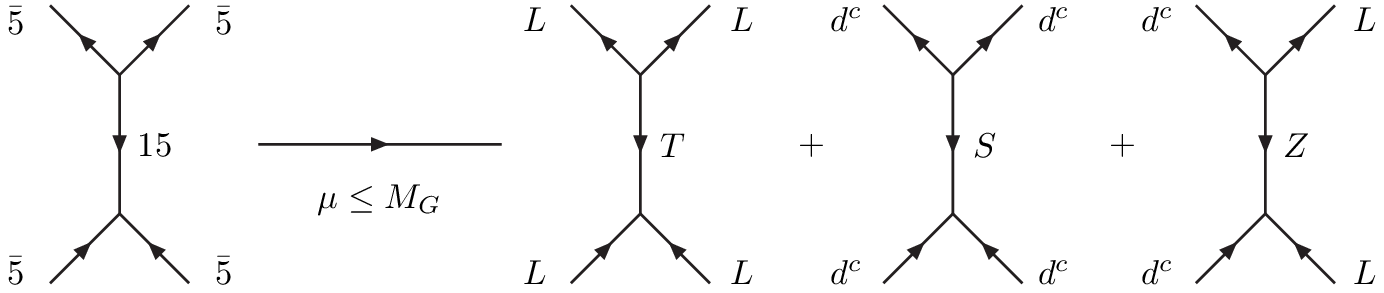}
\end{center}
\captions{\sf\small The $d=6$ supergraph from the $15$ exchange in
the $SU(5)$ symmetric phase (left). In the $SU(5)$ broken phase
($\mu \leq M_G$), the latter splits into the remaining three
supergraphes (right) which generate  the $B-L$ invariant $d=6$ K{\"
a}hler operators $(L L) ({L}^\dag {L}^\dag)$, $(d^c d^c) (d^{c\dag}
d^{c\dag})$   and $(d^c L) (d^{c\dag} {L}^\dag)$ at low-energy.}
\label{f2}
\end{figure}

Notice that we have relaxed the strict $SU(5)$ symmetry relations
for the Yukawa interactions and the mass term $M_5$ by assuming
$SU(5)$ breaking effects (due to insertions of the
24-representation), which are necessary to correct the relation
$\bY_e= \bY^T_d$~\cite{yeyd} and to solve the doublet-triplet
splitting problem~\cite{doubtrip}. 
Thus, beneath the scale $M_G$, the coloured partners $t$ and
$\bar{t}$ are considered to be decoupled in order to adequately
suppress dangerous $d=5$ baryon number violating operators. The only
$d=5$ operator, generated by the exchange of the 15-fragments, is
the $L$ violating neutrino mass operator (see Fig.~\ref{f1}).
Instead, Fig.~\ref{f2} shows that additional $d=6$ operators arise
from the exchange of $T, S$ and $Z$~\cite{ar}. However, these are
$B-L$ conserving and, being suppressed by $M_T^2$, are irrelevant
for low-energy phenomenology. Hence, the presence of $S$ and $Z$
does not introduce new sources of baryon-number violation which
could speed up the decay of the proton.

As mentioned above, the $B-L$ conserving superpotential of
Eq.~(\ref{su5}) contains the $M_5$-term from which the
$\mu$-parameter emerges in $W_0$. This mass parameter is not
predicted by our model and will be determined through the
requirement of EWSB (notice that the coupling $X H_1 H_2$ is
forbidden by $B-L$).

Also the coupling $\xi$ in Eq.~(\ref{su5}) could include
$24$-insertions, therefore allowing different masses for the
$15$-components. For simplicity, we consider the minimal case which
implies a common mass $M_S=M_Z=M_T=M_{15}$ at the GUT scale.
The Yukawa-unification condition
\be
\label{y15}%
\bY_S = \bY_T= \bY_Z = \bY_{15}\,,
\ee
can either hold or not at $M_G$, depending on the type and size of
the $SU(5)$ breaking effects. In this respect, we can discuss the
following two extreme scenarios:
\begin{description}
\item
{\bf (A)}~ Eq.~(\ref{y15}) holds at the GUT scale. As a consequence,
flavour violation is extended to all the couplings $\bY_S$, $\bY_T$
and $\bY_Z$, originating FV effects both in the lepton and quark
sector. In this work we shall focus on this general case.
\item
{\bf (B)}~ The states $S$ and $Z$ have negligible Yukawa couplings
[Eq.~(\ref{y15}) does not hold because of $SU(5)$ breaking]. In this
case, FV is confined to the lepton sector since the couplings in
$\bY_T$ are the only ones which exhibit a non-trivial flavour
structure. We have already studied this scenario in Ref.~\cite{af1}.
\end{description}

Beneath the messenger scale $M_T$, the particle content of our model
is that of the MSSM and so, the superpotential reduces to the sum of
the MSSM terms in $W_0$ and the effective neutrino mass operator
\be
\label{wmssm}
W_{<M_T} = W_0 + \frac{\la \bY_T}{2M_T}  (L H_2) (L
H_2) .
\ee
As mentioned above, the {\it vev} of $F_X$ induces the only
tree-level SSB terms which, below the GUT scale, read ($B_T=B_{15}$
at $M_G$)
\be
\label{ssb1}%
-{\cal L}^{>M_T}_{{\rm soft}}= B_{T} M_T (T\bar{T} +  Z\bar{Z}+
S\bar{S}) +{\rm h.c.}\,.
\ee
Such terms remove the mass degeneracy between the scalar and
fermionic messenger components. To avoid tachyonic scalar messengers
we require that $ \xi \langle F_X\rangle  < M_T^2$ (or $B_T < M_T$).
At the tree-level, the ordinary MSSM supermultiplets are degenerate
as they do not couple to the superfield $X$. Nevertheless, in the
presence of the bilinear terms given in Eq.~(\ref{ssb1}), the mass
splitting is radiatively generated at the scale $M_T$ through the
gauge and Yukawa interactions between the messenger states $S,\,T$
and $Z$ and the ordinary MSSM fields. Our scenario can be,
therefore, regarded as a gauge and Yukawa mediated SUSY breaking
realization of the triplet seesaw mechanism. This will become clear
in the next section where we present the complete MSSM SSB
lagrangian.
%
\section{The SSB mass parameters}
\label{SSBmass}

We now discuss the SSB terms of the MSSM lagrangian ${\cal L}^{\rm
MSSM}_{\rm soft}$ which are generated at the decoupling scale of the
heavy states $S,T$ and $Z$. Below $M_T$, the most general ${\cal
L}^{\rm MSSM}_{\rm soft}$ can be written as:
\bea
\label{Lsoft}%
-{\cal L}^{\rm MSSM}_{\rm soft}\!\!\! \!\!&=&\!\!\!\!\! \tl^\dagger
\bmm^2_{\tl} \tl +
 \te \bmm^2_{\te }  \te^\dagger+
\tq^\dagger \bmm^2_{\tq} \tq+ \td  \bmm^2_{\td} \td^\dagger + \tu
\bmm^2_{\tu} \tu^\dagger+
 m^2_{H_1} H^\dagger_1 H_1+ m^2_{H_2} H^\dagger_2 H_2  \nonumber \\
&& + ( H_1 \te  \bA_{e } \tl  +  H_1 \td \bA_{d } \tq+
 \tu  \bA_{u } \tq H_2 +
\frac12 M_a \la_a \la_a + B_H\mu H_2 H_1 + {\mbox h.c.})\,,
\eea
where we have adopted the standard notation for the slepton, squark
and Higgs soft masses, the trilinear couplings $\bA_{e,d,u}$, the
gaugino masses $M_a$ and the Higgs bilinear term $B_H$.

At the mass scale $M_T$, one-loop finite contributions are
generated\footnote{Since the RGE-induced splitting on the masses of
the 15-fragments is small, we decouple all the components at the
common threshold $M_T$.} for $\bA_{e,d,u}$, $M_a$ and $B_H$.
\begin{figure}
\begin{center}
\begin{tabular}{cc}
\includegraphics[width=6.2cm]{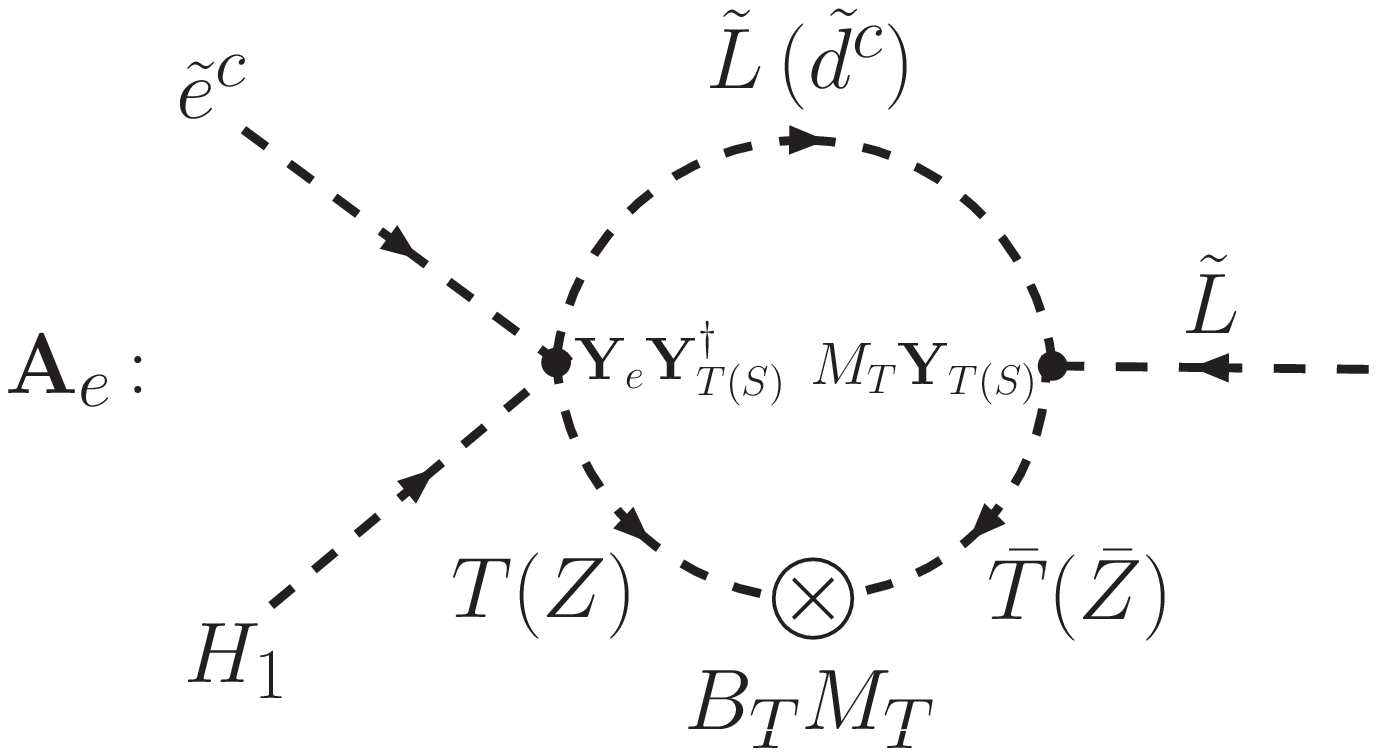}
&\includegraphics[width=6.2cm]{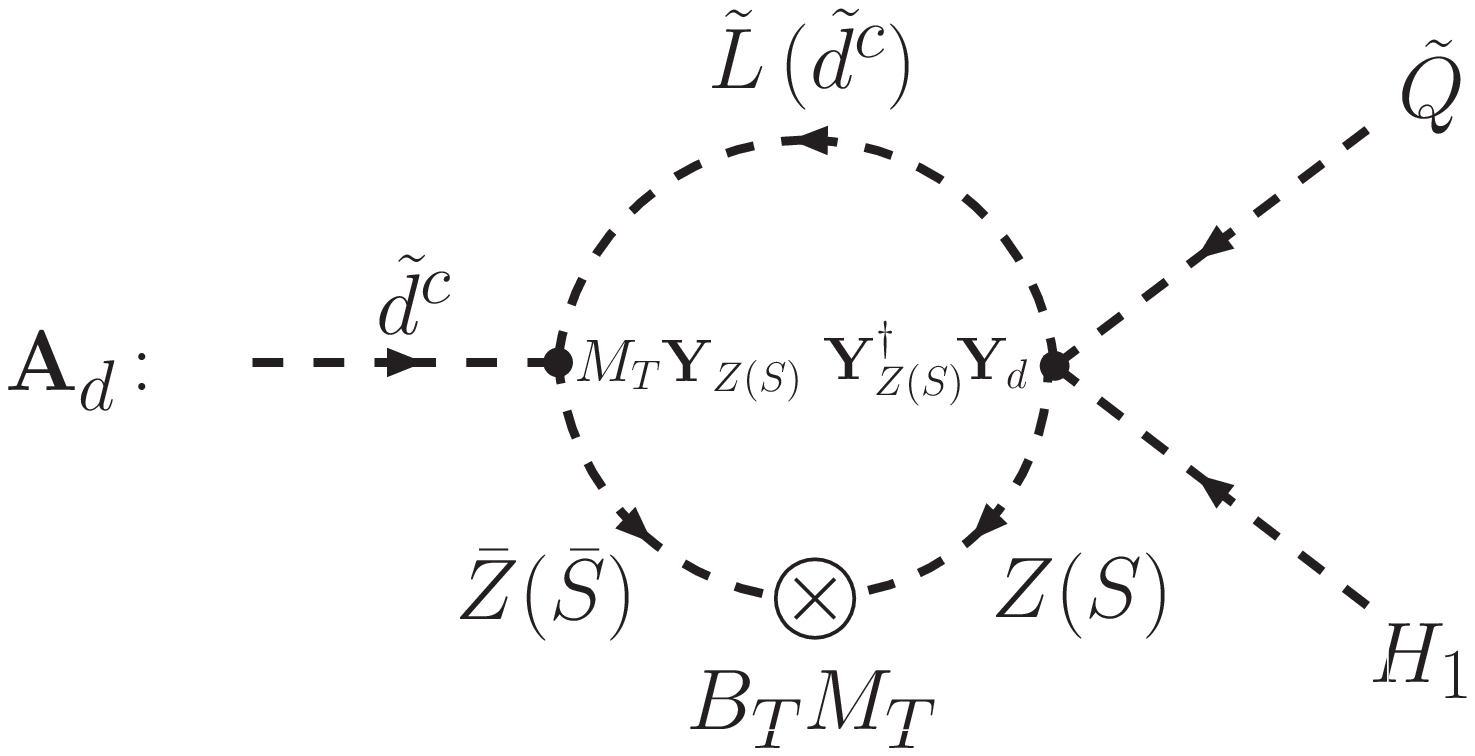}\vspace*{0.2cm}\\
 &\hspace*{-7cm}\includegraphics[width=6.2cm]{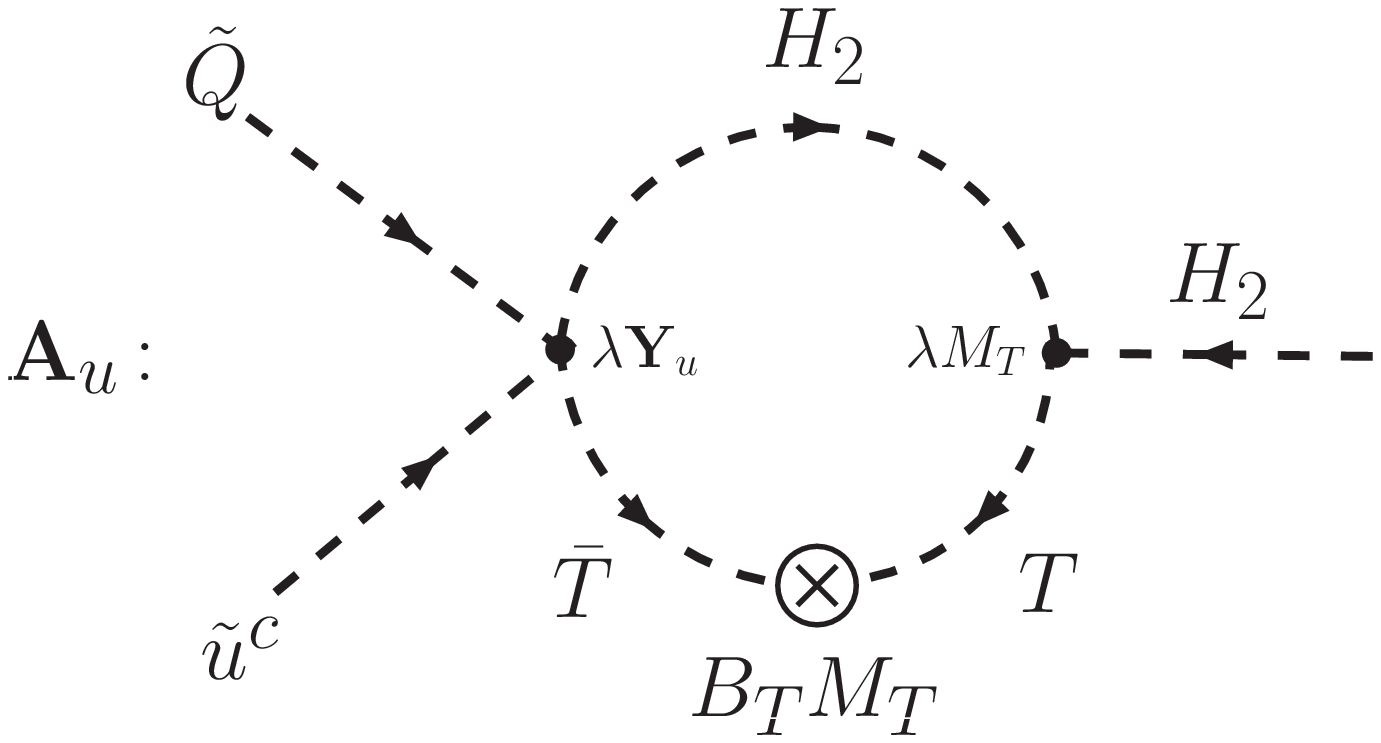} \\
\includegraphics[width=6.2cm]{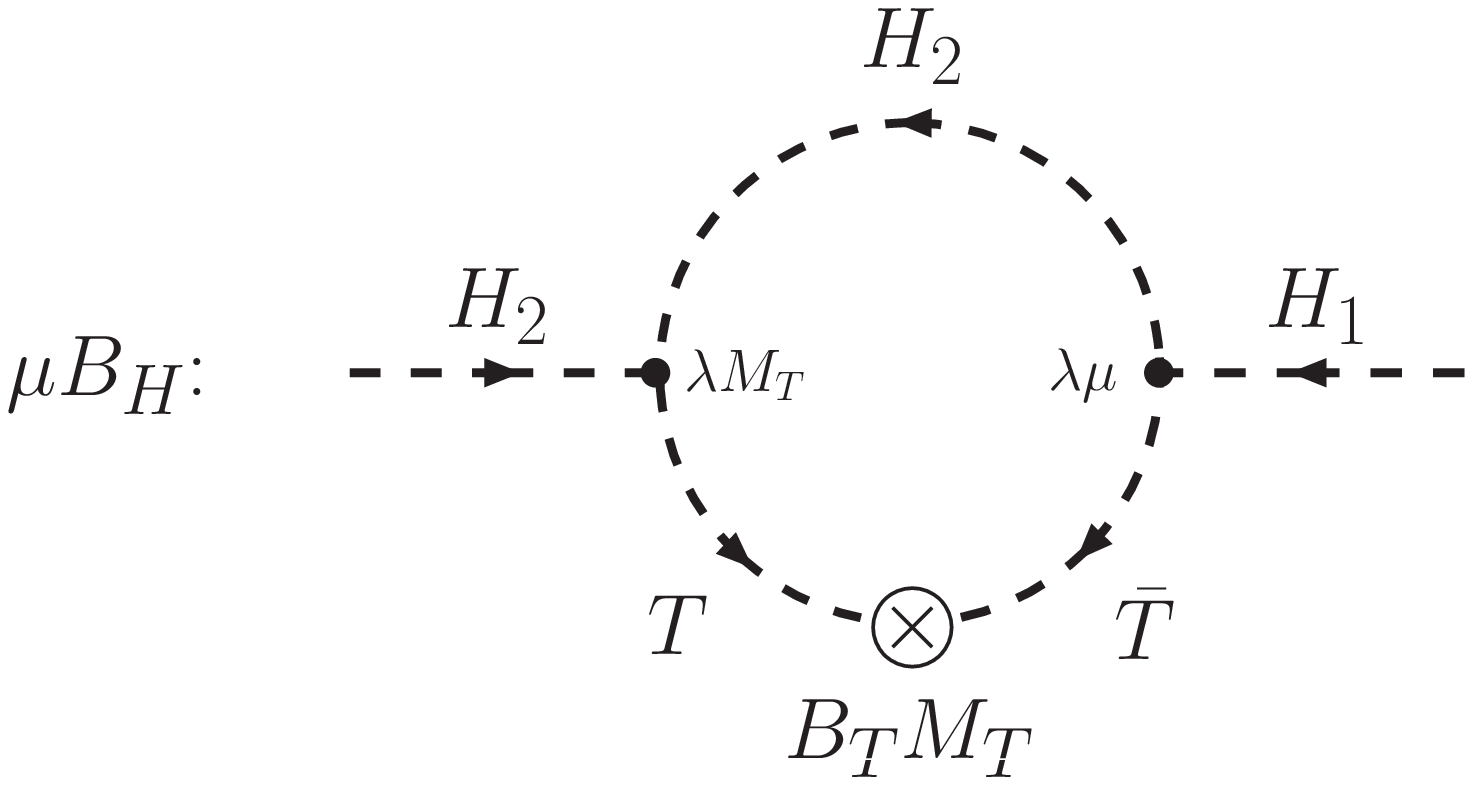} &\hspace{0.5cm}
\includegraphics[width=6.2cm]{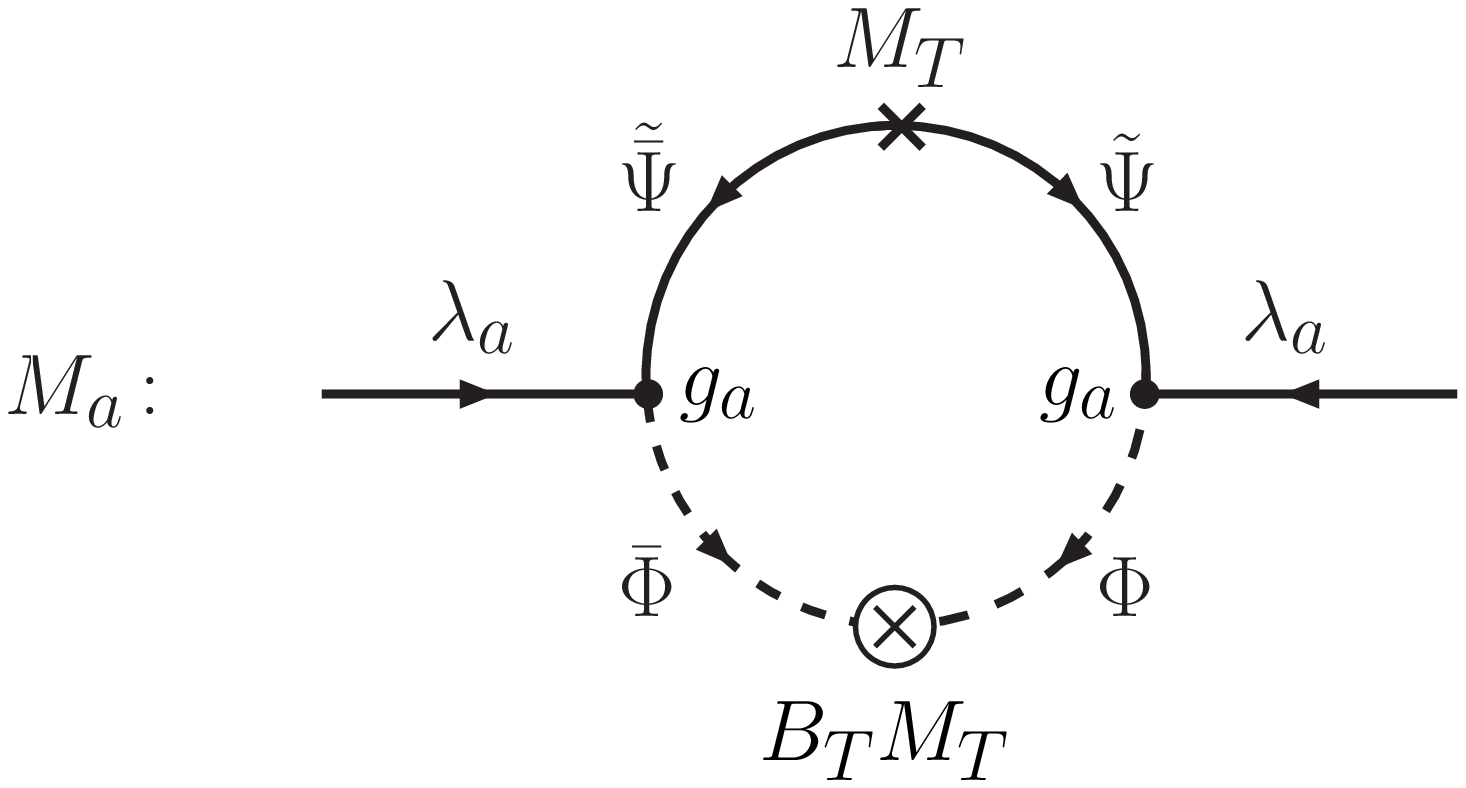}
\end{tabular}
\end{center}
\vspace*{-0.8cm} \captions{\small One-loop diagrams for the
trilinear terms ${\rm A}_e$, ${\rm A}_d$ and ${\rm A}_u$ (first and
second rows) and the bilinear Higgs term $\mu B_H$ and gaugino
masses $M_a$ (third row). The fields $\Phi(\bar{\Phi})$ and
$\tilde{\Psi}(\tilde{\bar{\Psi}})$ denote the scalar and fermionic
components of the $T(\bar{T}),Z(\bar{Z}),S(\bar{S})$ superfields in
such a way that $\Phi=(T,Z,S),(T,Z),(S,Z)$ and
$\tilde{\Psi}=(\tilde{T},\tilde{Z},\tilde{S}),(\tilde{T},\tilde{Z}),(\tilde{S},\tilde{Z})$
for $a=1,2,3$, respectively (the assignments for $\bar{\Phi}$ and
$\tilde{\bar{\Psi}}$ are obvious).} \label{f3}
\end{figure}
\begin{figure}
\begin{center}
\begin{tabular}{cc}
\includegraphics[width=6.0cm]{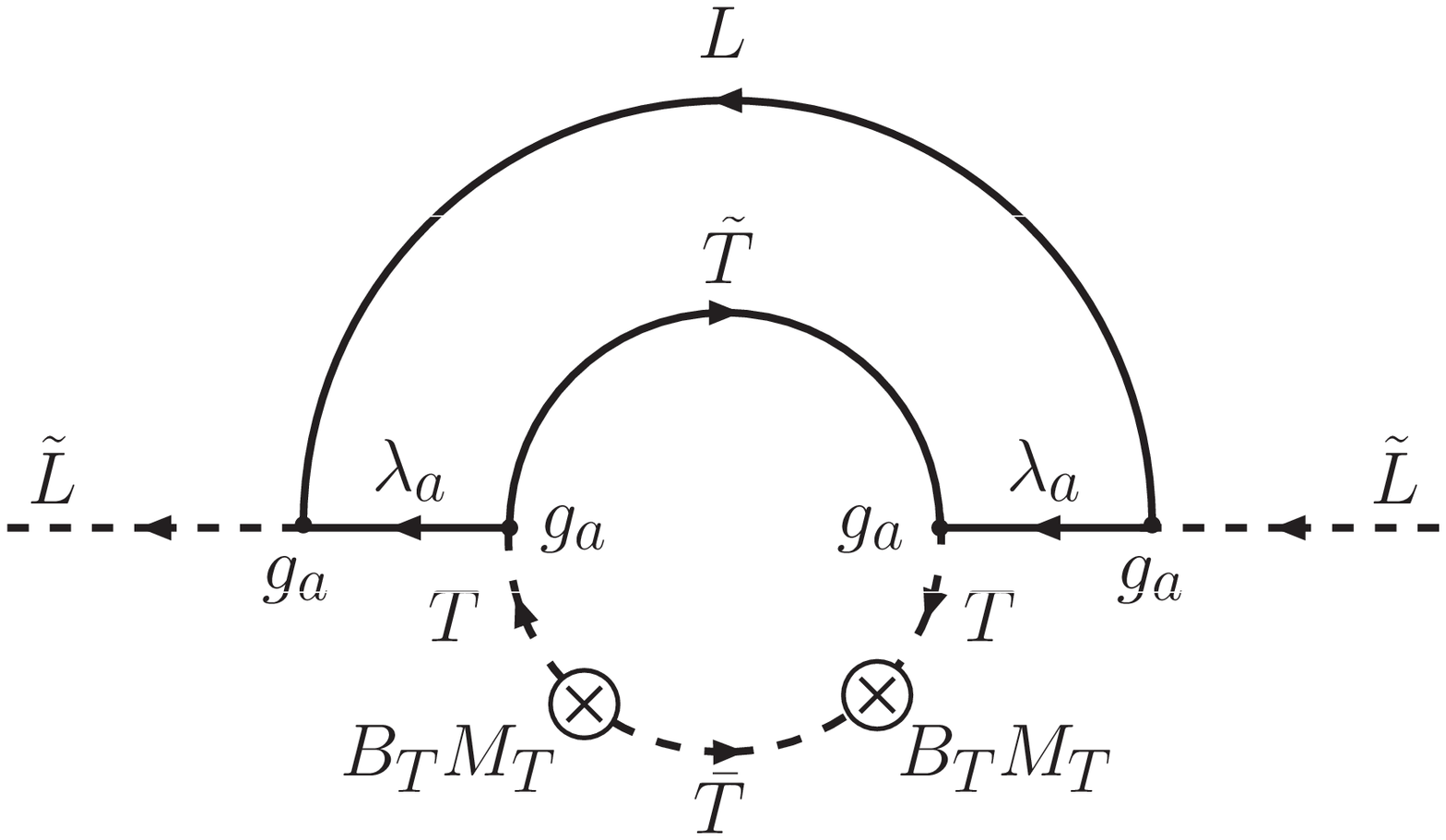}
&\hspace{0.8cm}\includegraphics[width=6.0cm]{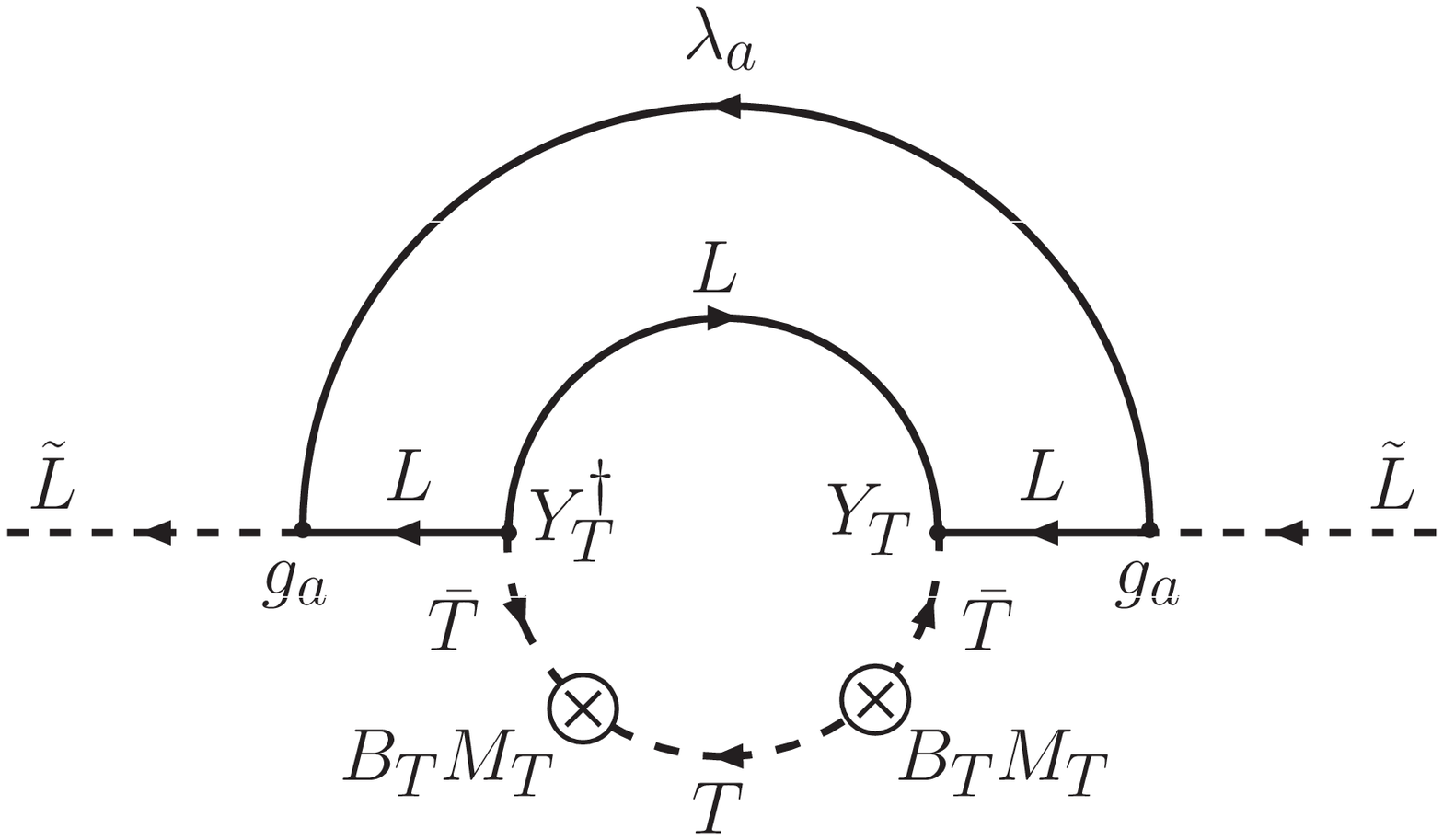}\vspace*{-0.3cm}\\
&\hspace{-6cm}\includegraphics[width=6.0cm]{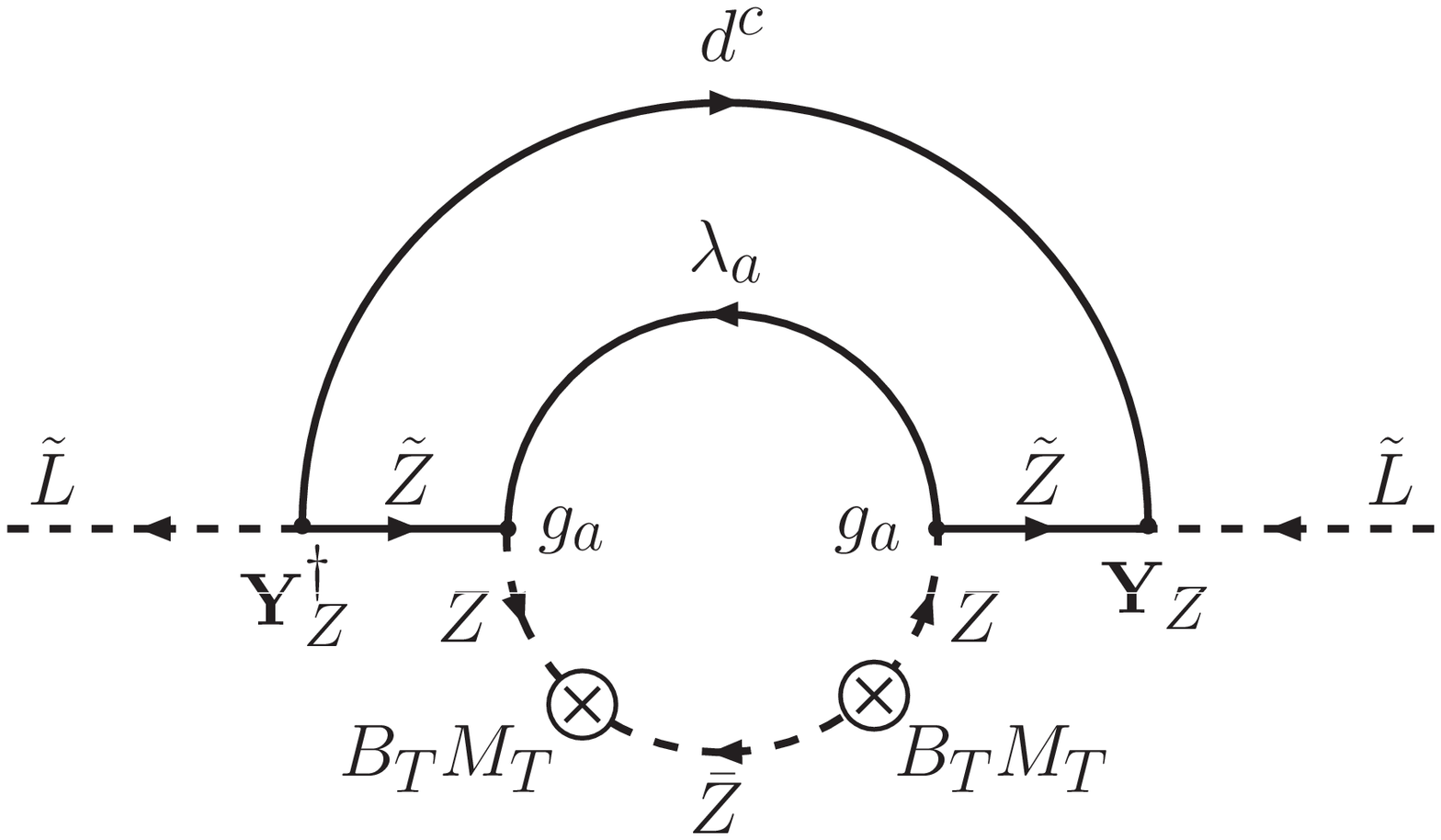}
\end{tabular}
\end{center}
\label{f4} \vspace*{-0.8cm}
\captions{\small Examples of two-loop diagrams which contribute to
${\rm m}_{\tilde{L}}^2$. The upper diagrams generate contributions
proportional to $g_{1,2}^4$ (left) and to $g_{1,2}^2{\rm
Y}_T^\dag{\rm Y}_T$ (right). Instead, the lower diagram leads to a
term proportional $g_a^2 \bY_Z^\dag \bY_Z$.}
\end{figure}
\noindent In Fig.~\ref{f3} we draw the one-loop diagrams for $\bA_e,
\bA_d$ and $\bA_u$ (first and second rows), $\mu B_H$ and $M_a$
(third row). The evaluation of these diagrams yields:
\bea
{\bA}_e & = & \frac{3 B_T}{16 \pi^2}
\bY_e (\bY^\dagger_T \bY_T + \bY^\dagger_Z \bY_Z)
\,,\nonumber \\
{\bA}_u &=&\frac{3B_T}{16 \pi^2} |\la|^2\bY_u  \,,\nonumber \\
{\bA}_d &=&\frac{ 2 B_T}{16 \pi^2} (\bY_Z\bY^\dagger_Z  + 2
\bY_S\bY^\dagger_S) \bY_d
\,, \nonumber \\
M_a &= &  \frac{7 B_T }{16 \pi^2}\,g_a^2\,,\nonumber \\
B_H & =& \frac{3 B_T }{16 \pi^2}\, |\la|^2 .
\label{soft1}%
\eea
In these expressions, $g_a (a=1,2,3)$ indicates the GUT normalized
gauge coupling constants ($g_1=g_2=g_3$ at the unification scale)
and the factor $N=7$ in the r.h.s. of the gaugino masses $M_a$
corresponds to twice the Dynkin index of the 15-representation. A
noteworthy feature of the above equations is that both the $A$-terms
and the Higgs doublet bilinear parameter $B_H$ are generated in our
scenario\footnote{The SSB terms $M_a, B_H$ and the trilinear
couplings require interactions which violate the $U(1)_R$ symmetry.
In this context, the messenger bilinear terms (\ref{ssb1}) are those
responsible for the $U(1)_R$ breaking.}, once the Yukawa couplings
and the $\mu$ parameter are present in the
superpotential\footnote{Although the $\mu$ parameter is not
predicted by our model, we nevertheless assume that whatever
mechanism generates $\mu$, it does not generate $B_H$.}.This is in
contrast with the minimal GMSB realizations, where only the gaugino
masses emerge at one-loop~\cite{gmall,GR-PR}. We remark that the
triplet states, being $SU(3)$ singlets, would not communicate SUSY
breaking to the gluino. This is one more motivation, besides the
aforementioned gauge coupling unification requirement, to consider
the GUT extension of the triplet seesaw with the 15 representation.

Regarding the SSB squared scalar masses, the leading
${\cal O}(F^2_X/M^2_T)={\cal O}(B^2_T)$ contributions
vanish at the one-loop level
due to a cancelation among the different terms~\cite{dns}.
For example, in the case of the soft masses $\bmm^2_{\tl}$, the
one-loop diagrams driven by the exchange of the $F$-component of $T$
cancel against those from the exchange of the $F$-component of $L$.
Non-vanishing ${\cal O}(F^2_X/M^2_T)$
contributions for the SSB squared scalar masses arise  at
the two-loop level\footnote{These  ${\cal O}(F^2_X/M^2_T)$ two-loop
contributions dominate over the ${\cal O}(F^4_X/M^6_T) ={\cal O}(B^4_T/M^2_T)$
one-loop ones for $M_T > (4\pi Y_T/g^2) B_T$.
In the following such a hierarchy is
fulfilled.}. They are finite and can be evaluated either by
diagrammatic computations or by means of a generalization of the
wave function renormalization method proposed in Ref.~\cite{GR,CP}.
For the sake of illustration, we have depicted in Fig.~\ref{f4} some
representative two-loop diagrams which generate contributions to
$\bmm^2_\tl$. The first diagram gives flavour-blind terms
proportional to $g^4_{1,2}$. In turn, the remaining two diagrams
generate LFV contributions proportional to $ g^2_{1,2}\bY^\dagger_T
\bY_T$ and to $ g^2_{1,2,3}\bY^\dagger_Z \bY_Z$.
In Appendix A we revisit the wave function renormalization approach
and provide general formulas to extract all the SSB terms just by
knowing the one-loop anomalous dimension matrices of the different
fields above and below the messenger scale $M_T$. Our computation
leads to the following result:
\bea
\label{soft2}%
\!\!\!\!\bmm^2_{\tl}& =& \left(\frac{|B_T|}{16 \pi^2}\right)^2 \left[
\frac{21}{10} g^4_1 + \frac{21}{2} g^4_2 - \left(\frac{27}{5} g^2_1
+21 g^2_2\right)\!\bY^\dagger_T\bY_T - \left(\frac{21}{15} g^2_1 +9
g^2_2 + 16 g^2_3\right)\!\bY^\dagger_Z\bY_Z\right. \no \\
&& \! \left. + 18  (\bY^\dagger_T\bY_T)^2 +15 (\bY^\dagger_Z
\bY_Z)^2+ 3{\rm Tr}(\bY^\dagger_T\bY_T) \bY^\dagger_T\bY_T + 12
\bY^\dagger_Z \bY_S \bY^\dagger_S \bY_Z
\right. \no \\
&& \! \left. + 3 {\rm Tr}(\bY^\dagger_Z\bY_Z)\bY^\dagger_Z \bY_Z + 9
\bY^\dagger_T \bY^T_Z \bY^*_Z \bY_T
+9 (\bY^\dagger_T \bY_T\bY^\dagger_Z\bY_Z +{\rm h.c.}) \right. \no \\
&& \!\left. + 3\bY^\dagger_T \bY^T_e \bY^*_e \bY_T +6 \bY^\dagger_Z
\bY_d \bY^\dagger_d \bY_Z
\right] \no \\
\!\! \bmm^2_{\te}& =& \left(\frac{|B_T|}{16 \pi^2}\right)^2
\left[\frac{42}{5} g^4_1 - 6 \bY_e( \bY^\dagger_T\bY_T+
\bY^\dagger_Z\bY_Z )\bY^\dagger_e \right]
\no \\
\!\!\bmm^2_{\tq}& = &\left(\frac{|B_T|}{16 \pi^2}\right)^2
\left[\frac{7}{30} g^4_1 + \frac{21}{2} g^4_2 + \frac{56}{3} g^4_3
-2 \bY^\dagger_d (\bY_Z\bY^\dagger_Z + 2 \bY_S\bY^\dagger_S)\bY_d -
3  |\la|^2 \bY^\dagger_u \bY_u
\right]\no \\
\!\!\bmm^2_{\tu} &= &\left(\frac{|B_T|}{16 \pi^2}\right)^2
\left[\frac{56}{15}
g^4_1 +\frac{56}{3} g^4_3- 6  |\la|^2 \bY_u \bY^\dagger_u\right]\no \\
\!\!\bmm^2_{\td}& = &\left(\frac{|B_T|}{16 \pi^2}\right)^2
\left[\frac{14}{15} g^4_1 +\frac{56}{3} g^4_3 - \left(\frac{16}{5} g^2_1
+48 g^2_3\right)\bY_S\bY^\dagger_S  - \left(\frac{14}{15}g^2_1+ 6 g^2_2
+\frac{32}{3}
g^2_3\right)\bY_Z\bY^\dagger_Z \right. \no \\
&& \! \left. +32 (\bY_S \bY^\dagger_S)^2 +4 {\rm
Tr}(\bY_S\bY^\dagger_S) \bY_S \bY^\dagger_S + 8 \bY_S \bY^*_Z\bY^T_Z
\bY^\dagger_S + 6\bY_Z \bY^\dagger_T
\bY_T\bY^\dagger_Z\right. \no \\
&& \! \left. +8(\bY_S \bY^\dagger_S\bY_Z \bY^\dagger_Z +{\rm h.c.})
+ 10 (\bY_Z\bY^\dagger_Z)^2
+ 2 {\rm Tr}(\bY_Z\bY^\dagger_Z)\bY_Z\bY^\dagger_Z \right. \no \\
&& \! \left. +8 \bY_S\bY_d^*\bY^T_d\bY^\dagger_S+2\bY_Z\bY_e^\dagger
\bY_e\bY^\dagger_Z\right]\no \\
%
%
\!\! m^2_{H_1} &= &\left(\frac{|B_T|}{16 \pi^2}\right)^2\left[
 \frac{21}{10} g^4_1 +   \frac{21}{2}  g^4_2
-3{\rm Tr}[(\bY^\dagger_T\bY_T
+\bY^\dagger_Z\bY_Z)\bY^\dagger_e\bY_e]
\right. \no \\
&& \! \left. -6{\rm Tr}[(\bY_Z\bY^\dagger_Z
+2\bY^\dagger_S\bY_S)\bY^\dagger_d\bY_d]
\right] \no \\
\!\! m^2_{H_2}& = &\left(\frac{|B_T|}{16 \pi^2}\right)^2\left[
\frac{21}{10} g^4_1 +   \frac{21}{2}g^4_2 - \left(\frac{27}{5} g^2_1
+21 g^2_2\right)|\la|^2 + 9 |\la|^2 {\rm Tr}(\bY_u\bY^\dagger_u) +
21 |\la|^4 \right].
\eea
One can immediately recognize that the diagonal gauge-mediated
contributions follow the expected form $\propto 2 k_a C^f_a N g^4_a
\, (k_1=3/5, k_2 =k_3 =1)$, where $C^f_a$ is the quadratic Casimir
of the $f$-particle [$C^f_1 =Y^2_f =(Q_f -T^3_f)^2$ and $C =
\frac{n^2 -1}{2n}$ for the fundamental representation of $SU(n)$].

The above expressions hold at the decoupling scale $M_T$ and are,
therefore, meant as boundary conditions for the SSB parameters which
then undergo (MSSM) RG running down to the low-energy scale
$\mu_{s}$. All the soft masses have the same scaling property
$m_{soft} \sim B_T/(16 \pi^2)$ which, on the basis of the
naturalness requirement $m_{soft} \sim {\cal O}(10^2-10^3~{\rm
GeV})$, implies that $B_T \gsim 10~{\rm TeV}$. In principle, the SSB
parameters also receive gravity mediated contributions of order the
gravitino mass $m_{3/2}\sim F/M_{PL}$ ($F^2= \langle |F_X|^2\rangle
+ \ldots$ is the sum of $F$-terms in the secluded sector). We assume
such contributions to be negligible, which is the case if $M_T \ll ~
10^{16}~{\rm GeV}\,\xi \langle F_X\rangle/F$. This also implies that
the gravitino is lighter than the MSSM sparticles.

It is apparent from  Eqs.~(\ref{soft1}) and (\ref{soft2}) (and
Figs.~\ref{f3} and \ref{f4}) that the gauge interactions participate
in the mediation of SUSY breaking in the gaugino and soft scalar
masses. In the particular case of the sfermion mass matrices, the
associated terms constitute the flavour blind contribution. Exactly
the same situation occurs in pure GMSB models where flavour
violation is automatically suppressed if the SUSY-breaking mediation
scale is lower than the flavour or GUT scale~\cite{gmall,GR-PR,bm}.
In our case, SUSY breaking is also mediated by the Yukawa
interactions $\bY_S, \bY_T, \bY_Z$ and $\la$, giving rise to the
bilinear parameter $B_H$, the trilinear couplings $\bA_{e,d,u}$ and
to additional contributions to the Higgs scalar masses and the
sfermion mass matrices\footnote{
Other examples of Yukawa mediated SUSY breaking can be
found {\it e.g.} in \cite{dns,DS,CP}.}.
In this way, FV is transmitted to the SSB
mass parameters. Since the origin of FV originally comes from the
couplings $\bY_{15}$ (felt by the $\bar{5}$ supermultiplets), one
expects that, due to the condition (\ref{y15}), flavour violation is
inherited with comparable size by the $\tl$ and $\td$ SSB parameters
$\bA_e$,$\bA_d$, $\bmm^2_{\tl}$ and $\bmm^2_{\td}$. We emphasize
that the FV entries of \eg, $(\bmm^2_{\tilde{L}})_{ij}$ $(i\neq j)$
show up as finite radiative contributions induced by $B_T$ at $M_T$,
and they are not significantly modified by the running evolution to
low-energy. This is different from a previous work~\cite{ar} where a
common SSB scalar mass $m_0 \sim {\cal O}(100~{\rm GeV})$ was
assumed at $M_G$ and the dominant LFV contributions to $\bmm^2_\tl$
were generated by RG evolution from $M_G$ down to the decoupling
scale $M_T$ [see Eq.~(\ref{susy1})]. In such a case, finite
contributions like those in Eqs.~(\ref{soft1}) and (\ref{soft2})
also emerge at $M_T$. Nevertheless, they are subleading with respect
to the RG corrections, since $B_T$ is of the same order as $m_0$.
Instead, in the present picture, there is a hierarchy between the
SSB parameter $B_T$ and the remaining ones [see Eqs.~(\ref{soft1})
and (\ref{soft2})], $B^2_T \gg (B_T g^2/16 \pi^2)^2  \sim
m^2_0$~\cite{af1}.

\section{Flavour structure in  $\bmm^2_\tl$ and $\bmm^2_\td$ from neutrino parameters}
\label{Flavour Structure}
In this section we discuss in detail the features of the flavour
structure which emerges in the mass matrices $\bmm^2_\tl$ and
$\bmm^2_\td$, and in the trilinear couplings $\bA_e$ and $\bA_d$.
These parameters depend on the Yukawa couplings $\bY_T, \bY_S,
\bY_Z$. The matrix $\bY_T$ is determined at $M_T$ according to the
matching expressed by Eq.~(\ref{T-mass}) {\it i.e.},
\be
\label{match}%
\bY_T= \frac{M_T}{\la v^2_2} \,\bmm_\nu\,,
\ee
where
the effective neutrino mass matrix $\bmm_\nu$
has undergone  the MSSM RG running from low-energy to $M_T$. The
matrices $\bY_S$ and  $\bY_Z$ are iteratively determined at $M_T$
under the unification constraint of Eq.~(\ref{y15}).
For the determination of the mass matrix $\bmm_\nu$ at low-energy
[cf. Eq.~(\ref{mix})] we consider three different types of neutrino
spectra:
\begin{enumerate}
\item
Normal hierarchy (NH): $m_1\ll m_2<m_3$, with
\be
\label{nh}%
m^2_2 = m^2_1 + \Delta m^2_{S} , ~~~ m^2_3=m^2_1 + \Delta m^2_{S} +
\Delta m^2_{A}\,,
\ee
\item
Inverted hierarchy (IH): $m_3\ll m_1 < m_2$, with
\be
\label{ih}%
m^2_1=m^2_3 -\Delta m^2_{S}  + \Delta m^2_{A} ,  ~~~ m^2_2 = m^2_3 +
\Delta m^2_{A}\,,  \ee
\item
Quasi degenerate (QD):   $m_1\approx m_2 \approx m_3$, with
\be
\label{qd} m^2_2 = m^2_1 + \Delta m^2_{S} , ~~~ m^2_3 = m^2_1 +
\Delta m^2_{S}+ \Delta m^2_{A} , ~~~ m^2_1 \gg \Delta m^2_{A}\,.
\ee
\end{enumerate}
The neutrino mass squared differences $\Delta m^2_{A} = |m^2_3-
m^2_2|$ and $\Delta m^2_{S} = m^2_2 -m^2_1$ are responsible for the
atmospheric  and solar neutrino oscillations, respectively, together
with the corresponding mixing angles $\theta_{23}$ and $\theta_{12}$
of the mixing matrix $\bV$ [see Eq.~(\ref{Um})]. From global
analysis of the neutrino oscillation data, the following best fit
values are available (with their allowed $3\sigma$
range)~\cite{fit}:
\bea
\label{bfp}%
\Delta m^2_{A} & = & 2.2^{+1.1}_{-0.8}\times 10^{-3}~{\rm eV}^2
\,\,\,,\,\,\,\sin^2\theta_{23} = 0.5^{+0.18}_{-0.16} \no \\
\Delta m^2_{S}  & = & 7.9^{+1.0}_{-0.8}\times 10^{-5}~{\rm eV}^2,
\,\,\,,\,\,\,\sin^2\theta_{12} = 0.3^{+0.1}_{-0.6}
\eea
For the lepton mixing angle $\theta_{13}$  the following  upper
bound has been settled (at $3 \sigma$)~\cite{fit}:
\be\label{13}%
\sin^2\theta_{13}< 0.043\,.
\ee
It is worth noticing that, for a NH spectrum, the largest entries of
$\bmm_\nu$ (and, hence, of $\bY_T$) are those in the $\tau-\mu$
block, where the largest neutrino mass $(\Delta m_A^2)^{1/2}$
dominates\footnote{These considerations are valid for vanishing
CP-phase, $\delta=0$. From this point forward we restrict ourselves
to this limit. The discussion about the impact of $\delta\neq 0$ on
our results is postponed to Section~\ref{numanalysis}.}. Moreover,
they are of comparable size due to maximal atmospheric mixing. On
the other hand, the remaining matrix elements are smaller by one or
two orders of magnitude, depending on the value of $\theta_{13}$
(which introduces the dependence on $(\Delta m_A^2)^{1/2}$ to these
entries). In the IH case, while $(\bmm_\nu)_{ee}$ and the $\tau-\mu$
block are of the same order, $(\bmm_\nu)_{\mu e}$ and
$(\bmm_\nu)_{\tau e}$ are generically smaller. Finally, when the
neutrino mass spectrum is of the QD type, the leading mass $m_1$
renders the matrix pattern less hierarchical and enhances the
overall magnitude of $\bmm_\nu$ (or, equivalently, of $\bY_T$).

Let us now turn to the flavour structure of the SSB parameters which
are generated at the messenger scale $M_T$. For simplicity of our
discussion, we temporarily assume that the unification condition
(\ref{y15}) remains approximately valid at $M_T$, $\bY_T \approx
\bY_S \approx \bY_Z$. In addition, we disregard the terms
proportional to the Yukawa couplings $\bY_{e, d}$ in the expressions
of $\bmm^2_\tl$ and $\bmm^2_\td$ given in Eqs.~(\ref{soft2}). Under
these assumptions, we can write:
\bea
\label{fva}%
(\bmm^2_\tl)_{ij} &\!\! \approx \!\!& -
\left(\frac{{B_T}}{16\pi^2}\right)^2 \kappa \left[\bV(\bmm^{D
}_\nu)^2 \bV^\dagger\right]_{ik} \left[ c^e_a \,g^2_a
\delta_{kj}
- 72\, \kappa \left[\bV(\bmm^{D }_\nu)^2 \bV^\dagger\right]_{kj}
-6\, \kappa {\rm Tr}(\bmm^{D}_\nu)^2 \delta_{kj}\right] ,
 \no \\
(\bA_e)_{ij} & \approx & \frac{{3B_T}}{8\pi^2}\, \kappa
\,(\bY_e)_{ii}\left[\bV
(\bmm^{D }_\nu)^2 \bV^\dagger\right]_{ij} , \no \\
(\bmm^2_\td)^*_{ij} &\!\!\approx \!\!& -
\left(\frac{{B_T}}{16\pi^2}\right)^2 \kappa \left[\bV(\bmm^{D
}_\nu)^2 \bV^\dagger\right]_{ik} \left[ c^d_a \,g^2_a
 \delta_{kj}
- 72 \, \kappa \left[\bV(\bmm^{D }_\nu)^2 \bV^\dagger\right]_{kj}
-6\, \kappa {\rm Tr}(\bmm^{D}_\nu)^2 \delta_{kj}\right] ,
 \no \\
(\bA_d)^*_{ij} &\approx & \frac{{3B_T}}{8\pi^2} \kappa \left[\bV
(\bmm^{D }_\nu)^2 \bV^\dagger\right]_{ij}(\bY_d)_{jj}\quad\quad ,
\quad\quad \kappa = \left(\frac{M_T}{\la v^2_2}\right)^{\!2}\,,
\eea
where $c^e_a, c^d_a$ are: $c^e_1 =\frac{102}{15}\, , c^e_2 = 30\, ,
c^e_3 = 16$,  $c^d_1 =\frac{62}{15}~, c^d_2 = 6$ and $ c^d_3
=\frac{176}{3}$ (summation of repeated indices is implicit). The
flavour indices are $i,j =e, \mu, \tau \,(d, s, b)$ for the slepton
(squark) matrices $\bmm^2_\tl$ and $\bA_e$ ($\bmm^2_\td$ and
$\bA_d$). Wherever necessary, the correspondences $e\leftrightarrow
d, \mu \leftrightarrow s, \tau \leftrightarrow b$ are understood.

From Eqs.~(\ref{T-mass})-(\ref{Um}), and considering the three
different kinds of neutrino mass spectra shown in
(\ref{nh})-(\ref{qd}), we have:
\bea
\label{yy1}%
\left[\bV (\bmm^{D }_\nu)^2 \bV^\dagger\right]_{i
j}=\left\{\begin{array}{l}%
m^2_1 \delta_{i j} +\Delta m^2_{S}\bV_{i2}\bV^*_{j 2}  +(\Delta
m^2_{A} +\Delta m^2_{S} ) \bV_{i 3}\bV^*_{j 3}\;\;\;{\rm (NH, QD)}
\\ \\%
m^2_3 \delta_{i j} +\Delta m^2_{A} \bV_{i 2}\bV^*_{j 2} +(\Delta
m^2_{A} - \Delta m^2_{S} ) \bV_{i 1}\bV^*_{j 1}\;\;\;{\rm (IH)}
\end{array}\right.\,,
& &
\eea
\\
which manifestly shows that the FV ($i\neq j$) entries are always
independent of the lightest neutrino mass. The explicit expressions
in terms of the neutrino parameters read:
\\
\bea \label{yy2}%
\left[\bV (\bmm^{D }_\nu)^2 \bV^\dagger\right]_{\tau \mu} &=& \pm
{\Delta m^2_{A}}\left[c^2_{13}c_{23} s_{23}\pm\rho\,\left( c_{23}
s_{23} (s^2_{12}- c^2_{12}s^2_{13}) + c_{12} s_{12}s_{13} (s^2_{23}
-c^2_{23}) \right) \right] \no \\ \no \\
\left[\bV (\bmm^{D }_\nu)^2 \bV^\dagger\right]_{\mu e} & = & \pm
\Delta m^2_{A} c_{13}\left[ s_{13}s_{23}  \pm\rho\, c_{12}\left(
c_{23}s_{12} + c_{12}s_{23}s_{13} \right) \right]\, \no \\
&& \no \\
\left[\bV (\bmm^{D }_\nu)^2 \bV^\dagger\right]_{\tau e}
&=&\pm{\Delta m^2_{A}}c_{13}\left[c_{23}s_{13} \mp \rho
\,c_{12}\left( s_{23} s_{12}-  c_{12}c_{23}s_{13}\right) \right]\,,
\eea
\\
where the  upper (lower) sign applies for either the NH or QD (IH)
neutrino spectrum, and $\rho= \Delta m^2_{S} /\Delta m^2_{A} \approx
0.04$. Due to the smallness of the parameter $\rho$, the quantities
$|(\bY^\dagger_T \bY_T)_{i j}| \propto |[\bV (\bmm^{D }_\nu)^2
\bV^\dagger]_{i j}|$ are not sensitive to the type of neutrino mass
spectrum, unless $s_{13} \approx \rho\, c_{12} s_{12}c_{23}/c_{23}
\,({\rm or}\,\rho \,c_{12} s_{12}c_{23}/s_{23}) \approx 0.02$. For
such values of $s_{13}$, the $\mu e$ and $s d$ ($\tau e$ and $b d$)
matrix elements in Eqs.~(\ref{fva}) would be strongly suppressed or
even vanish for a IH (NH) mass pattern. Instead, the entries $\tau
\mu\, (b s)$  do not change significantly with $\theta_{13}$, since
the terms proportional to $s_{13}$ are suppressed by $\rho\cos
(2\theta_{23})\approx 0$. The quartic combinations $\left[\bV
(\bmm^{D }_\nu)^2 \bV^\dagger\right]^2$ can be approximately
obtained from the corresponding quadratic ones of Eq.~(\ref{yy2}) by
performing the following replacements:
\bea
\label{yy4}%
{\rm NH}&& \!\!\!\!\!\!\!\!:\,\,\Delta m^2_A \rightarrow (\Delta
m^2_A)^2 (1 + 2\rho)\;\;\quad,\quad\;\; \rho \rightarrow \rho^2 (1 -
2\rho)\,,\no
\\{\rm \medskip} {\rm IH}&&\!\!\!\!\!\!\!\!: \,\,\Delta m^2_A \rightarrow
(\Delta m^2_A)^2\;\;\quad\quad\quad\;\;\quad,\quad\;\;
\rho \lto 2 \rho\left(1 - \frac{\rho}{2}\right)\,,\no \\
{\rm QD}&&\!\!\!\!\!\!\!\!: \,\,\Delta m^2_A \lto 2 \,m^2_1 \,\Delta
m^2_A\;\;\quad\quad\;\;\,\;,\quad\;\; \rho \rightarrow \rho\,.
\eea
\\
It is then straightforward to obtain the explicit expressions for
the r.h.s. of Eqs.~(\ref{fva}), taking into account Eqs.~(\ref{yy2})
and (\ref{yy4}). However, it is more instructive to consider the
results for certain ranges of $\theta_{13}$. Let us focus on
$\bmm^2_\tl$, as the result for $\bmm^2_\td$ will follow directly.
Distinguishing the three types of neutrino mass neutrino spectra, we
find:
\\

$\bullet\,$ NH spectrum, for $s_{13} \ll \rho^2 c_{23}c_{12}
s_{12}/s_{23} \approx 8\times 10^{-4}$:
\bea
\label{NH1}%
(\bmm^2_{\tl})_{\tau \mu} & \!\!\! \approx \!\!\! &
-\left(\frac{B_T}{16\pi^2}\right)^2 \kappa \,
\left[\bV(\bmm^{D}_\nu)^2 \bV^\dagger\right]_{\tau \mu } \,\left[
c^e_a g^2_a -78\, \kappa \Delta m^2_A (1 +2 \rho) \right] ,
\no \\
& & \no \\
(\bmm^2_{\tl})_{\mu e,\tau e} &\!\!\! \approx \!\!\! & -
\left(\frac{B_T}{16\pi^2}\right)^2 \kappa \, \left[\bV(\bmm^{D
}_\nu)^2 \bV^\dagger\right]_{\mu e,\tau e} \, \left[c^e_a g^2_a - 6
\, \kappa \Delta m^2_A (1 +14 \rho) \right]\,,
\eea

while for $s_{13} \gg \rho c_{23}c_{12} s_{12}/s_{23} \approx0.02$:

\be
\label{NH2}%
(\bmm^2_\tl)_{ ij }\approx - \left(\frac{{B_T}}{16\pi^2}\right)^2
\kappa \, \left[\bV(\bmm^{D }_\nu)^2 \bV^\dagger\right]_{ ij }
\,\left[c^e_a g^2_a  -78\, \kappa \Delta m^2_A (1 +2 \rho) \right]
\ee

$\bullet\,$ IH spectrum for $s_{13} \ll \rho c_{23}c_{12}
s_{12}/s_{23} \approx0.02$:
\bea \label{IH1}%
(\bmm^2_{\tl})_{\tau \mu} & \!\!\! \approx \!\!\! & -
\left(\frac{B_T}{16\pi^2}\right)^2 \kappa \, \left[\bV(\bmm^{D
}_\nu)^2 \bV^\dagger\right]_{\tau \mu } \,\left( c^e_a g^2_a
 - 84 \, \kappa \Delta
m^2_A
\right) ,    \no \\
& & \no \\
(\bmm^2_{\tl})_{\mu e,\tau e} &\!\!\! \approx \!\!\!& -
\left(\frac{B_T}{16\pi^2}\right)^2 \kappa \, \left[\bV(\bmm^{D
}_\nu)^2 \bV^\dagger\right]_{\mu e\,\tau e} \, \left( c^e_a g^2_a
-156 \, \kappa \Delta m^2_A \right) ,\no \\
&& \eea%

while for $s_{13} \gg 2\rho c_{23}c_{12} s_{12}/s_{23} \approx
0.04$:
\be \label{IH2} (\bmm^2_\tl)_{ ij } \approx -
\left(\frac{{B_T}}{16\pi^2}\right)^2 \kappa \, \left[\bV(\bmm^{D
}_\nu)^2 \bV^\dagger\right]_{ ij } \,\left( c^e_a g^2_a  - 84 \,
\kappa \Delta m^2_A  \right) . \ee

$\bullet\,$ QD spectrum (for any value of $s_{13}$)
\be
\label{QD1}%
(\bmm^2_\tl)_{ ij } \approx - \left(\frac{{B_T}}{16\pi^2}\right)^2
\kappa \, \left[\bV(\bmm^{D }_\nu)^2 \bV^\dagger\right]_{ ij }
\left[ c^e_a g^2_a - 162 \, \kappa \,  m^2_1 \left(1+  \frac{\Delta
m^2_A}{ 27 m^2_1}\right) \right] .
\ee
The above expressions exhibit the relevant flavour violating factors
$[\bV (m^D_\nu)^2 \bV^\dagger]_{i j}$ separated from the piece
composed by terms proportional to the gauge couplings $g_a^2$ and
terms proportional to $\kappa \,\Delta m^2_A$ (NH and IH) or $\kappa
\,m^2_1$ (QD). For a given effective SUSY breaking scale $B_T$, the
overall size of those entries is controlled by $\kappa \propto
(M_T/\la)^2$. It is interesting to notice that, for a certain value
of $M_T/\la $, the $g_a^2$-terms are canceled and thus all the three
off-diagonal entries can be vanishing. If $s_{13} \gg 0.02$, this
peculiar flavour-blind suppression happens when $M_T/\la \sim v^2_2
(\frac{c^e_ag^2_a}{ 78 \Delta m^2_A})^{1/2}$ and $M_T/\la \sim v^2_2
(\frac{c^e_a g^2_a }{ 84\Delta m^2_A})^{1/2}$, for the NH and IH
neutrino spectrum, respectively. Instead, for the QD case, the
suppression occurs for $M_T/\la \sim v^2_2 (\frac{c^e_a g^2_a }{162
m^2_1})^{1/2}$, irrespective of $s_{13}$. If $s_{13}$ is tiny and
the neutrino spectrum is either NH or IH, the entry $(\tau \mu)$ is
suppressed for $M_T/\la$ smaller (larger) by $\approx 1/3$ ($\approx
1.4$), with respect to the entry $(\mu e)$ [$(\tau e)$] for the NH
(IH) spectrum. We can summarise this analysis saying that the
off-diagonal entries of $\bmm^2_\tl$ and $\bmm^2_\td$ can undergo a
strong reduction in two cases: 1)\,when $s_{13}\approx 0.02$ and so
the quantities $[\bV(\bmm^D_\nu)^2 \bV^\dagger]_{\tau e}$ and
$[\bV(\bmm^D_\nu)^2 \bV^\dagger]_{\mu e}$ are vanishing for the NH
and IH spectrum, respectively; 2)\,when there is a cancelation
between the quadratic and quartic contributions, which depends on
the parameters $M_T$ and $\la$ and on the neutrino spectrum.

In the parameter range where the FV entries $(\bmm^2_\tl)_{ij}$  are
dominated by either the quadratic or the quartic Yukawa terms, the
relative size of LFV (QFV) in the $\mu-\tau$ ($b-s$) and $\mu-e$
($s-d$) sectors, does not depend of the ratio $M_T/\la$, and can be
approximately predicted in terms of only the low-energy
observables\footnote{We recall that, for low/moderate values of
$\tan\beta$, the off-diagonal entries (\ref{fva}) are not
substantially modified by the RG running from the high scale $M_T$
to low-energies. Therefore, in Eq.~(\ref{LFratio2}) we have
disregarded the renormalization effects on the neutrino mass [see
Eq.\,(\ref{match})]. However, this does not alter the form of
$\bmm_\nu$, since it amounts to an overall correction factor, except
possibly for the entries $(\bmm_\nu)_{\tau i}$ which receive extra
(overall) corrections in the regime of large $\tan\beta$.
\label{ft-mnu}} $\bV$ and $\bmm^D_\nu$, as:
\bea
\label{LFratio2}%
R_{23/12}= \frac{ ( \bmm^{2 }_{\tilde{L}})_{\tau \mu}} {( \bmm^{2
}_{\tilde{L}})_{\mu e} }\sim \frac{ ( \bmm^{2 }_{\td})_{b  s}} {(
\bmm^{2 }_{\td})_{s d} } & \approx & \frac{\left[\bV (\bmm^{D
}_\nu)^2\bV^\dagger\right]_{\tau \mu}}
{\left[\bV (\bmm^{D }_\nu)^2\bV^\dagger\right]_{\mu e}} ,  \no \\
R_{13/12}=\frac{ ( \bmm^{2 }_{\tilde{L}})_{\tau e}}{( \bmm^{2
}_{\tilde{L}}) _{\mu e} } \sim \frac{ ( \bmm^{2 }_{\td})_{b  d}}{(
\bmm^{2 }_{\td})_{s d} } & \approx& \frac{\left[\bV
(\bmm^{D}_\nu)^2\bV^\dagger\right]_{\tau e}} {\left[\bV (\bmm^{D
}_\nu)^2\bV^\dagger\right]_{\mu e}}\,.
\eea
Plugging the results of Eqs.~(\ref{yy2}) into the above expressions,
and taking the central values (\ref{bfp}) for the neutrino
parameters, we have:
\bea \label{ratios}
R_{23/12}|_{s_ {13}=0}& \approx & 40 \,(-40) ,
\,\,\,\,\,\,\,\,\,\,~~~~~
R_{23/12}|_{s_ {13}=0.2}\approx  3.2\, (3.8) , \no \\
R_{13/12}|_{s_ {13}=0}& \approx & -1\,(-1) \,  \,\,\,~~~~~~~~~~
R_{13/12}|_{s_ {13}=0.2}\approx  0.8 \,(1.2)\, ,
\eea
for the NH (IH) spectrum. The QD case gives the same results as the
NH one.

\section{Phenomenological viability}
\label{phenviab}
In the previous sections, we have set up a theoretical framework
consisting of the superpotential (\ref{wmssm}) and the SSB
lagrangian (\ref{Lsoft}), beneath the energy scale $M_T$, whose
relevant soft mass parameters (\ref{soft1}, \ref{soft2}) emerge at
the messenger energy scale. Next, we describe the criteria employed
to constrain the parameter space spanned by $M_T, B_T$ and $\la$.
With the purpose of investigating the phenomenological viability of
our scenario, a detailed numerical analysis is carried out in
Section~\ref{paramspace}. The results are then thoroughly discussed.

\subsection{The phenomenological constraints}
\label{phenconst}
Our approach to relate the low-energy measured parameters with the
high-energy quantities, such as the Yukawa couplings, follows a
bottom-up perspective. In particular, this allows us to determine
the following quantities:
\begin{itemize}
\item
The matrix $\bY_T$ (as well as $\bY_S$ and $\bY_Z$) is determined at
$M_T$ as already explained in the previous section [see
Eq.~(\ref{match})].
\item
The Yukawa matrices $\bY_e, \bY_u, \bY_d$ are extracted from the
corresponding charged fermion masses, modulo $\tan\beta$.

\item
The remaining  low-energy input parameters $\mu (\mu_{s})$ (with its
sign) and $\tan\beta$ are determined by the EWSB conditions:
\bea
\label{ewsb}%
\mu^2=\frac{-\tan^2\beta \ov{m}^2_{H_2} +
\ov{m}^2_{H_1}}{\tan^2\beta-1}
-\frac{M^2_Z}{2}\;\quad\quad,\quad\quad\; \sin 2\beta = \frac{2\mu
B_H}{ \ov{m}^2_{H_1} +  \ov{m}^2_{H_2}+ 2\mu^2}\,.
\eea
We have included the one-loop tadpole corrections $t_{1,2}$ through
the redefinition of the Higgs scalar masses, {\it i.e.}
$\ov{m}^2_{H_1}= {m}^2_{H_1} - \frac{t_1}{\sqrt{2} v \cos\beta},
\ov{m}^2_{H_2}= {m}^2_{H_2} - \frac{t_2}{\sqrt{2} v \sin\beta}$.
According to the standard practice, the above minimisation
conditions are imposed at the scale $\mu_{s} = \sqrt{m_{\tilde{t}_1}
m_{\tilde{t}_2} }$.
\end{itemize}

Hence, we are left with three free parameters, $M_T, B_T$ and the
coupling $\la$, which can all be taken to be real, without loss of
generality.
%
\begin{table}
\renewcommand{\tabcolsep}{1.2pc}
\begin{center}\begin{tabular}{lrc}
\hline \hline\noalign{\smallskip} BR & Present limits & Future
sensitivity
\\  \hline \noalign{\smallskip}
$\mu^{-} \to e^{-} \ga$ &  $1.2\times 10^{-11}$ \cite{exp} &
$10^{-14}$ \cite{MEG}
\smallskip\\\hline \noalign{\smallskip}
$\tau^{-}\to \mu^{-} \ga $ & $6.8\times 10^{-8}$ \cite{exp-rad}&
$10^{-9}~\cite{taumu:fut}$
\smallskip\\ \hline \noalign{\smallskip}
$\tau^{-} \to e^{-} \ga$  & $1.1\times 10^{-7}$ \cite{exp-rade} &
$10^{-9}~\cite{taumu:fut}$
\smallskip\\ \hline \noalign{\smallskip}
$\mu^{-} \to e^{-} e^{+} e^{-} $&  $1.0\times 10^{-12}$
\cite{sindrum}  & $10^{-14}$~\cite{mu3e}
\smallskip\\ \hline \noalign{\smallskip}
$\tau^{-} \to \mu^{-}\mu^{+} \mu^{-} $& $1.9\times 10^{-7}$
\cite{babar}  & $10^{-9}~\cite{taumu:fut}$
\smallskip\\ \hline \noalign{\smallskip}
$\tau^{-} \to \mu^{-} e^{+}e^{-}  $&  $1.9\times 10^{-7}$
\cite{belle} & $10^{-9}~\cite{taumu:fut}$
\smallskip\\\hline \noalign{\smallskip}
$\tau^{-} \to e^{-} e^{+} e^{-}  $& $2.0\times 10^{-7}$ \cite{list}
& $10^{-9}~\cite{taumu:fut}$
\smallskip\\ \hline \noalign{\smallskip}
$\tau^{-} \to e^{-} \mu^{+} \mu^{-}  $& $3.3\times 10^{-7}$
\cite{list} & $10^{-9}~\cite{taumu:fut}$
\smallskip\\ \hline \noalign{\smallskip}
CR($\mu \to e\,$;\,Ti ) & $1.7\times 10^{-12}$ \cite{sindrum2} &
$10^{-18}~\cite{prime}$
\smallskip
 \\
\hline \hline
\end{tabular}\end{center}
\captions{Present limits and future sensitivities for the branching
ratios (BR) of several LFV processes. The bound on the $\mu\to e$
conversion rate (CR) in Ti is also shown.}\label{tb1}
\end{table}
%
The parameter space spanned by these parameters is constrained as
follows.
\begin{itemize}
\item
We impose the constraint coming from the decay $\mu \to e \ga$,
using the results in \cite{HMTY}. The experimental upper bound for
the branching ratio of this decay is shown in Table~\ref{tb1}. The
relevant LFV entry is $(\bmm^2_\tl)_{\mu e}$ [see Eqs.~(\ref{fva})
and (\ref{yy2})].

\item
A conservative limit upon the lightest Higgs boson mass, $m_h >
110~{\rm GeV}$, is considered on the basis of the negative LEP2
direct searches~\cite{lep}. Our predictions on $m_h$ include the
low-energy radiative corrections which are obtained linking our code
to {\tt FeynHiggs}~\cite{FH}.

\item
We also require that the sparticle masses (which pass the tachyonic
test) respect the experimental lower bounds set by Tevatron and LEP
direct searches~\cite{PDG}.

\item
The SUSY contribution to the anomalous magnetic moment of the muon
$\delta a_\mu$~\cite{HMTY} has been considered and subjected to the
following constraint at $99\%\,$ C.L.~\cite{MP}:
\be%
\label{gmu} - 13\times 10^{-10} < \delta a_\mu < 57 \times
10^{-10}\,.
\ee
However, we will see that this requirement does not imply an
important constraint on the parameter space.
\item
We have checked that all the coupling constants remain in the
perturbative regime up to the scale $M_G$. In particular, while the
presence of a complete GUT representation below $M_G$ does not alter
the value of $M_G$, the gauge coupling $g_G$ gets an additional
contribution $\delta g^{-2}_G = \frac{N}{8 \pi} {\rm
ln}({M_T}/{M_G})$. The requirement of perturbativity implies
$\frac{N}{2 \pi} {\rm ln}({M_G}/{M_T}) \lsim 24$ which, for $N=7$,
gives $M_T\gsim 10^{7}~{\rm GeV}$. The parameter space is further
restricted by imposing the perturbativity requirement on the
coupling constants $\la$, $\bY_{T}, \bY_{S}$ and $\bY_{Z}$.
\end{itemize}

As it will be shown in the next section, the most stringent
constraints come from the sparticle spectrum, $\mu\to e \ga$ decay,
the Higgs boson mass and the requirement of radiative EWSB and
perturbativity.

\subsection{Parameter space analysis}
\label{paramspace}

\begin{figure}
\begin{center}
\begin{tabular}{c}
\includegraphics[width=12cm]{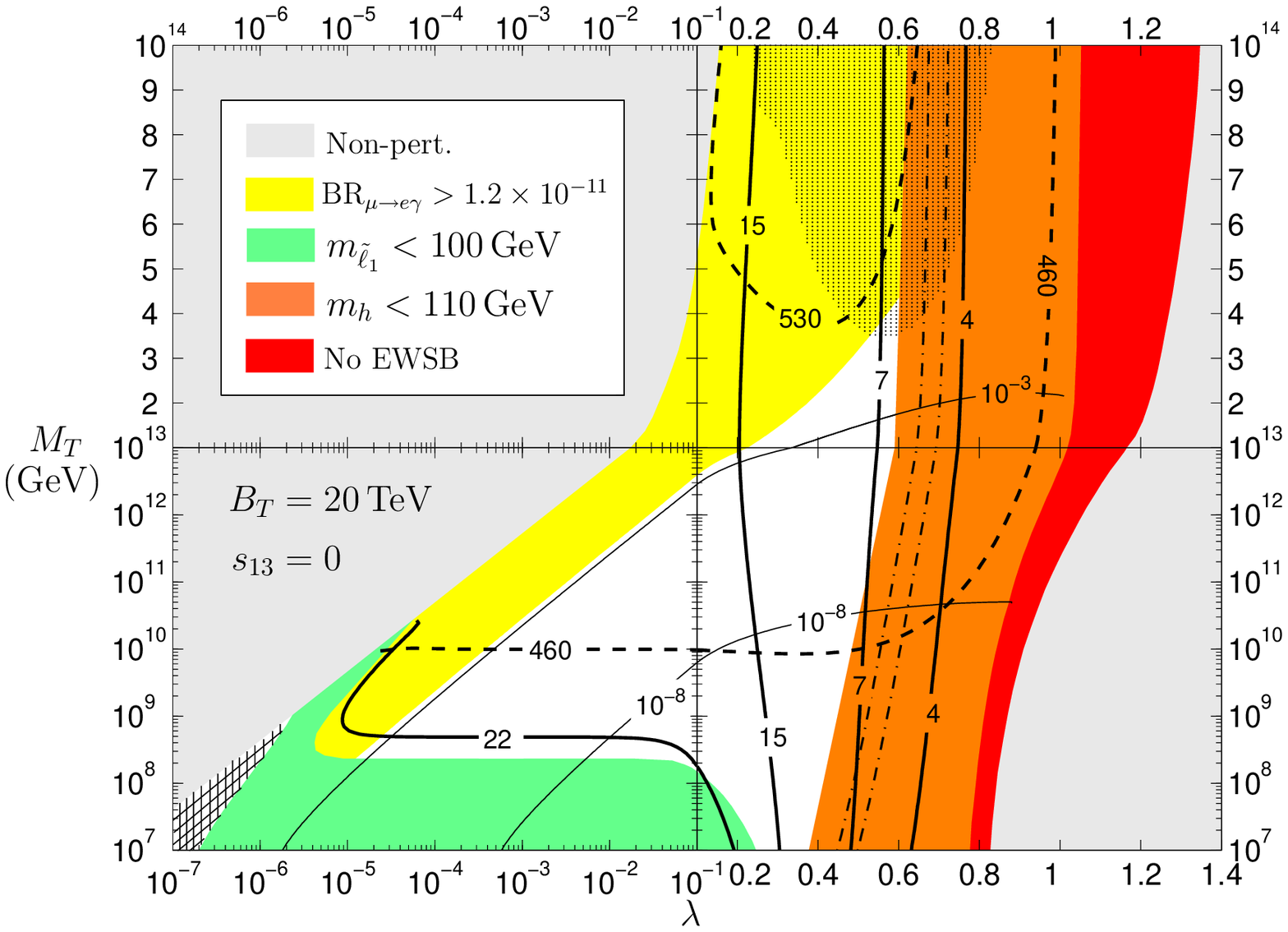}\\
\includegraphics[width=12cm]{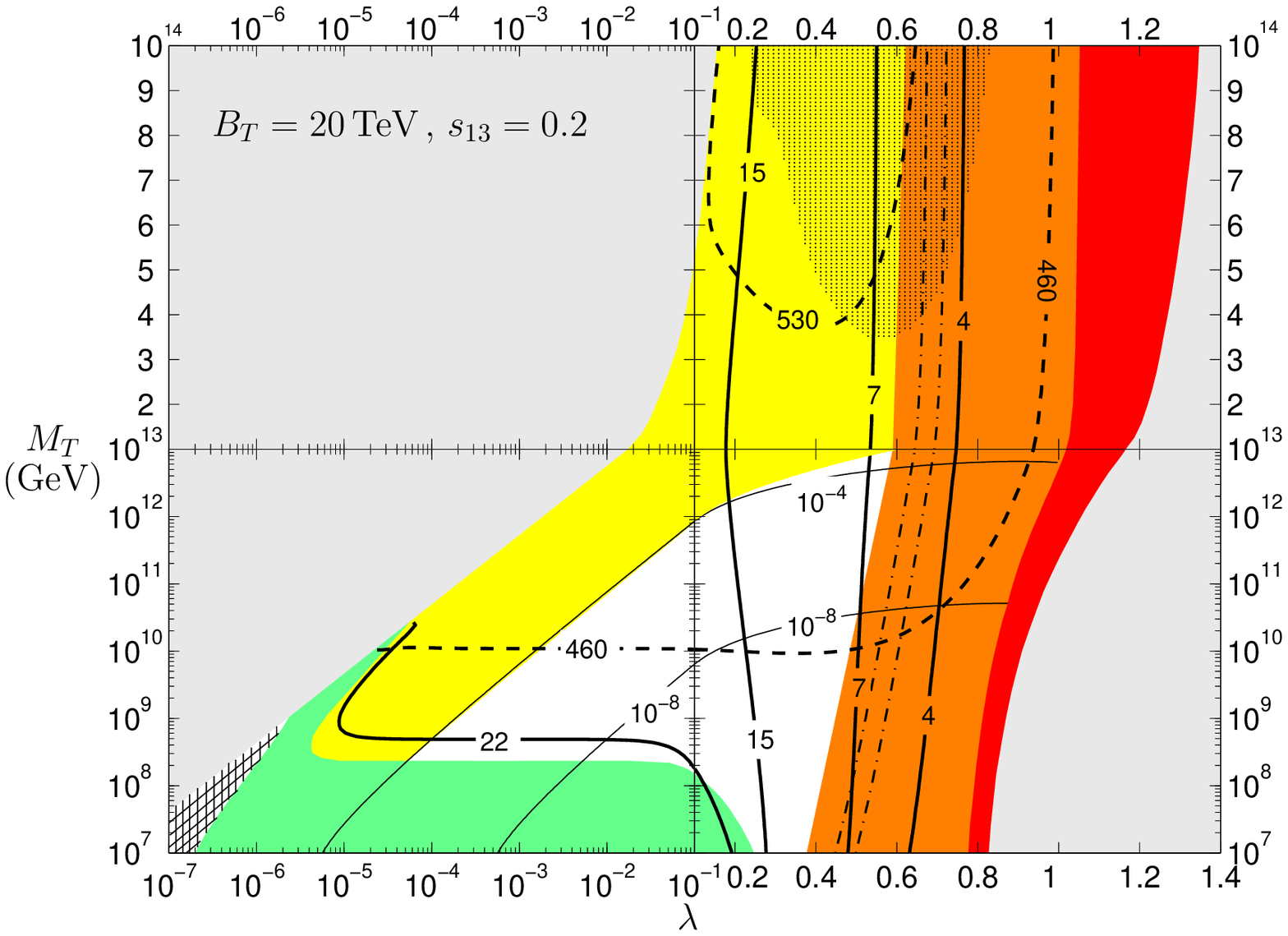}
\end{tabular}
\captions{\small The $(\la, M_T)$-parameter space configuration for
$B_T=20$~TeV and $s_{13}=0~(s_{13}=0.2)$ in the upper (lower) panel.
Perturbativity and EWSB exclude the light-grey and red regions (see
legend in the upper panel), respectively.  The orange region is
excluded by the Higgs mass bound $m_h<$~110 GeV for $m_t=170.2$~GeV,
while the left-most (right-most) dash-dotted line delimit the same
region for $m_t=172.5\,(174.8)$~GeV. Inside the yellow and green
areas, $\BR(\mu\to e \ga)$ is above the present experimental upper
bound and the lightest slepton mass is below 100~GeV, respectively.
In the dotted and hatched areas, the neutralino $\tilde{\chi}^0_1$
is the lightest MSSM sparticle and $\tilde{\ell}_1$ is tachyonic,
respectively. The thick-solid (dashed) lines correspond to the
isocontours of $\tan\beta$ ($\mu$ in GeV) for $m_t=172.5$~GeV. The
thin solid lines refer to the FV parameter $\delta^d_{bs}$ [defined
in Eq.~(\ref{df})]. See text for more details.\vspace*{0.cm} }
\label{f5}
\end{center}
\end{figure}

In Fig.~\ref{f5} the constraints imposed on the parameter space
$(\la, M_T)$ are shown for $B_T=20~{\rm TeV}$ and $s_{13}=0 (0.2)$
in the upper (lower) panel, taking the NH neutrino spectrum (later
we will comment on the other cases). The light-grey regions are
excluded by the perturbativity requirement. For each value of $M_T$,
there is a minimum value of $\la$, scaling as $\sim 2\times
10^{-4}(M_T/10^{11}~{\rm GeV})$, below which the couplings $\bY_T$
and/or $\bY_{S,Z}$ reach the Landau pole between $M_T$ and $M_G$.
Similarly, there is an upper bound for $\la$, above which $\la$
itself blows up.
\begin{figure}
\begin{center}
\begin{tabular}{c}
\includegraphics[width=13cm]{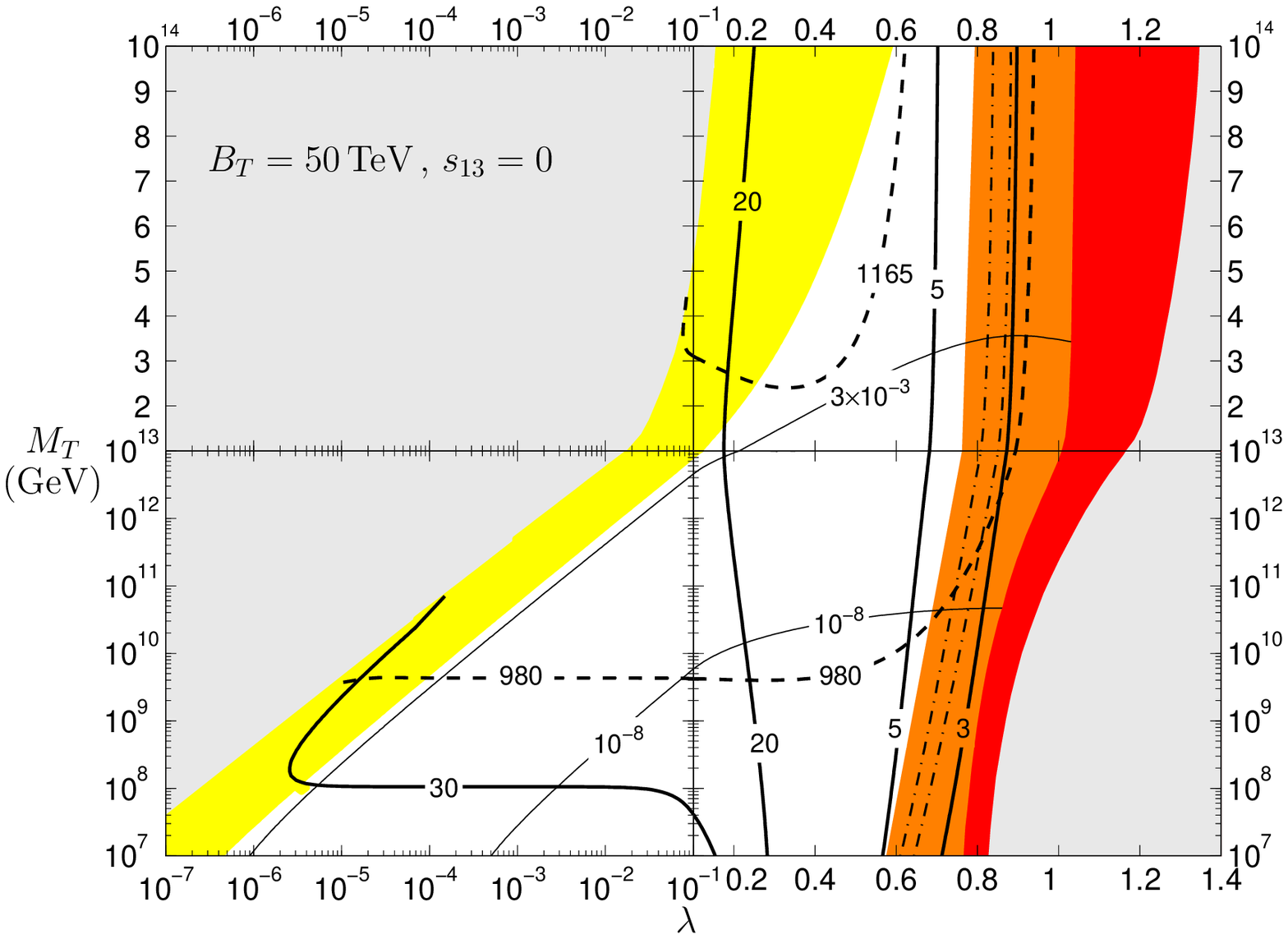}\\
\includegraphics[width=13cm]{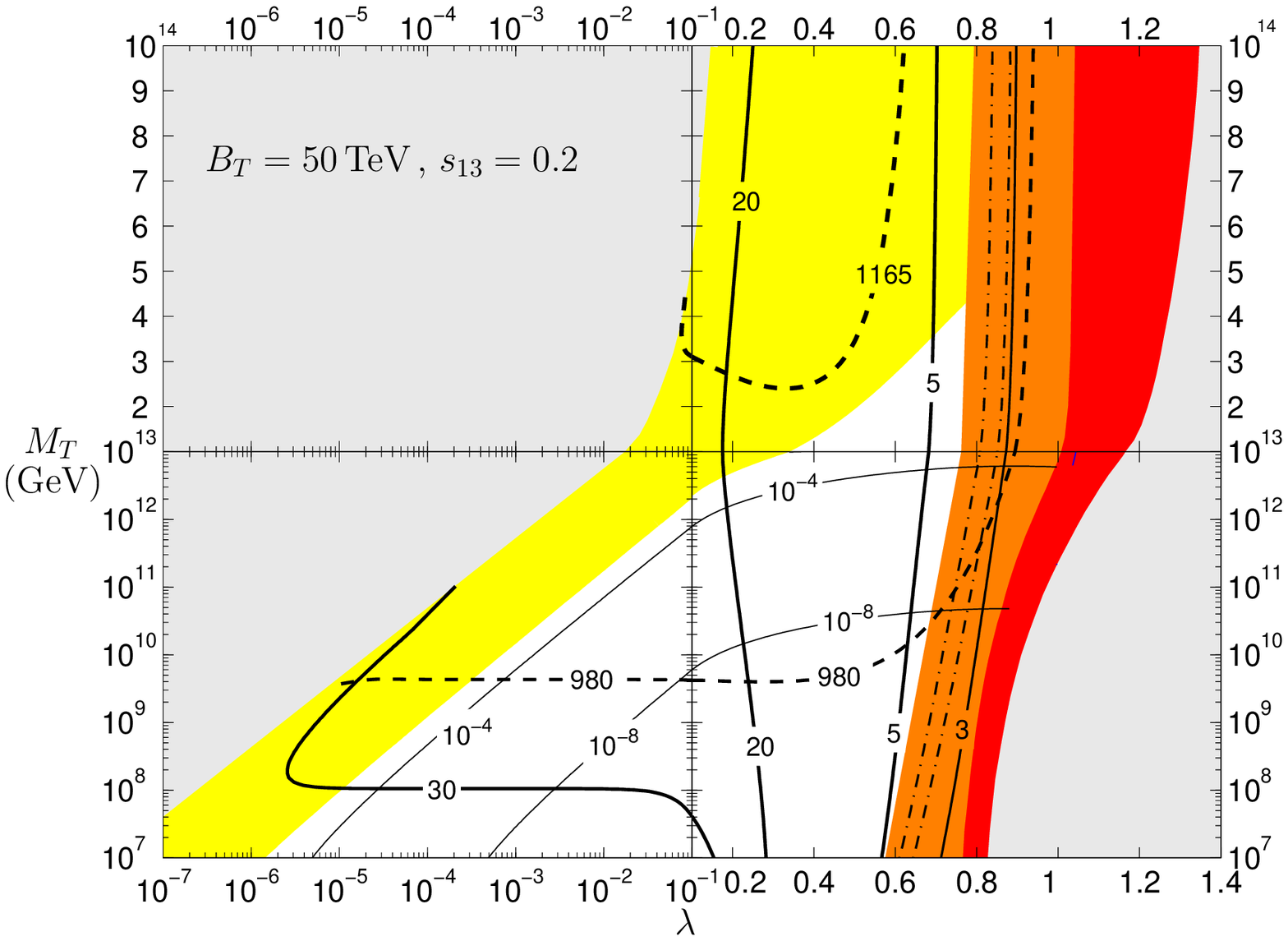}
\end{tabular}
\captions{\small The same as in Fig.~5 for $B_T = 50~{\rm TeV}$.
\vspace*{0.cm} } \label{f6}
\end{center}
\end{figure}

\begin{figure}
\begin{center}
\begin{tabular}{cc}
\includegraphics[width=8.0cm]{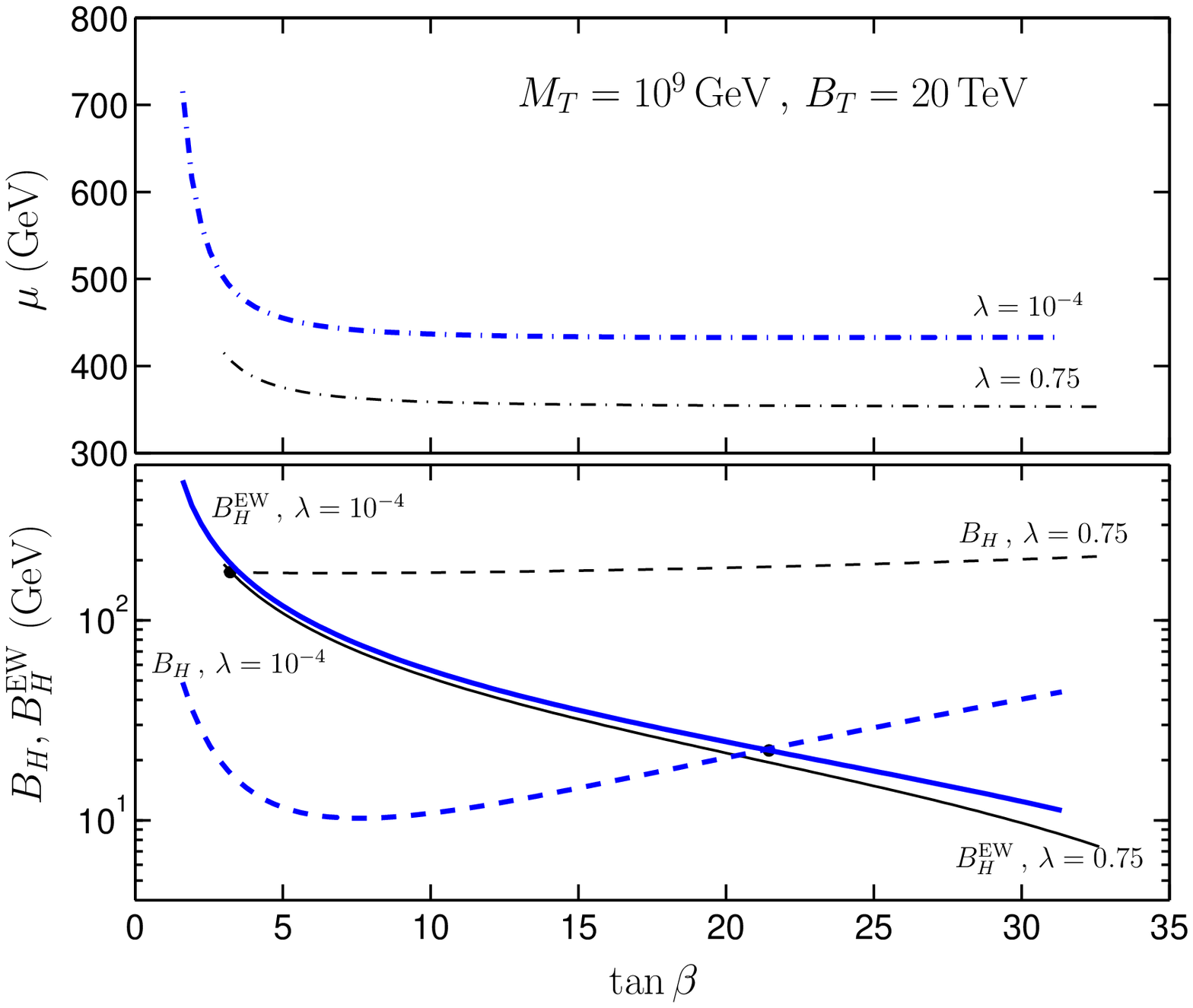} &
\includegraphics[width=8.0cm]{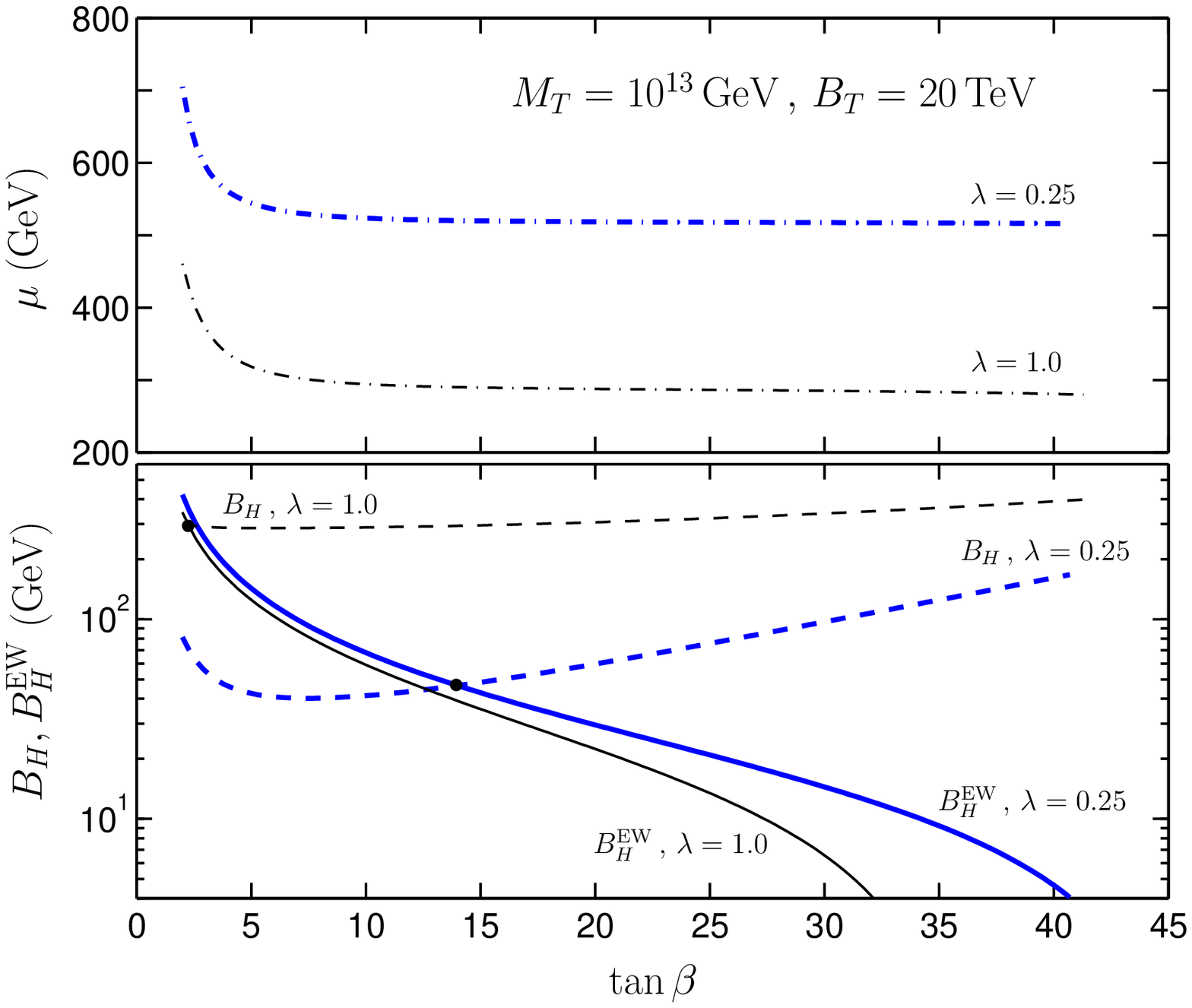}
\end{tabular}
\captions{Left panel: Curves of $B_H$ (dashed lines) obtained at
$\mu_s$ from the running of $B_H(M_T)$, and of $B_H^{EW}$ (solid
lines) determined by the minimization conditions (\ref{ewsb}), as a
function of $\tan\beta$ for $B_T=20$~TeV, $M_T=10^9$~GeV and
$\lambda=10^{-4},0.75$ (lower plot). In the upper plot, the
corresponding values of the $\mu$ parameter are shown. Right panel:
The same as in the left one but for $M_T=10^{13}$~GeV and
$\lambda=0.25,1.0$.}
 \label{f7}
\end{center}
\end{figure}
The EWSB constraint excludes the red region (which covers the range
with $\la \sim 1$ along the whole interval of $M_T$) limited on the
left by the least achievable value of $\tan\beta\approx 2.5$. We
have studied the sensitivity of our results to the top mass,
considering the $1\sigma$ allowed range $m_t=(172.5\pm
2.3)$~GeV~\cite{cdfd0}. The most relevant effects of varying $m_t$
are the ones induced by the top-Yukawa term in the RG running of the
soft mass $m^2_{H_2}$. However, this variation does not affect our
results considerably and, therefore, we will mainly take the central
value $m_t=172.5$~GeV.

The EWSB conditions select values of $\tan\beta$ (thick
solid-contours) up to moderate ones\footnote{As already mentioned
above, varying $m_t$ in its $1\,\sigma$ interval does not lead to
appreciable changes on the values of $\mu$ and $\tan\beta$. Thus, we
have only displayed the $\tan\beta$ and $\mu$ iso-contours for the
central value of $m_t$.}, ($\tan\beta \lsim 25$). Notice that the
largest values of $\tan\beta$ are achievable only for $M_T\lsim
10^{9}~{\rm GeV}$ and small values of $\la \lsim 0.2$. In fact,
Eq.~(\ref{ewsb}) shows that larger values of $\tan\beta$ (or smaller
$\sin2\beta$) require suppressed $B_H(\mu_{s})$, which can be
obtained by shortening the running energy interval (and so, smaller
$M_T$ are needed). We also display the iso-contours of the $\mu
(\mu_{s})$ parameter (dashed-lines) as obtained by the EWSB
conditions. One can realize that this parameter slightly increases
with $M_T$ due to the enhancement of the RG factor which affects
$m^2_{H_2}(\mu_{s})$ in the minimisation condition of
Eq.~(\ref{ewsb}). The whole allowed parameter space covers the range
$\mu \sim (450 - 550) ~{\rm GeV}$.

To better understand the behavior of the $\mu$ and $\tan\beta$
contours, it may be useful to consider Fig.~\ref{f5} together with
Fig.~\ref{f7}. Indeed, the latter displays the solutions of
Eq.~(\ref{ewsb}) for $\mu$ (upper plots) and $B_H$ (lower plots),
when $M_T=10^9~(10^{13})~{\rm GeV}$ in the left (right) panel. We
have chosen four distinct points in the $(\lambda,M_T)$ parameter
space and shown, for each case, the predicted parameter $B_H(\mu_s)=
B_H(M_T) +\Delta B_H$, where $B_H(M_T)$ is determined from
Eqs.~(\ref{soft1}) and $\Delta B_H$ englobes the RG running between
$M_T$ and $\mu_s$. The curves of $B_H(\mu_s)$ have to be compared
with those of  $B_H$ as extracted from Eq.~(\ref{ewsb}) (denoted by
$B^{EW}_{H}$ in the plots). The crossing of these two curves
(indicated by a black dot) signals the presence of a solution at the
corresponding value of $\tan\beta$. For instance, the left-panel of
Fig.~\ref{f7} shows that for $B_T=20$~TeV, $M_T=10^9$~GeV and
$\lambda=10^{-4}(0.75)$, EWSB occurs for $\tan\beta\approx 22(3)$
with $\mu \approx 450(400)$~GeV. A similar example is illustrated in
the right-panel of Fig.~\ref{f7} for
$B_T=20$~TeV, $M_T=10^{13}$~GeV and $\lambda=0.25$ and 1.\\

Consider now the constraint imposed by  the lower bound on $m_h$.
The orange region in Fig~{\ref{f5}} shows the portion of the
parameter space forbidden by the condition $m_h > 110$~GeV, taking
the lower limit of the $1\sigma$ range for the top-mass
($m_t=170.2$~GeV). The dependence on $m_t$ comes from the low-energy
radiative corrections $\propto \frac{m^4_t}{m^2_W} \ln
(\mu^2_s/m^2_t)$. As $m_t$ increases, the region shrinks, being
delimited by the left-most (right-most) dashed-line when
$m_t=172.5\,(174.8)$~GeV. It is also clear that, for each value of
$M_T$, the upper bound on $\lambda$ is set by the Higgs mass
constraint \eg, $\lambda<0.45-0.55$ when $M_T=10^9$~GeV. Since the
dependence of the Higgs mass on $\tan\beta$ mostly comes from the
tree-level contribution $|\cos 2\beta |M_Z$, the iso-contours of
$m_h$ closely follow those of $\tan\beta$, setting
in this way the least allowed value of $\tan\beta\approx 7$.\\

In our framework, the lightest MSSM supersymmetric particle is
tipically the lightest slepton $\tilde{\ell}_1$, except inside the
dotted region where $m_{\tilde{\chi}^0_1 }< m_{\tilde{\ell}_1}$
(which is almost entirely excluded by the Higgs and $\mu\rightarrow
e \gamma$ constraints). The mass of $\tilde{\ell}_1$ turns out to be
below the LEP2 lower bound of about $100$ GeV in the region of the
parameter space filled in green, where the large values of
$\tan\beta \gsim 22$ reduce $m_{\tilde{\ell}_1}$ through the
left-right mixing at the electroweak-symmetry breaking (see also
Section~\ref{sparspec}). In the hatched region (lower-left corner),
where $\tan\beta \approx 30$, the lightest slepton is tachyonic.\\

All the constraints discussed so far, being related to `unflavoured'
observables, are not sensitive to the angle $\theta_{13}$, as it is
shown by the comparison between the upper ($\sin\theta_{13} =0$) and
lower ($\sin\theta_{13} =0.2$) panels of Fig.~\ref{f5}. On the
contrary, the size of LFV strongly depends on it [see
Eq.~(\ref{yy2})] and hence, the region excluded by the bound on
$\BR(\mu \to e\ga)$ changes when different values of $\theta_{13}$
are considered. For $M_T\gsim 2\times 10^{8}~{\rm GeV}$, the
$\mu\rightarrow e \gamma$ constraint provides the most restrictive
lower bound on $\la$. This stems from the fact that the size the LFV
entry $(\bmm^2_\tl)_{\mu e}$ scales as $(M_T/\la)^2$
[Eq.~(\ref{fva})]. Consequently, the allowed $\la$ range is wider
for lower values of $M_T$, closing up for $M_T \sim 4\times
10^{13}~{\rm GeV}$.  By switching on $s_{13}$ (lower panel), the size
of $\mu e$ LFV  is enhanced as
\be
\frac{ (\bmm^{2}_{\tilde{L}})_{ \mu e}|^{s_{13}=0.2} } {(\bmm^{2
}_{\tilde{L}})_{\mu e}|^{s_{13}=0} } \approx 1 + \left( \frac{\Delta
m^2_{A} }{\Delta m^2_{S}  }\right)^2 \frac{s_{13}}{s_{12} c_{12}}
\sim 12
\ee
and, correspondingly, the $\mu \to e \ga$ limit implies that the
lower bound on $\la$ increases by a factor of $\sim(140)^{1/4}
\approx 3.5$ and so, the allowed parameter space shrinks. Taking
into account all the constraints considered above, one concludes
that values of $M_T \gsim 2\times 10^{13}{\rm GeV}$ are excluded for
$B_T=20$~TeV, independently of the value of $\lambda$.

An equivalent analysis is presented in Fig.~\ref{f6} for
$B_T=50$~TeV. The comparison with the previous case indicates that
the regions excluded by the perturbativity and EWSB requirements are
not significantly affected. On the other hand, the constraint on the
Higgs mass implies a slightly different upper bound on $\la$ ($\la
\lsim 0.6$) and a decrease of the minimum allowed value of
$\tan\beta$ ($\tan\beta \gsim 4$). Indeed, the sparticle spectrum is
now heavier and thus, the radiative corrections $\sim {\rm ln}
(\frac{\mu_{s}^2}{m_t^2})$ to $m_h$ are larger. For this reason,
smaller values of $\tan\beta$ are tolerated in the tree-level
contribution $\sim M_Z |\cos 2\beta|$. Contrarily to the previous
case, $\tll_1$ is never tachyonic and its mass lies above the LEP
bound. Consequently, for $M_T =10^7-10^8\,{\rm GeV}$ the allowed
$\la$-range is much more extended with respect to the case with $B_T
=20~{\rm TeV}$. Moreover, larger values of $\tan\beta$ are now
possible ($\tan\beta\lsim 40$). The heavier spectrum also makes the
$\mu \to e \gamma$ constraint weaker, reducing the excluded yellow
region. In conclusion, the allowed parameter space enlarges when
$B_T$ increases. Also for this case, the effect of non-zero
$\theta_{13}$ (lower panel) produces similar results as for smaller
$B_T$, by raising the lower bound of $\la$.

Before concluding this section, a comment is in order about the
influence of the type of neutrino spectrum on the allowed parameter
space. In the IH case, the perturbativity constraint on the Yukawa
couplings $\bY_{T, S, Z}$ would just imply a slightly larger minimum
of $\la$, for each $M_T$. All other constraints would instead be
unaffected. For a QD spectrum, the effect from the perturbativity
requirement would be much stronger (depending on the magnitude of
the overall neutrino mass $m_1$) since, as already mentioned, all
the $\bY_{T, S, Z}$ entries increase when compared to the NH or IH
cases. Therefore, the light-grey region would mostly cover the
yellow area excluded by the $\BR(\mu \to e \ga)$ bound. In
conclusion, either for the IH or QD spectrum, the resulting allowed
parameter space would not be much different from the NH case
displayed in Figs.~\ref{f5} and \ref{f6}.

\section{Phenomenological predictions}
\label{phenaspects}

After describing the main phenomenological constraints imposed on
the $(\lambda,M_T)$ parameters space, we will now go through the
specific features of the sparticle and Higgs spectra
(Section~\ref{sparspec}) of our scenario, and to the implications
for several low-energy LFV processes (Section~\ref{LFVQFV}).

\subsection{Sparticle and Higgs spectroscopy}
\label{sparspec}%

The spectrum of the superpartners is determined by the finite
radiative contributions to the SSB parameters at $M_T$ (see
Section~\ref{SSBmass}), acting as boundary conditions, and by the
subsequent MSSM RG running from $M_T$ to $\mu_s$.
The physical scalar masses are obtained by taking into account the
latter effect at one-loop level~\cite{rgeeq} and the low-energy $D$
and $F$-term contributions.
To get some intuition on the main features of the physical spectrum,
we present some qualitative arguments in addition to the complete
numerical results.

In the present framework, the boundary conditions for the gaugino
and sfermion masses are not universal at $M_T$, even when $M_T$ is
not far from $M_G$. This is due to the different gauge quantum
numbers of the MSSM field representations [see Eq.\,(\ref{soft2})].
At lowest order, the gaugino masses $\ov{M}_a$  at the messenger
scale\footnote{In this section, we denote by overbar any quantity
evaluated at the scale $M_T$.} are in proportion to the gauge
coupling squared, $\ov{M}_1: \ov{M}_2: \ov{M}_3= \ov{\alpha}_1 :
\ov{\alpha}_2 : \ov{\alpha}_3$ ($\alpha_a = g^2_a/4 \pi$). This
relation is maintained at low-energy, like in unified SUGRA
scenarios. The low-energy gaugino masses are given by:
\be\label{ma}%
M_a (\mu_{s}) = \frac{B_T }{16 \pi^2}\,N\,g_a^2(\mu_{s}) , ~~~~~
a=1,2,3\,,
\ee
which leads to:  
$M_2 \approx 1.86\, M_1 \approx 0.35\, M_3$ (taking $\mu_s \sim
700\,{\rm GeV})$.
 The most interesting aspect comes from the fact that the scalar
and gaugino masses are mutually related at the messenger scale. For
the sake of our discussion, let us disregard for the moment the
contributions proportional  to the Yukawa couplings in the
expressions of the scalar masses in Eqs.~(\ref{soft2}), as well as
the Yukawa effects in the (1-loop) renormalization. In such a case,
the low-energy sfermion masses have the form:
\bea\label{rel1}
\bmm^2_{\tf}(\mu_{s}) &=& \frac{2 k_a C^f_a}{N}~ \ov{M}^2_a + \Delta
R_a \;\;\;,\;\;\; \no \\
\Delta R_a &=&  \ov{M}^2_a \frac{2 k_a C^f_a (1 - x_a)}{b_a}\;
\;\;,\;\;\;x_a = \frac{g^4_a(\mu_{s})}{\bar{g}^4_a} ,
\eea
where $b_1= 33/5, b_2=1, b_3=-3$. The first term in
$\bmm^2_{\tf}(\mu_{s})$ corresponds to the high-energy boundary
contribution while the second one,  $\Delta R_a$, accounts for the
RG effects induced by the gaugino masses. The squark masses $m_\tq$
receive the main RG correction from the gluino mass term, which
amounts to a positive shift on $m_\tq^2$. The term $\Delta R_3$ is
larger than the boundary condition value (since $|1-x_3| > |b_a|/N =
3/7$), by a factor of approximately $8~(2)$ for $M_T =
10^{14}~(10^{7})~ {\rm GeV}$. Notice that, in the minimal GMSB model
($N=1$), the dominance of $\Delta R_3$ holds only for a messenger
scale above $\sim 10^{12}~{\rm GeV}$ ~\cite{gsmspec}. Therefore, we
expect the low-energy ratio $m_{\tq}/M_3$ to be given as
\be\label{rel2}
\frac{m_\tq}{M_3} \approx \left\{ \frac{2 C^q_3}{b_3}
\frac{1-x_3}{x_3} \,\left[1 + \frac{b_3}{N
(1-x_3)}\right]\right\}^{1/2}\,,
\ee
which lies within  $0.9 - 0.8$  for $M_T\sim 10^{14} - 10^{7} ~{\rm
GeV}$. Hence, the gluino is the heaviest of the coloured sparticles.
\begin{figure}
\begin{center}
\begin{tabular}{cc}
\includegraphics[width=8cm]{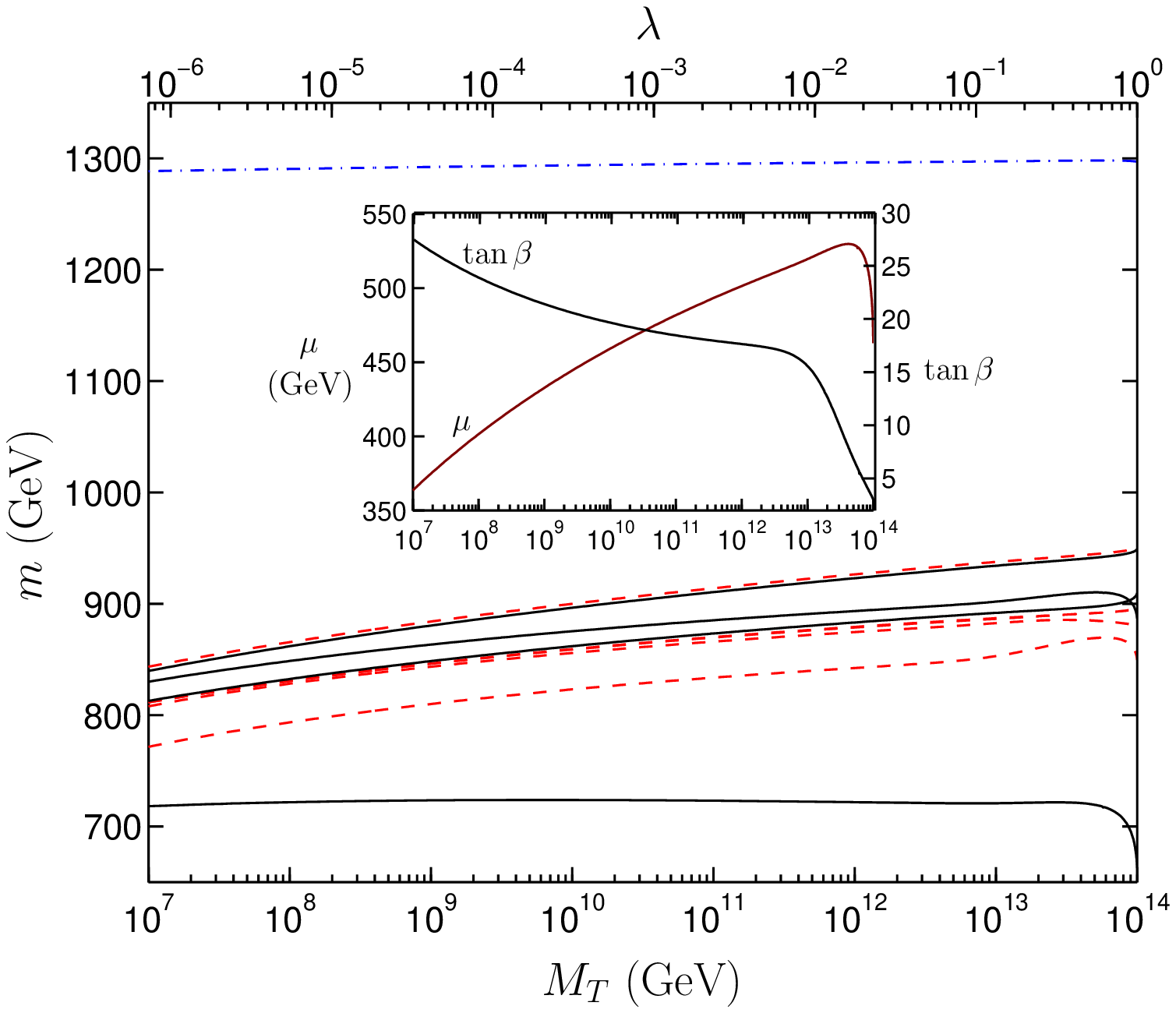} &
\includegraphics[width=8cm]{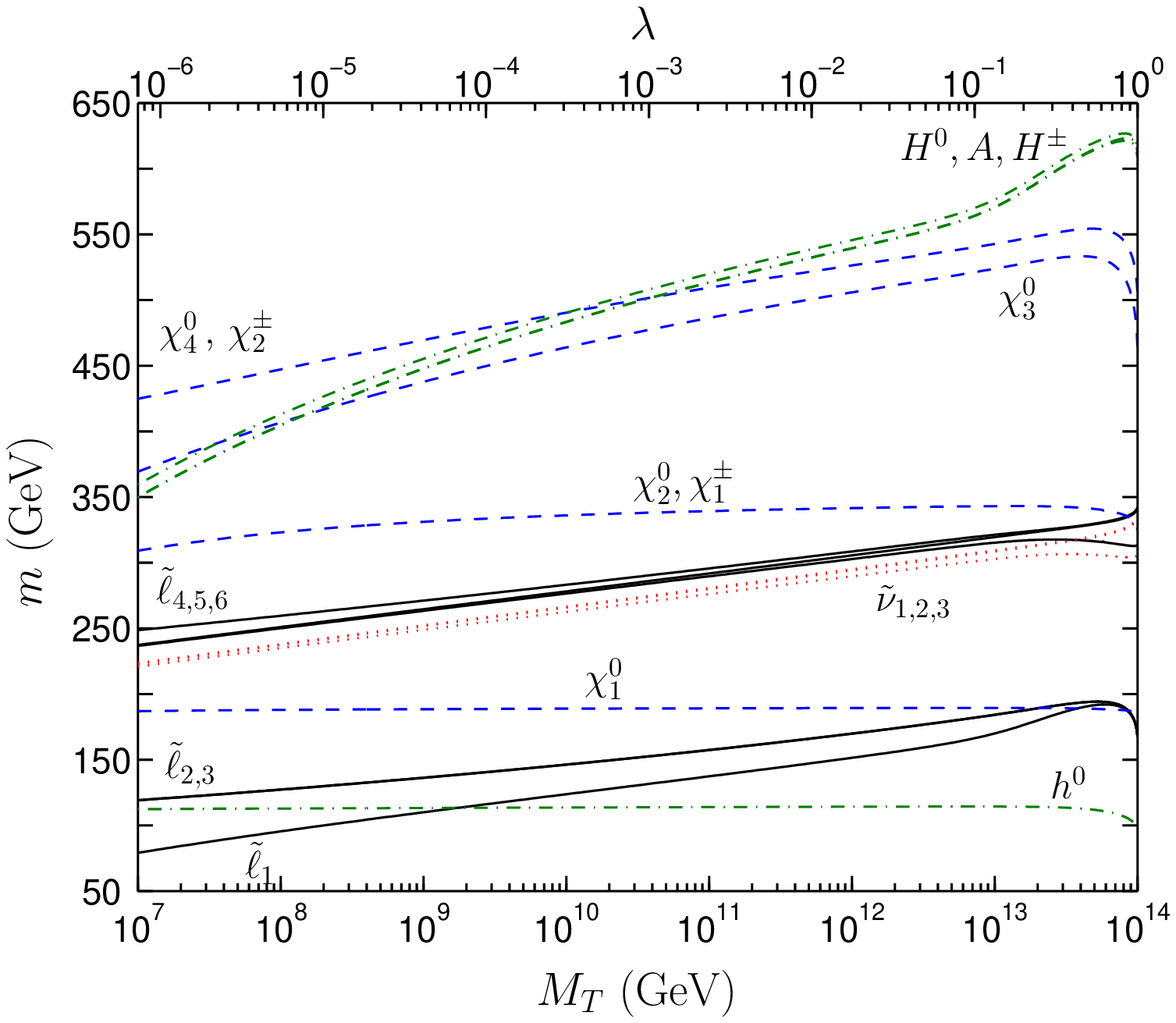}
\end{tabular}
\captions{\small The sparticle and Higgs spectrum for $B_T=20~{\rm
TeV}$ as a function of $M_T$ and correspondingly taking $\la$ (as
indicated in the upper horizonal axis) along the isocontour
$\BR({\mu \rightarrow e \gamma})=1.2\times 10^{-11}$. Left panel:
physical squark masses, $m_{\tilde{u}}$ (black solid lines),
$m_{\tilde{d}}$ (red dashed) and the pole gluino mass
$m_{\tilde{g}}$ (blue dash-dotted). Right panel: charged slepton
masses $m_{\tilde{\ell}}$ (solid black lines), sneutrino masses
$m_{\tilde{\nu}}$ (dotted blue), neutralino and chargino masses
$m_{\tilde{\chi}^0}$ and $m_{\tilde{\chi}^{\pm}}$ (dashed blue). The
dash-dotted green lines correspond to the Higgs boson masses $m_h,
m_H, m_A$ and $m_{H^\pm}$. In the small inner plot (on the left
panel) we show the behavior of $\tan\beta$ (red line) and $\mu$ as
obtained by the EWSB conditions.}
 \label{f8}
\end{center}
\end{figure}
%

In Fig.~\ref{f8} we display the sparticle and Higgs spectra for
$B_T=20~{\rm TeV}$. The values of $\lambda$ (shown in the upper
horizontal axis) and $M_T$, are the ones along the border which
delimits on the right the $\mu\to e\ga$ excluded region\footnote{The
sparticle and Higgs boson spectra do not depend much on $\lambda$
and so, the results shown in Fig.~\ref{f8} are quite representative
of the whole parameter space at $B_T=20$~TeV.} (see upper panel of
Fig.~\ref{f5}). The gluino pole mass includes the low-energy finite
corrections (see {\it e.g.}~\cite{BPMZ}), which ammount to
$20-30\,\%$ of the tree-level value $M_3$. As anticipated the gluino
is the heaviest superparticle. The first and second generation
squarks $\tilde{d}$ (dashed curves) and $\tilde{u}$ (solid) (mainly
composed by left-handed squarks) are the heaviest. Instead, the
lightest squark $\tilde{t}_1$ is mainly a right-handed $\tilde{t}^c$
whose mass is pushed down by a negative shift, driven mainly by  the
top Yukawa-induced renormalization. This effect, together with  the
left-right squark mixing, is enhanced for $M_T \gsim 10^{13}~{\rm
GeV}$ where $\tan\beta$ becomes smaller (see the inner panel where
we have drawn the behavior of $\tan\beta$ and $\mu$ versus $M_T$).
This analysis has shown that for $B_T=20~{\rm TeV}$ in the allowed
$(\la, M_T)$ portion, where the $\mu\to e \ga$ is close to the
present bound, the squark spectrum lies in the range $800 - 950
~{\rm GeV}$. This mass range will be soon explored by the Large
Hadron Collider (LHC)~\cite{LHC} with a luminosity
$\mathcal{L}=100$~pb$^{-1}$ and a center-of-mass energy
$\sqrt{s}=14$~TeV.

The left-handed slepton masses can instead be compared with the
$SU(2)_W$ gaugino mass $M_2$. In this case the renormalization
factor in Eq.~(\ref{rel1}) is comparable to the boundary
contribution. At low-energy, the ratio $m_{\tl}/M_2$ lies in the
range $1 - 0.5$ for $M_T= 10^{14} - 10^{7}~{\rm GeV}$. In the right
panel of Fig.~\ref{f8} we show the physical spectrum of the
electroweak states. The heaviest charged sleptons
$\tilde{\ell}_{4,5,6}$ (which are mainly left-handed sleptons) have
masses around $250 - 320~{\rm GeV}$ in the allowed range $M_T \sim
10^{8} - 10^{13}~{\rm GeV}$. The sneutrino masses are splitted from
the latter states mainly by the $SU(2)_W$ $D-$term,
$m^2_{\tilde{\ell}} - m^2_{\tilde{\nu}}= m^2_W|\cos 2\beta|$.

The right-handed slepton masses  receive contributions only from the
$U(1)_Y$ gauge interactions and, therefore, are smaller than the
other sfermion masses. In the range $M_T = 10^{14} - 10^{7}~{\rm
GeV}$, the high energy contribution is larger than the
renormalization term $\Delta R_1$ by a factor of $\sim 1 - 3$. The
low-energy ratio $m_{\te}/M_1$ is approximately in the range $1
-0.7$ for $M_T = 10^{14} - 10^{7}~{\rm GeV}$. To these estimates one
has to add the negative shift from the $Y_\tau$-induced
renormalization and the effect from the left-right slepton mixing at
the electroweak-symmetry breaking. Both these contributions are
important for large $\tan\beta$ and lower the physical
$\tilde{\tau}_1$ mass below $m_{\tilde{\tau}^c}$. In Fig.~\ref{f8}
(right panel) we can see that for $M_T \lsim 10^{8}~{\rm GeV}$ the
$\tilde{\tau}_1$ mass is pushed below $100~{\rm GeV}$ and, in the
range $M_T \sim  10^{8} - 10^{13}~{\rm GeV}$, $\tilde{\tau}_1$ is
indeed the lightest MSSM sparticle. For larger values of $M_T$,
corresponding to small $\tan\beta$, the neutralino
$\tilde{\chi}^0_1$ becomes the lightest MSSM sparticle. Hence,
either $\tilde{\tau}_1$ or $\tilde{\chi}^0_1$ would decay into the
gravitino, which is in fact the lightest SUSY particle in our
framework. In conclusion, the slepton masses lie in the range $\sim
100-320$~GeV (for $B_T=20$~TeV) and, therefore, are within the
discovery potential of the LHC.

Concerning the physical charginos and neutralinos, the inner plot
(right panel) shows that the parameter $\mu$ comes out to be larger
than $M_Z$ and $\mu^2 - M^2_{1,2} > M^2_Z$ hold for most of the
parameter space. These hierarchies imply that the lightest
neutralino $\tilde{\chi}^0_1$ is mainly a $B$-ino and has mass
$m_{\tilde{\chi}^0_1}\approx M_1$ , while $\tilde{\chi}^0_2$ and the
lightest chargino $\tilde{\chi}^{\pm}_1$ are almost degenerate and
are mainly $W$-inos with mass $\approx M_2$~\cite{GR}. Therefore,
both these masses do not exhibit a significant dependence on $M_T$,
as seen in Fig.~\ref{f8} (right panel). The  heaviest chargino
$\tilde{\chi}^{\pm}_2$ and neutralinos $\tilde{\chi}^0_{3,4}$ are
mostly higgsinos with mass set by the $\mu$ parameter, increasing
therefore with $M_T$.

Notice that, increasing the value of $B_T$, all the sparticles
become linearly heavier since they scale as $B_T$.

Finally, also the Higgs boson masses can be predicted in our
scenario. As already mentioned in Section~\ref{phenconst}, the mass
of the lightest CP-even Higgs boson ($m_h$ ) has been computed by
including the low-energy radiative corrections. In the allowed
parameter space of Fig.\ref{f5}, $m_h$ turns out to be in the range
$110-120~{\rm GeV}$, thus being testable in the near future  at the
LHC (mainly through the Higgs decay into 2 photons~\cite{higgs}).
The heaviest CP-even ($H$), CP-odd ($A$) and charged Higgs bosons
are much heavier. At tree-level, $ m_A = \mu B_H/\sin 2\beta$ and
for $m_A \gg M_Z$ all these states are almost degenerate, $m_H \sim
m_{H^\pm} \sim m_A$. For $B_T = 20~{\rm TeV}$, Fig.~\ref{f8} shows
that $m_A, m_H, m_{H^\pm} \approx 400 - 470 ~{\rm GeV}$. For larger
values of $B_T$ the masses of such non-standard Higgs bosons
increase, while $m_h$ increases by a few GeV due to the logarithmic
sensitivity to $B_T$. The Higgs sector is therefore characterized by
a decoupling regime with a light SM-like Higgs boson ($h$) and the
three heavy states ($H, A, H^\pm$).

\begin{figure}
\begin{center}
\includegraphics[width=8.1cm]{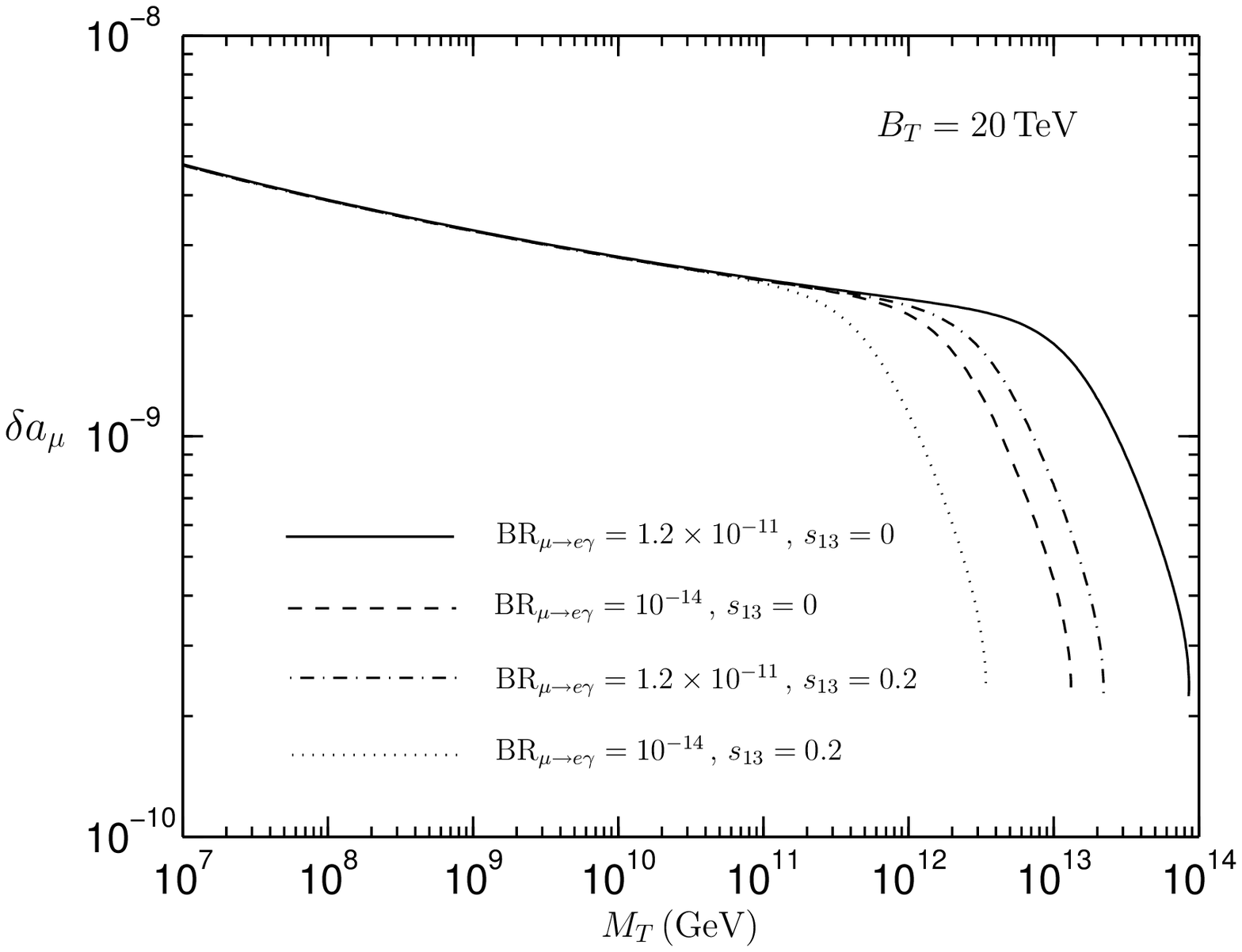}
\includegraphics[width=8.1cm]{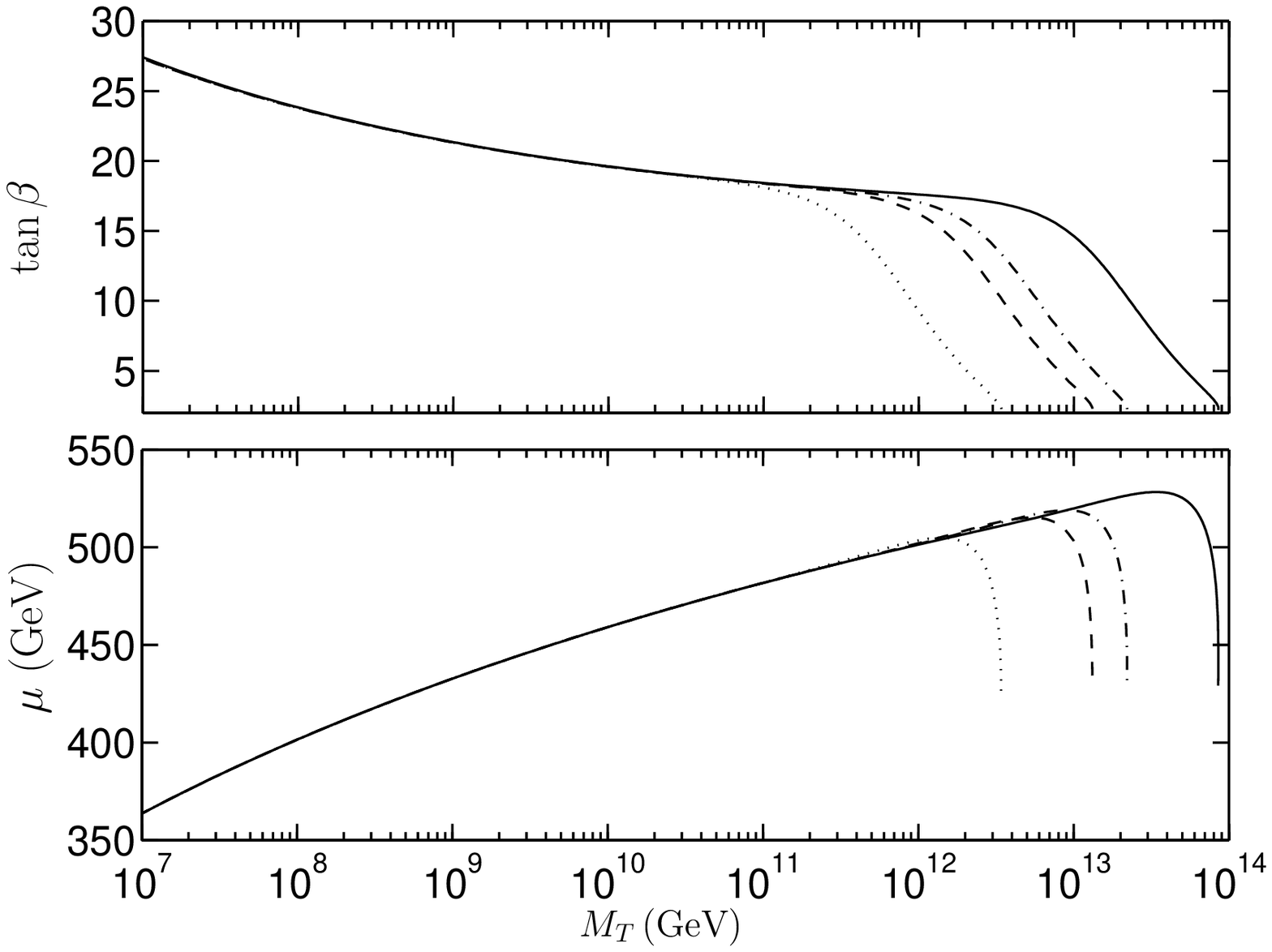}
\captions{\small Left panel: SUSY muon anomalous magnetic moment
contribution $\delta a_\mu$ as a function of $M_T$ (taken with its
corresponding $\la$ along the lines of constant ${\rm
BR}({\mu\rightarrow e \gamma})=1.2\times10^{-11},10^{-14}$) for
$s_{13}=0,0.2$ and $B_T=20$~TeV. Right panel: curves of $\mu$ (lower
plot) and $\tan\beta$ (upper plot), for the same inputs considered
in the left-panel.}
 \label{fg2}
\end{center}
\end{figure}

We conclude this section with a comment on the SUSY contribution to
the muon anomalous magnetic moment $\delta a_\mu$. In Fig.~\ref{fg2}
we show the $\delta a_\mu$ behavior as a function of $M_T$ in
correspondence with different values of ${\rm BR}({\mu\rightarrow e
\gamma})$ and for $B_T=20$~TeV. It is well known that $\delta a_\mu$
is induced by the dipole operator whose dominant contributions go as
$\sim\frac{g^2m_\mu^2 \mu\tan\beta}{16\pi^2m^3_{soft}}  $. Since the
sparticle spectrum does not change significantly with $M_T$
(Fig.~\ref{f8}), $\delta a_\mu$ directly reflects the demeanor of
$\tan\beta$ (cf. left and right panels). The sign of $\delta a_\mu$
is the same as the one of $\mu$ and, in the allowed range $M_T\gsim
2\times10^{8}$~GeV, $\delta a_\mu \lsim 4 \times10^{-9}$ respects
the constraint (\ref{gmu}). As $\delta a_\mu$ scales as $1/B_T^2$,
it decreases for larger $B_T$.

\subsection{LFV: Model Independent Predictions}
\label{LFVQFV}
We have already described the structure and parameter dependence of
the flavour violating SSB mass parameters in Section~\ref{Flavour
Structure}. Here, we intend to further investigate the
phenomenological implications by considering several LFV processes
(besides $\mu \to e \ga$) such as, $\mu \to e e e, \tau \to \mu \ga,
\tau \to e \ga, \tau \to \mu\mu\mu, \tau \to eee, \tau \to \mu e
e\,\tau \to  e \mu\mu$ and $\mu\to e$ conversion in $^{22}_{48}{\rm
Ti}$. The present experimental upper bounds and future sensitivities
for the BRs of these decays are collected in Table~\ref{tb1}.

Let us briefly recall some points related to the computation of such
processes. The radiative decays $\ell_j \to \ell_i +\gamma$ are
induced by the effective dipole operator:
\be
\label{dipole}%
e m_j (i D^L_{j i} \bar{\ell}_i\bar{\sigma}^{\rho
\sigma}\bar{\ell}^c_j + i D^R_{j i} {\ell}^c_i {\sigma}^{\rho
\sigma} \ell_j + {\rm h.c.}) F_{\rho\sigma} \,,
\ee
where $e$ and $F_{\rho\sigma}$ are the electric charge and the
electromagnetic field strength, respectively. The corresponding
branching ratios are:
\be
\label{br-rad}%
\BR(\ell_j \to \ell_i +\gamma) = \frac{48 \pi^3\alpha}{G^2_F}
\left[|D^L_{ji}|^2 + |D^R_{ji}|^2 \right]~ \BR(\ell_j \to \ell_i
\nu_j\bar{\nu}_i)\,,
\ee
where $\alpha = e^2/(4 \pi)$, $G_F$ is the Fermi constant, $\BR(\mu
\to e \nu_\mu \bar{\nu}_e)\approx 1,\, \BR(\tau \to \mu \nu_\tau
\bar{\nu}_\mu)\approx 17\%$ and $\BR(\tau \to e \nu_\tau
\bar{\nu}_e)\approx 18\%$. 
Since LFV resides in the parameters $\bmm^2_\tl$ and $\bA_e$, the
coefficient $D^L$ is the dominant one. In the mass-insertion
approximation its parametric dependence is:
\be
\label{pd1}%
D^L_{ji} \sim \frac{g^2}{16 \pi^2} \frac{1}{m^2_{soft}}
\left[\frac{(\bmm^2_\tl)_{ji}}{m^2_{\rm soft}} ~~{\rm or}~~
\frac{(\bmm^2_\tl)_{ji}}{m^2_{\rm soft}} \tan\beta ~~ {\rm or} ~~
\frac{(\bA_e)_{ji}}{(\bY_e)_{jj}m_{\rm soft}} \right] ,
\ee
where the coupling $g$ can be either $g_2$ or $g'=\sqrt{3/5}g_1$. In
the MSSM framework, the dipole coefficient $D^L$  receives
contributions from three different diagram topologies which, under
certain features of the sparticle spectrum, could cancel against
each other. In the latter case, these decays would be strongly
suppressed even if the sparticle spectrum were not too
heavy~\cite{br}. This can be realized if, for instance, the A-terms
$\bA_e$ are large, or the gaugino mass parameters $M_1$ and $M_2$
have opposite sign. In our scenario such peculiar situations do not
occur, so the suppression of the dipole operators can only arise
from large mass parameters ({\it i.e.} by increasing $B_T$) or from
cancelations inherent to the $\bmm^2_\tl$ off-diagonal entries, as
discussed in Section~\ref{Flavour Structure}.

The strict predictions regarding the ratios (\ref{LFratio2}) and
(\ref{ratios}) can be translated into predictions for the ratios of
the BRs~\cite{ar} once Eqs.~(\ref{br-rad}) and (\ref{pd1}) are taken
into account:
\bea
\label{brs1}%
\frac{\BR(\tau \to \mu\ga)}{\BR(\mu \to e\ga)} & \approx &
\left[\frac{(\bmm^2_{\tl})_{ \tau \mu}}{(\bmm^2_{\tl})_{\mu
e}}\right]^2 \frac{\BR(\tau \to \mu \nu_\tau \bar{\nu}_\mu)}
{\BR(\mu \to e \nu_\mu \bar{\nu}_e)} \approx\left\{\begin{array}{l}
300\quad\quad\quad \quad [s_{13}=0]\, \\
\\
2\, (3)\!\!\quad\quad\quad \quad  [s_{13}=0.2]
\end{array} \right.\,\no  \\
&& \no \\
&& \no \\
\frac{\BR(\tau \to e\ga)}{\BR(\mu \to e\ga)} & \approx &
\left[\frac{(\bmm^2_{\tl})_{ \tau e}}{(\bmm^2_{\tl})_{\mu
e}}\right]^2 \frac{\BR(\tau \to e \nu_\tau \bar{\nu}_e)} {\BR(\mu
\to e \nu_\mu \bar{\nu}_e)} \approx \left\{\begin{array}{l}
0.2\quad\quad\quad \quad \,\,\,\, [s_{13}=0] \\
\\
0.1 ~(0.3)\,\,\quad \quad  [s_{13}=0.2]
\end{array}\right.\,,
\no \\
&&
\eea
where the results on the r.h.s hold for all the three types of
neutrino mass spectrum and the numbers in parenthesis apply for IH
case, whenever this is different from the others.

Regarding the 3-body decays $\ell_j \to \ell_i\ell_k\ell_k$ and
$\mu\to e$ conversion in Ti, these receive contributions from $\ga$-
(dipole $D^{L,R}$ and monopole $C^{L,R}$ operators), $Z$- (monopole
$A^{L,R}$ operator) and Higgs- (monopole $\Delta^{L,R}$ operator)
exchange diagrams, as well as from box ($B^{LL,RR}, \, B^{LR,RL}$)
diagrams. The branching ratios are given by:
\bea
\label{br3body}%
\BR(\ell_j^- \to \ell_i^- \ell_k^+ \ell_k^-) &=& \frac{1}{8
G^2_F}\left\{ a_k|F^{LL}_{ji}|^2 + |F^{LR}_{ji}|^2 +|F^{RL}_{ji}|^2
+
a_k|F^{RR}_{ji}|^2 + \right. \no \\
&& \left. +4 e^2 {\rm Re}\left[D^L_{ji}(a_k\bar{F}^{LL}_{ji} +
\bar{F}^{LR}_{ji})
+D^R_{ji}(\bar{F}^{RL}_{ji} + a_k\bar{F}^{RR}_{ji})\right]\right. \no \\
&&  \left. +8 e^4 ( |D^L_{ji}|^2 + |D^R_{ji}|^2)\left( {\rm
log}\frac{m^2_j}{m^2_k} - b_k \right)\right\} \BR(\ell_j^- \to
\ell_i^- \bar{\nu}_i \nu_j)\,,
\eea
where $a_k = 1 (2), ~b_k = 3 (11/4) $ for $k\neq i\, (k=i)$, and the
coefficients $F^{N P} \,(N, P = L,R)$ are combinations of $A^{N},
B^{N P}, C^{N}$ and $ \Delta^{N}$ (see {\it e.g.} \cite{br}).
We have found some discrepancies between our analytical computations
concerning the coefficients $A^{L,R}$ with those published by some
authors~\cite{HMTY}. In Appendix~2 we have reported and discussed
our results for $A^{L,R}$. Moreover, in  Appendix~3 we present the
results for the box coefficients $B^{M N}$ relevant for the decays
$\tau \to \mu e e $ and $\tau \to e \mu \mu $, calculated to all
order in the electroweak-breaking effects. For the sake of brevity,
we refer to the existing literature ({\it e.g.} \cite{HMTY} and
\cite{br}) for the explicit formulas of all the other coefficients,
including the one relevant for the CR($\mu \to e\,$;\,Ti). Like for
the dipole $D^L$, the presence of LFV in the left-handed sector
implies that only the coefficients $A^L, B^{LL}, B^{LR}, C^L,
\Delta^L$ are important. We just recall their parametric dependence:
\be
\label{pd2}
 A^L_{ji}\,,\, \frac{B^{LL (LR)}_{ji}}{g^2}\,,\,
C^L_{ji} \,\sim\, \frac{g^2}{16 \pi^2} \frac
{(\bmm^2_\tl)_{ji}}{m^4_{soft}}\quad\quad , \quad\quad\Delta^L_{ji}
\sim \frac{g^2}{16 \pi^2}  \frac {(\bmm^2_\tl)_{ji}}{m^2_{soft}} .
\ee
From what has been commented above about suppressing $D^L$ in our
scenario, and by comparing Eq.~(\ref{pd1}) with (\ref{pd2}), one can
realize that, whenever the coefficients $D^L$ get suppressed, $A^L,
B^{LL, LR}$ and $ C^L$ undergo the same fate. Regarding $\Delta^L$,
one has that $\Delta^L_{ji} \to 0$ only if $(\bmm^2_\tl)_{ji}\to 0$
since these coefficients are insensitive to an overall mass scale
increasing. Consequently, the $D^L$ contributions to the 3-body BRs
[CR$(\mu\to e)$] are dominant with respect to those from the
remaining operators, due to the $\tan\beta$ and phase-space
logarithmic-factor [$\tan\beta$] enhancement.

Although the Higgs-exchange diagram contribution also benefits from
a $\tan\beta$ enhancement, its numerical relevance requires the
Higgs bosons $A$ and $H$ to be significantly lighter than the
sleptons, charginos and neutralinos~\cite{Hexc,brh,br,AH}. As
already discussed in Section~\ref{sparspec}, in our scenario this
does not occur, thus the Higgs-mediated contributions come out to be
subleading.

In the aforementioned dipole-dominance situation, the LFV processes
under consideration can be directly compared with the radiative
decays:
\bea \label{br3}
\frac{\BR(\mu \to e e e)}{\BR(\mu \to e \ga)}& \approx &
\frac{\alpha}{3 \pi}
\left(\ln\frac{m^2_\mu}{m^2_e} -\frac{11}{4}\right) \approx 6\times 10^{-3},\no\\
\frac{\BR(\tau \to e e e)}{\BR(\tau \to e \ga)}& \approx &
\frac{\alpha}{3 \pi}
\left(\ln\frac{m^2_\tau}{m^2_e} -\frac{11}{4}\right) \approx 10^{-2},\no\\
\frac{\BR(\tau \to e \mu \mu )}{\BR(\tau \to e \ga)}& \approx &
\frac{\alpha}{3 \pi}
\left(\ln\frac{m^2_\tau}{m^2_\mu} -3\right) \approx 2\times 10^{-3} , \no\\
\frac{\BR(\tau \to \mu \mu \mu)}{\BR(\tau \to \mu \ga)}& \approx &
\frac{\alpha}{3 \pi} \left(\ln\frac{m^2_\tau}{m^2_\mu}
-\frac{11}{4}\right)
\approx 2.2 \times 10^{-3} , \no\\
\frac{\BR(\tau \to \mu e e)}{\BR(\tau \to \mu \ga)}& \approx &
\frac{\alpha}{3 \pi} \left(\ln\frac{m^2_\tau}{m^2_e} -3\right)
\approx 10^{-2}
, \no\\
\frac{\CR(\mu \to e\,;\,{\rm Ti})}{\BR(\mu \to e \ga)}& \approx &
16\,\alpha^4 Z_{\rm eff}^4 Z |F(q)|^2\frac{\Gamma_{\mu}}{\Gamma_{\rm
capt}^{\rm Ti}} \approx 5\times 10^{-3}\,,
\eea
where $Z=22$, $Z_{\rm eff}=17.6$, $|F(q)|^2\approx 0.54$,
$\Gamma_{\rm capt}^{\rm Ti}\approx 2.6\times10^6\,{\rm s}^{-1}$ is
the muon capture width in Ti~\cite{HMTY} and $\Gamma_\mu \approx
4.5\times 10^{5}\,{\rm s}^{-1}$ is the total muon decay width. In
addition to these correlations, we also find that the 3-body decays
can be related to each other by using the ratios (\ref{ratios}) as
follows:
\bea
\label{br4}%
\frac{\BR(\tau \to \mu \mu \mu)}{\BR(\mu \to e e e )}& \approx &
\left[\frac{(\bmm^2_{\tl})_{ \tau \mu}}{(\bmm^2_{\tl})_{\mu
e}}\right]^2 \left(\frac{ {\rm ln}\frac{m^2_\tau}{m^2_\mu}
-\frac{11}{4}} { {\rm ln}\frac{m^2_\mu}{m^2_e} -\frac{11}{4}}
\right) \frac{\BR(\tau \to \mu \nu_\tau \bar{\nu}_\mu)} {\BR(\mu \to
e \nu_\mu \bar{\nu}_e)} \approx \left\{\begin{array}{ll}
100 & [s_{13}=0] \\
\\
0.6 ~(0.9)&[s_{13}=0.2]
\end{array}\right.\no \\
&& \no \\
\frac{\BR(\tau \to \mu e e)}{\BR(\mu \to e e e )}& \approx &
\left[\frac{(\bmm^2_{\tl})_{ \tau \mu}}{(\bmm^2_{\tl})_{\mu
e}}\right]^2 \left(\frac{ {\rm ln}\frac{m^2_\tau}{m^2_e} -3} { {\rm
ln}\frac{m^2_\mu}{m^2_e} -\frac{11}{4}} \right) \frac{\BR(\tau \to
\mu \nu_\tau \bar{\nu}_\mu)} {\BR(\mu \to e \nu_\mu \bar{\nu}_e)}
\approx \left\{\begin{array}{ll}
550 &\quad\,\,\,\,[s_{13}=0] \\
\\
3~(4) &\quad\,\,\,\,[s_{13}=0.2]
\end{array}\right.\no \\
&& \no\\
\frac{\BR(\tau \to e e e)}{\BR(\mu \to e e e )}& \approx &
\left[\frac{(\bmm^2_{\tl})_{ \tau e}}{(\bmm^2_{\tl})_{\mu
e}}\right]^2 \left(\frac{ {\rm ln}\frac{m^2_\tau}{m^2_e} -
\frac{11}{4} } { {\rm ln}\frac{m^2_\mu}{m^2_e} -\frac{11}{4}}
\right) \frac{\BR(\tau \to e \nu_\tau \bar{\nu}_e)} {\BR(\mu \to e
\nu_\mu \bar{\nu}_e)} \sim \left\{\begin{array}{ll}
0.3 &\,\,[s_{13}=0] \\
\\
0.2~(0.4)&\,\,[s_{13}=0.2]
\end{array}\right.
\no \\
&&
\no \\
\frac{\BR(\tau \to e \mu \mu )}{\BR(\mu \to e e e )}& \approx &
\left[\frac{(\bmm^2_{\tl})_{ \tau e}}{(\bmm^2_{\tl})_{\mu
e}}\right]^2 \left(\frac{ {\rm ln}\frac{m^2_\tau}{m^2_\mu} - 3 } {
{\rm ln}\frac{m^2_\mu}{m^2_e} -\frac{11}{4}} \right) \frac{\BR(\tau
\to e \nu_\tau \bar{\nu}_e)} {\BR(\mu \to e \nu_\mu \bar{\nu}_e)}
\approx \left\{\begin{array}{ll}
0.06 &[s_{13}=0] \\
\\
0.05~(0.07) &[s_{13}=0.2]
\end{array}\right.\,,
\no \\
&&
\eea
where the parenthesis enclose the results for the IH spectrum (when
different from the other cases). In Table~\ref{tb2} we present a
synoptic view of the correlation pattern predicted in our context,
assuming that the present bound on $\mu \to e \ga$ (Table~\ref{tb1})
is saturated, choosing $s_{13}=0\,(0.2)$ and setting all the
remaining neutrino parameters at their best fit points (\ref{bfp}).

\begin{table}
\renewcommand{\tabcolsep}{2.2pc}
\begin{center}\begin{tabular}{lrc}
\hline \hline\noalign{\smallskip} Expectation & $s_{13}=0$ &
$s_{13}=0.2$
\\  \hline \noalign{\smallskip}
$\BR(\tau^{-}\to \mu^{-} \ga) $ & $3\times 10^{-9}$ & $2\,(3) \times
10^{-11}$ \smallskip
\\\hline \noalign{\smallskip}
$\BR(\tau^{-} \to e^{-} \ga)$  & $2 \times10^{-12}$ &$
1\,(3)\times10^{-12}$
\smallskip\\ \hline
\noalign{\smallskip}$\BR(\mu^{-} \to e^{-} e^{+} e^{-}) $&  $6\times
10^{-14}$ & $6\times 10^{-14}$\smallskip\\ \hline
\noalign{\smallskip}%
$\BR(\tau^{-} \to \mu^{-}\mu^{+} \mu^{-}) $&  $7\times 10^{-12}$
&$4\,(6)\times 10^{-14}$
\smallskip\\ \hline \noalign{\smallskip}
$\BR(\tau^{-} \to \mu^{-} e^{+} e^{-} ) $&  $3\times 10^{-11}$ &
$2\,(3)\times 10^{-13}$
\smallskip\\ \hline \noalign{\smallskip}
$\BR(\tau^{-} \to e^{-} e^{+} e^{-} ) $ &  $2\times 10^{-14}$ &
$1\,(3)\times 10^{-14}$
\smallskip \\ \hline\noalign{\smallskip}
$\BR(\tau^{-} \to e^{-} \mu^{+} \mu^{-})  $&  $3\times 10^{-15}$ &
$2\,(4)\times 10^{-15}$
\smallskip \\ \hline\noalign{\smallskip}
$\CR(\mu\to e\,;\,{\rm Ti})  $&  $6\times 10^{-14}$ & $6\times
10^{-14}$
 \\
\hline \hline
\end{tabular}\end{center}
\captions{\small Expectations for the various LFV processes from
Eqs.~(\ref{brs1}), (\ref{br3}) and (\ref{br4}), assuming $\BR(\mu
\to e\ga) = 1.2\times 10^{-11}$. The results in parenthesis apply to
the case of the IH neutrino spectrum, whenever these are different
from those obtained for NH and QD.} \label{tb2}
\end{table}

\subsection{LFV processes: Numerical analysis}
\label{numanalysis}%

We are now in position to analyse our numerical results regarding
the predictions on the LFV decay branching ratios. At this point, we
are interested in knowing how and to what extent the LFV processes
can probe the parameter space of our framework. In other words, will
the upcoming experimental sensitivities be enough to test the
allowed parameter space of Figs.~\ref{f5} and \ref{f6}?
Instead of plotting the contours relative to the various $\BR$s in
the allowed $(\lambda,M_T)$ space, we have fixed $B_T=20~{\rm TeV}$,
selected some values of $M_T$ and performed the analysis along one
direction, which can be either $s_{13}$ (and the phase $\delta$) or
$\la$. This is representative enough and allows us to properly
display the main features.

\begin{figure}
\begin{center}
\begin{tabular}{cc}
\includegraphics[width=7.5cm]{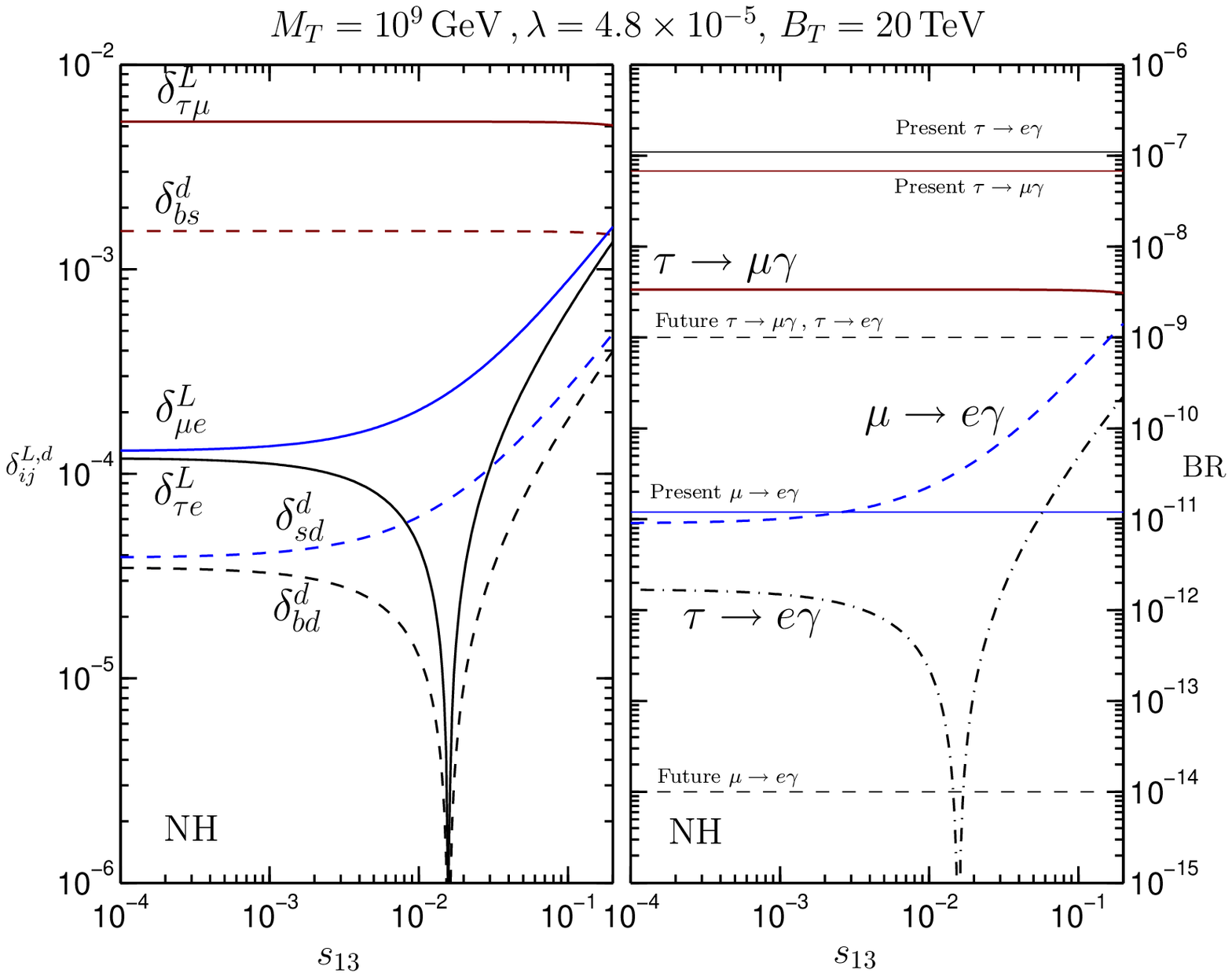} &
\includegraphics[width=7.5cm]{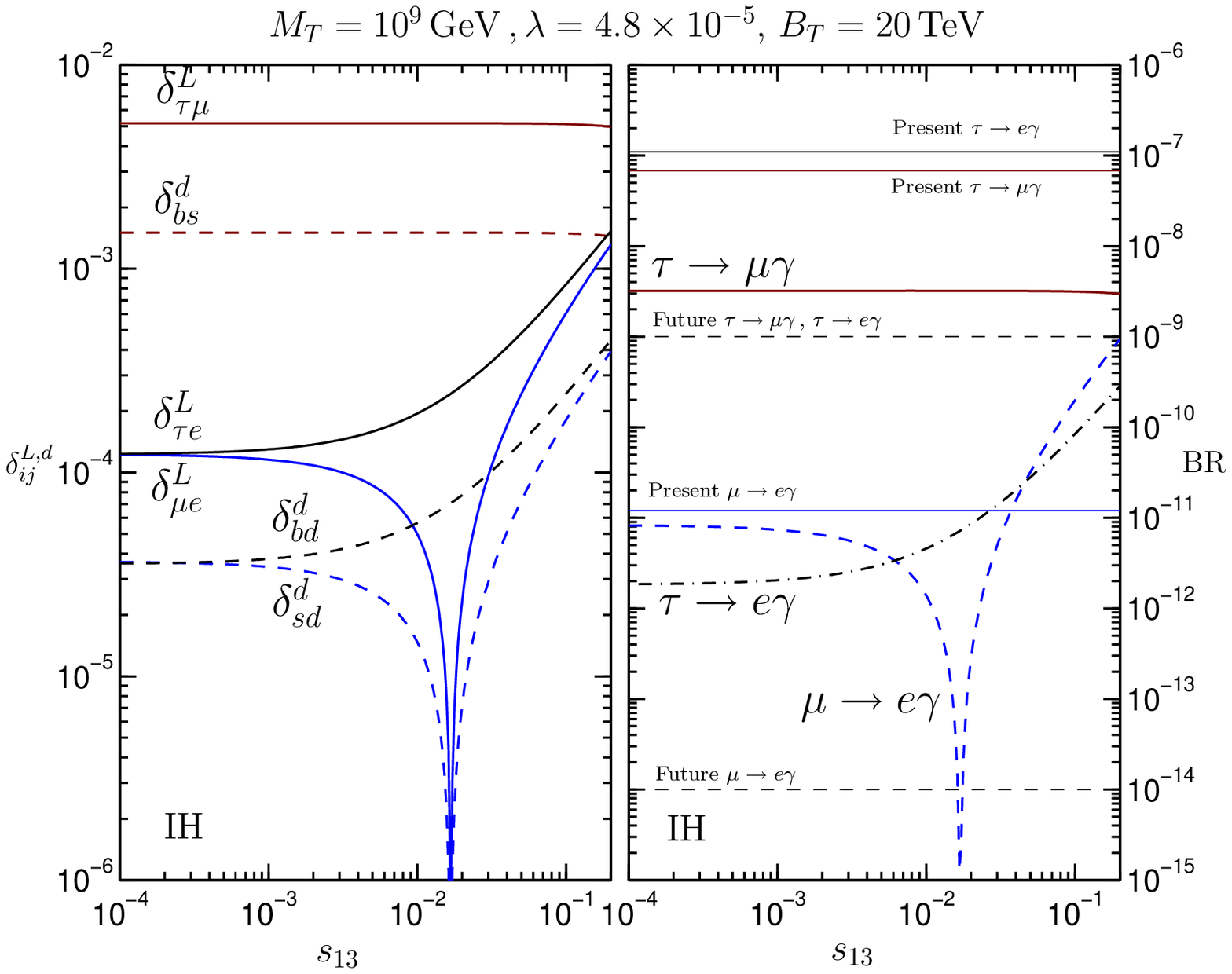}\\
  &\hspace*{-7.7cm}\includegraphics[width=7.5cm]{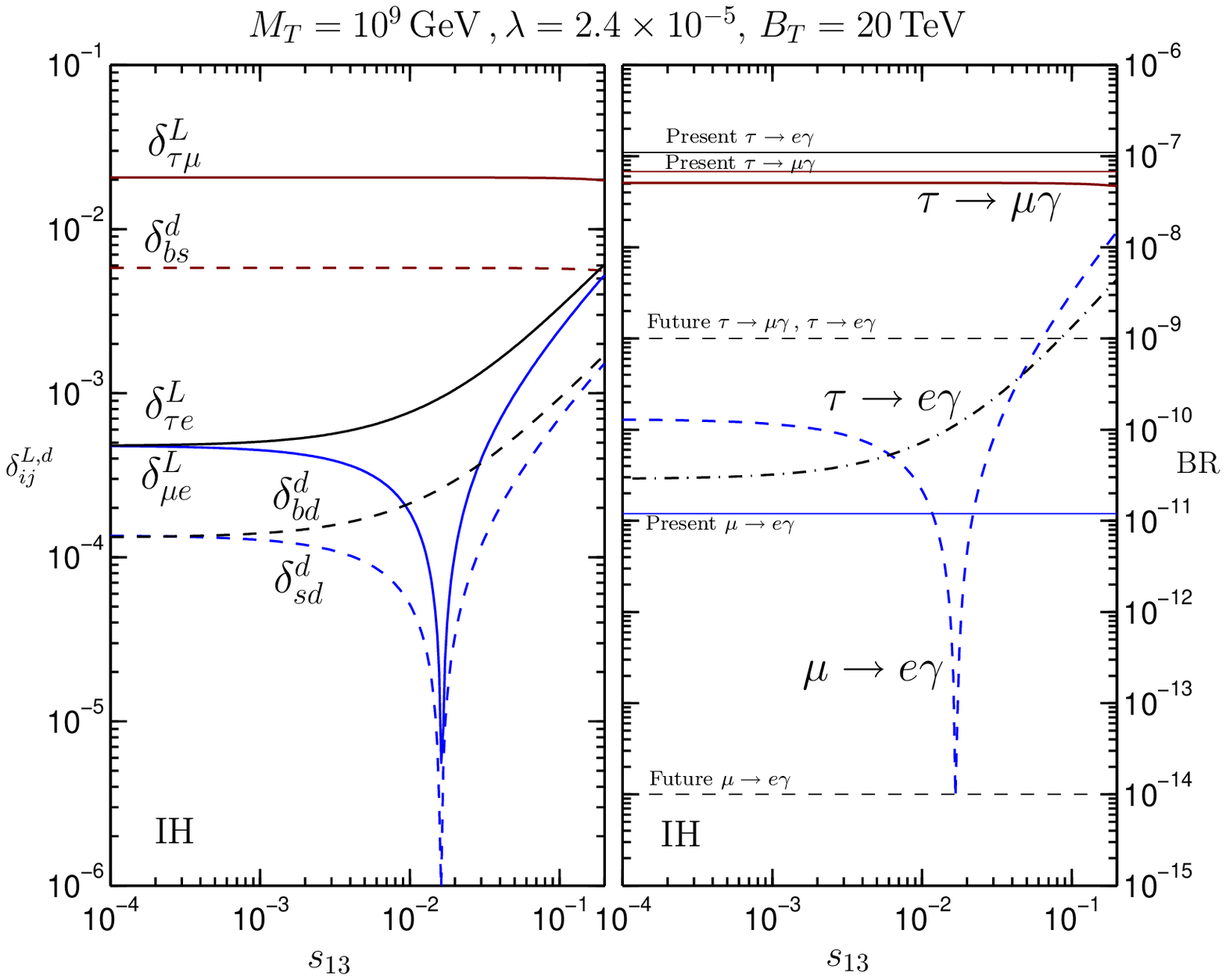}
\end{tabular}
\captions{\small FV parameters $\delta^L_{ij}, (i,j =e, \mu, \tau)$,
$\delta^d_{ij}, (i,j = d, s, b)$ and $ \BR(\ell_j\rightarrow \ell_i
\ga)$ as a function of $s_{13}$ for the NH (left upper-panel) and IH
(right upper-panel and lower panel). The parameters $M_T$, $B_T$ and
$\la$ are fixed as indicated on the top of each panel. For each
radiative decay, the present upper bound (horizontal solid-lines)
and future experimental sensitivity (horizontal dashed-lines) for
the BRs are shown.} \label{f9}
\end{center}
\end{figure}

Consider the effective size of LFV and QFV, which can be
parameterized by the following dimensionless
parameters~\cite{univers,lfv2}:
\be
\label{df}%
\delta^{L}_{ij}= \frac{|(\bmm^2_{\tl})_{ij}|}{m^2_{\tl}} ,
\,\,\,(i,j =e,\mu,\tau) \quad\quad,\quad\quad \delta^{d}_{ij}=
\frac{|(\bmm^2_{\td})_{ij}|}{m^2_{\td}}  ,\,\, \, (i,j =d,s,b)\,,
\ee
where $m^2_{\tl}$ and  $m^2_{\td}$ are the average $\tl$ and $\td$
masses, respectively.
The parameters $\delta^{L, d}$ are independent of $B_T$, while their
overall size is determined by the ratio  $M_T/\la$. In Fig.~\ref{f9}
we have plotted $\delta^L_{ij}$ and $\delta^d_{ij}$ as a function of
$s_{13}$ for two selected points of the parameter space shown in
Fig.~\ref{f5}: $(\la, M_T)=  (4.8\times 10^{-5}, 10^{9}~{\rm GeV})$
(Fig.~\ref{f9}, upper panels) and $(\la, M_T)=  (2.4\times 10^{-5},
10^{9}~{\rm GeV})$ (Fig.~\ref{f9}, lower panel).
The first point lies in the allowed portion of the parameter space
studied in Fig.~\ref{f5} for $s_{13} =0$ (very close to the region
delimited by the  present $\mu \to e\ga$ bound), whereas it is
excluded for $s_{13}=0.2$. The second point falls into the region
excluded  by  the $\mu \to e \ga$ bound, for the NH (and QD)
spectrum, irrespective of $s_{13}$ (but it is allowed in the IH case
for some values of $s_{13}$).

The behavior of $\delta^{L, d}$ just reflects that of the relevant
FV structure $\bV (\bmm^D_\nu)^2 \bV^\dagger$ in Eq.~(\ref{yy2}).
For example, notice that they scale as $1/\la^2$ outside the
cancelation `dip'. This figure clearly shows  that $\delta^L_{\tau
\mu}, \delta^d_{b s}$ are insensitive either to $s_{13}$ or to the
type of neutrino spectrum (cf. the upper panels), while
$\delta^L_{\mu e} (\delta^d_{s d})$  gets exchanged with
$\delta^L_{\tau e} (\delta^d_{b d})$ when passing from the NH to the
IH case. This latter feature is due to the fact that $\theta_{23}=
\pi/4$ and so the flavours $\mu$ and $\tau $ (or $b$ and $s$) are
indistinguishable. The relative ratios $\delta^L_{\tau \mu}/
\delta^L_{\mu e}, \, \delta^d_{b s}/ \delta^d_{s d}$ and
$\delta^L_{\tau e}/ \delta^L_{\mu e} , \,\delta^d_{b d}/ \delta^d_{s
d}$ just reproduce the absolute values of $R_{23/12}$ and
$R_{13/12}$ in Eq.~(\ref{ratios}), respectively. As we have already
deduced from (\ref{yy2}), this constant-ratio rule is violated for
$s_{13} \approx 0.02$,  where $\delta^L_{\tau e}, \, \delta^d_{b d}$
and $\delta^L_{\mu e}, \, \delta^d_{s d}$ are strongly suppressed
for the NH and IH spectrum, respectively. All the above
peculiarities are common to the related curves of
$\BR(\ell_j\to\ell_i\ga)$, plotted on the right of each panel. In
particular, the ratios (\ref{brs1}) are remarkably reproduced,
except for $s_{13} \approx 0.02$ where $\BR(\tau \to e \ga)$ and
$\BR(\mu \to e \ga)$ undergo a sharp suppression for the NH and IH
spectrum, respectively. Hence, the point with $\la= 4.8\times
10^{-5}$ and  $s_{13}\sim 0.02$ is excluded by the $\BR(\mu \to
e\ga)$ bound in the NH case, whereas it is allowed in the IH with
$\BR(\tau \to \mu \ga) \sim 3\times 10^{-9}$, $\BR(\tau \to e \ga)
\sim {\cal O}(10^{-11})$. The point with $\la=2.4\times 10^{-5}$
(lower panel) is phenomenologically viable if the neutrinos have IH
masses and $s_{13} \sim 0.02$. In such a case, $\BR(\mu \to e
\ga)\ll 10^{-11}$, $\BR(\tau \to \mu \ga) \sim 5\times 10^{-8}$ and
$\BR(\tau \to e \ga)\sim {\cal O}(10^{-10})$. This is an example
where $\BR(\tau\to \mu\ga)$ is close to the present bound and
$\mu\to e \ga$ might be unobservable.

In the examples considered above, the size of the $\delta^d$'s in
each family sector is smaller by a factor of $\sim 3.5$  than the
one of the corresponding $\delta^L$'s. This results from different
compensating effects. Namely, the squark masses are about three
times larger than the slepton ones (due to the gluino-induced
renormalization effect; see also Fig.~\ref{f8}), while the
$(\bmm_\td^2)_{ij}$ are larger than the $(\bmm_\tl^2)_{ij}$ at the
messenger scale $M_T$ (because of the major strong coupling
contribution). In the first example, $\delta^d_{b s}~\sim 1.5\times
10^{-3}$ (left panel). This is the maximal value attainable by
$\delta^d_{b s}$ in the allowed parameter space of Fig.~\ref{f5}. To
perceive the phenomenological relevance of such $\delta^d_{b s}$, we
need to confront it with the gluino and squarks masses. Recall that
for $B_T=20~{\rm TeV}$ we got $m_\tq \sim 900~{\rm GeV}$ and $M_3
\sim 1.3~{\rm TeV}$ (see Fig.~\ref{f8}). Then, $\delta^d_{b s}\sim
10^{-3}$ is well below the experimental bound posed by the measured
$\BR(b\to s \ga)$~\cite{deltas}. The predicted value $\delta^d_{s
d}\sim 5\times 10^{-4}$ (obtained for $s_{13}=0.2$) also lies below
the bound inferred from
$K^0-\bar{K}^0$ mixing~\cite{deltas}.\\

\begin{figure}
\begin{center}
\includegraphics[width=8.0cm]{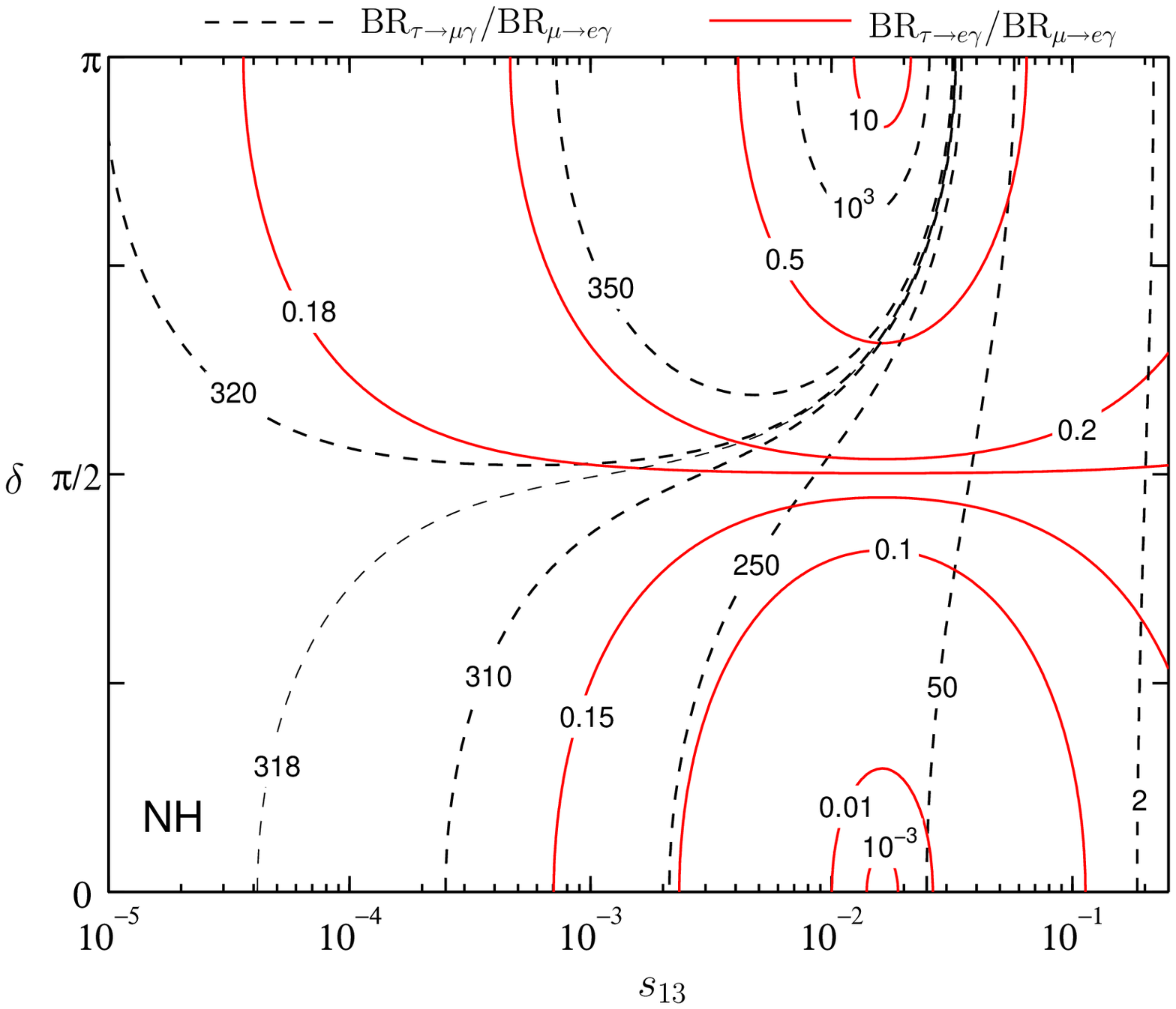}
\includegraphics[width=8.0cm]{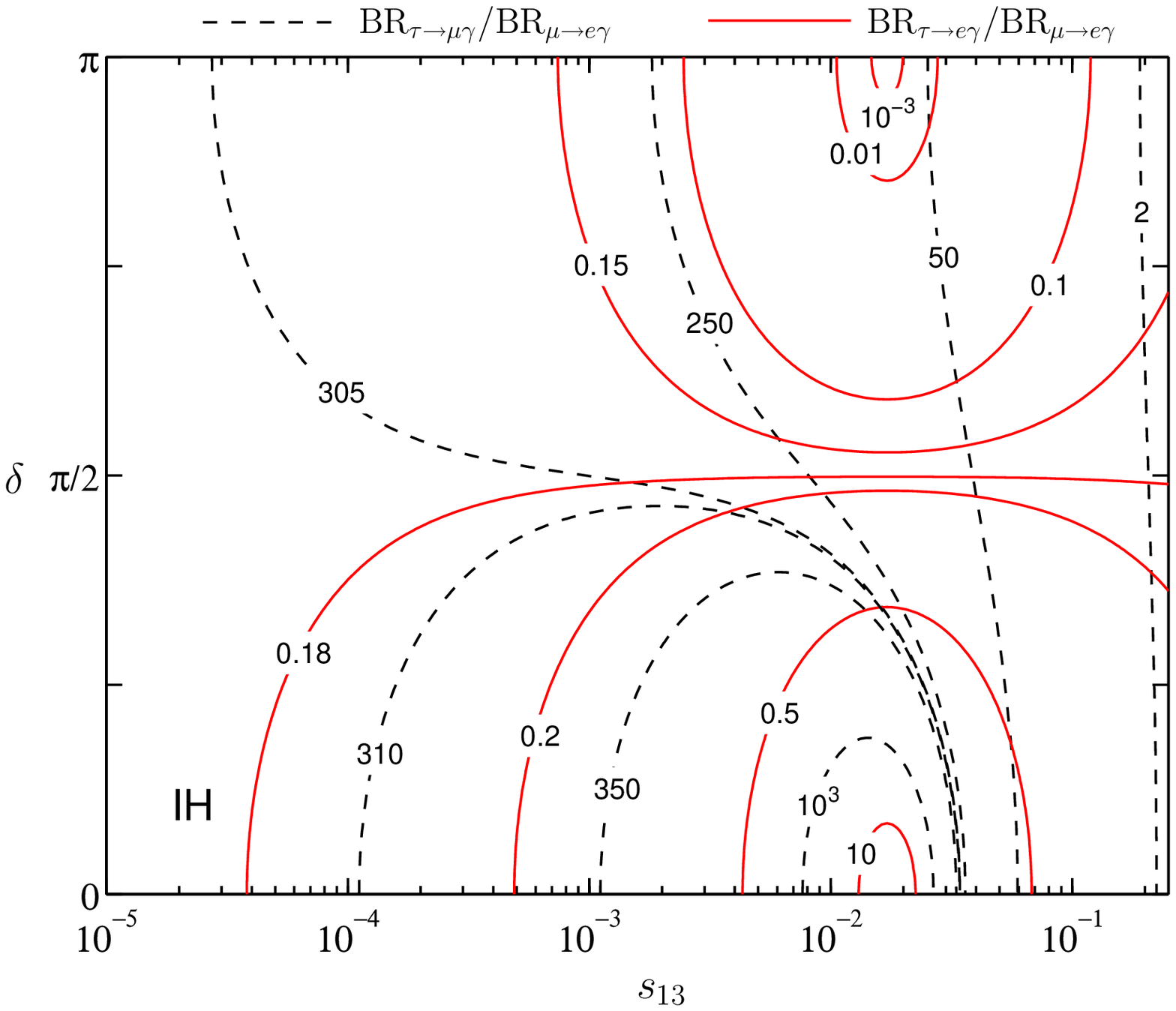}
\captions{\small Contours of the ratios ${\rm BR}{(\tau \to \mu
\ga)}/{\rm BR}{(\mu\to e \ga)}$ (dashed lines) and  ${\rm BR}{(\tau
\to e\ga)}/{\rm BR}{(\mu\to e \ga)}$ (solid lines) in the
$(s_{13},\delta)$-plane, for the NH (left panel) and IH (right
panel) neutrino mass spectra.}
 \label{f10}
\end{center}
\end{figure}

Fig.~\ref{f10} is aimed to address  the dependence of the ratios
$\BR(\tau \to  \mu \ga)/ \BR(\mu \to e \ga)$ and $\BR(\tau \to e
\ga)/ \BR(\mu \to e\ga)$ on $s_{13}$ and the CP-violating phase
$\delta$ for a NH (left panel) and IH spectrum (right panel). The
results do not depend on the ratio $M_T/\la$ as far as either the
quadratic or the quartic terms dominate in the LFV entries
$\bmm^2_\tl$. Regarding the ratio $\BR(\tau \to  \mu \ga)/ \BR(\mu
\to e \ga)$, it slightly decreases (increases) with increasing
$s_{13}$ for $\delta < \pi/2$ ($\delta>\pi/2$).
However, this ratio blows up at $s_{13}\sim 0.02$ and $\delta
\approx \pi \,(0)$ for the NH (IH) spectrum. The $\BR(\tau \to e
\ga)/ \BR(\mu \to e \ga)$ is also sensitive to both $s_{13}$ and
$\delta$. In fact, for a given value of $s_{13}$, the increase of
this ratio with $\delta$ is quite modest in most of the $s_{13}$
range. Still, for $s_{13}\approx 0.02$, $\BR(\tau \to e \ga)/
\BR(\mu \to e \ga)$ goes to zero (infinity) when $\delta=0\,(\pi)$
for the NH case. Instead, the opposite occurs when the IH pattern is
considered since there is an interchange of the roles played by
$\BR(\mu \to e \ga)$ and $\BR(\tau \to  e \ga)$ (as already
discussed). In short, barring the range around $s_{13}\approx 0.02
$, the predictions (\ref{brs1}), (\ref{br3}) and (\ref{br4}) are not
substantially altered when the
effects of $\delta\neq 0$ are included.\\

\begin{figure}
\begin{center}
\begin{tabular}{cc}
\includegraphics[width=7.8cm]{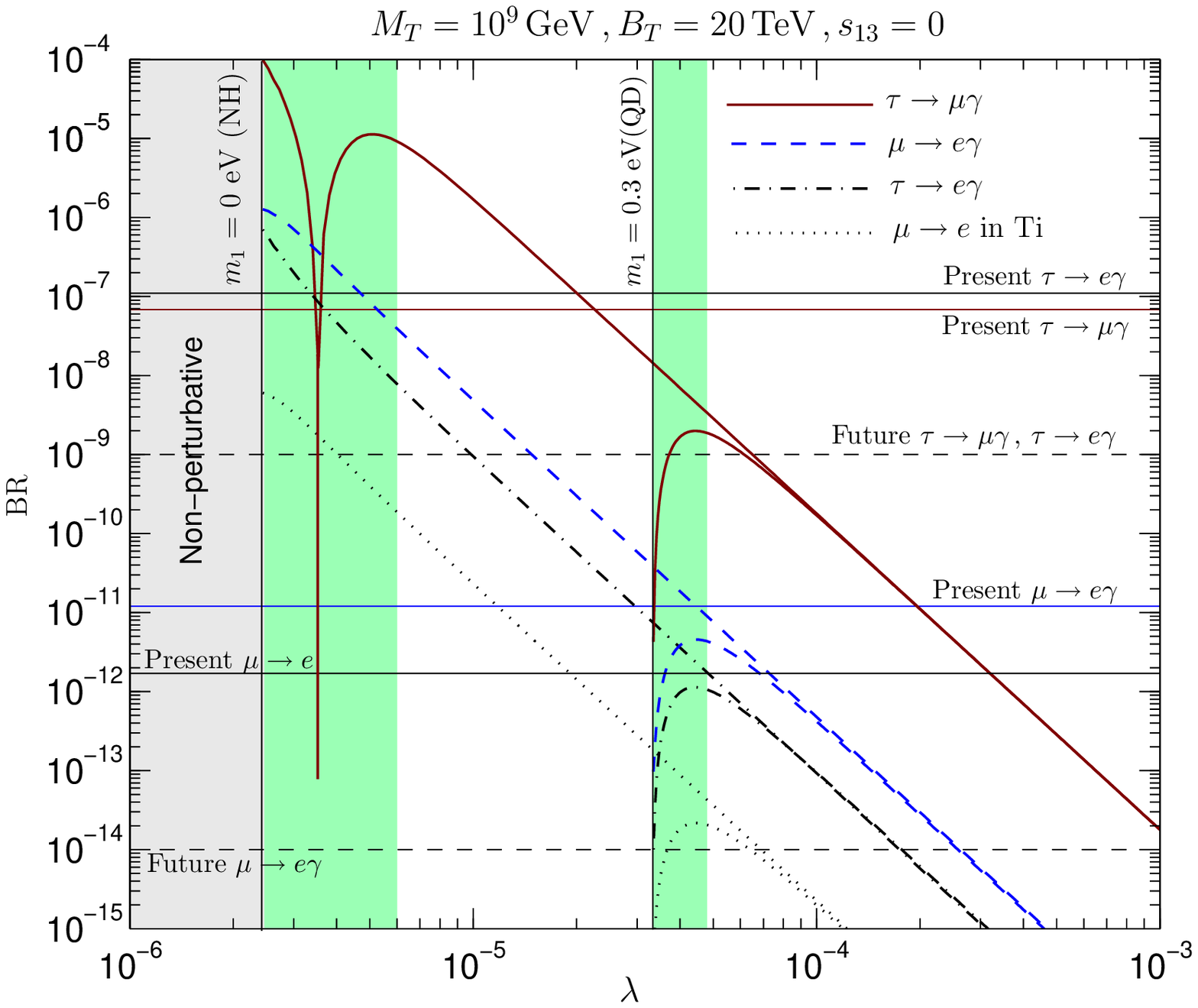} &
\includegraphics[width=7.8cm]{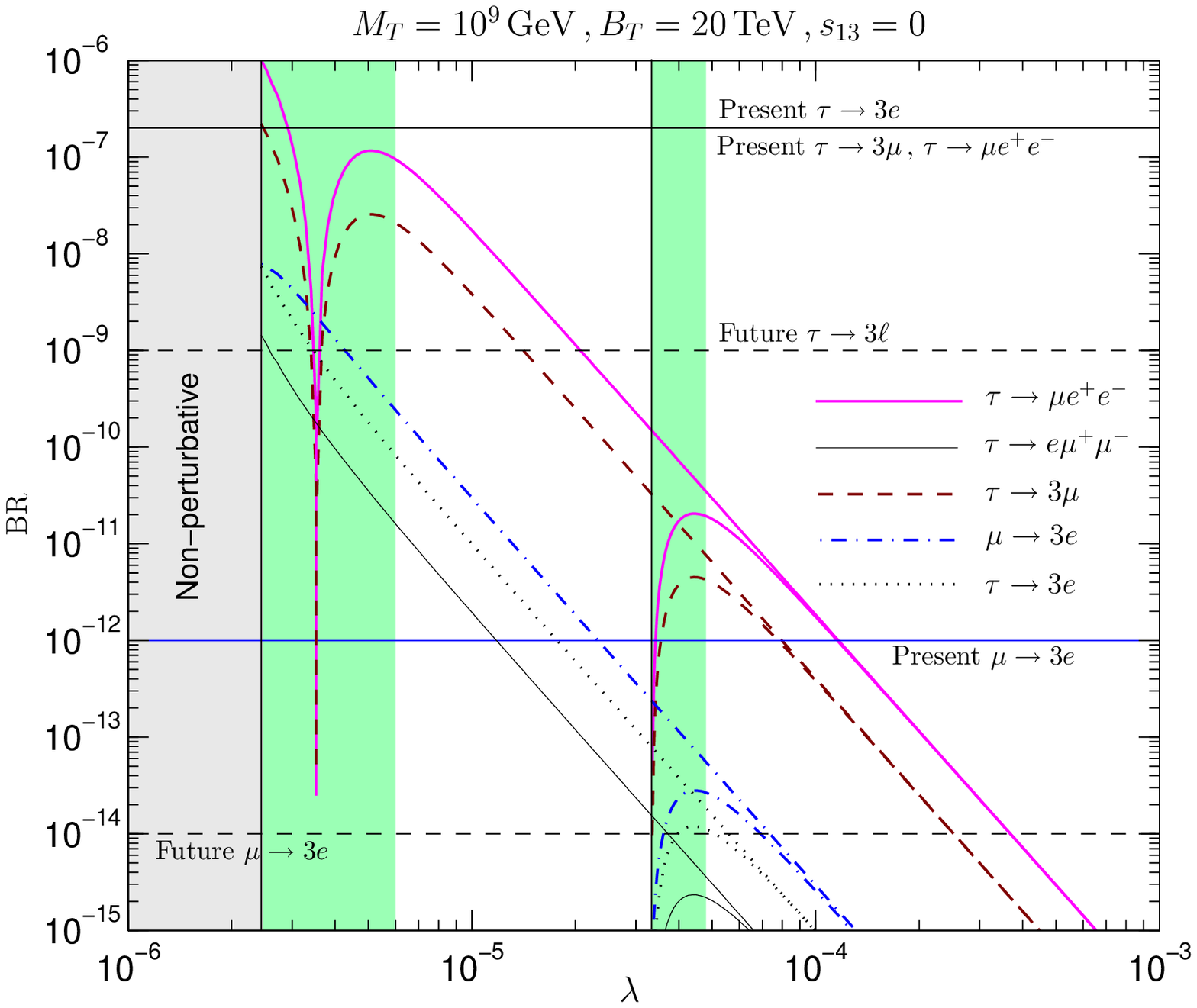}\\
\includegraphics[width=7.8cm]{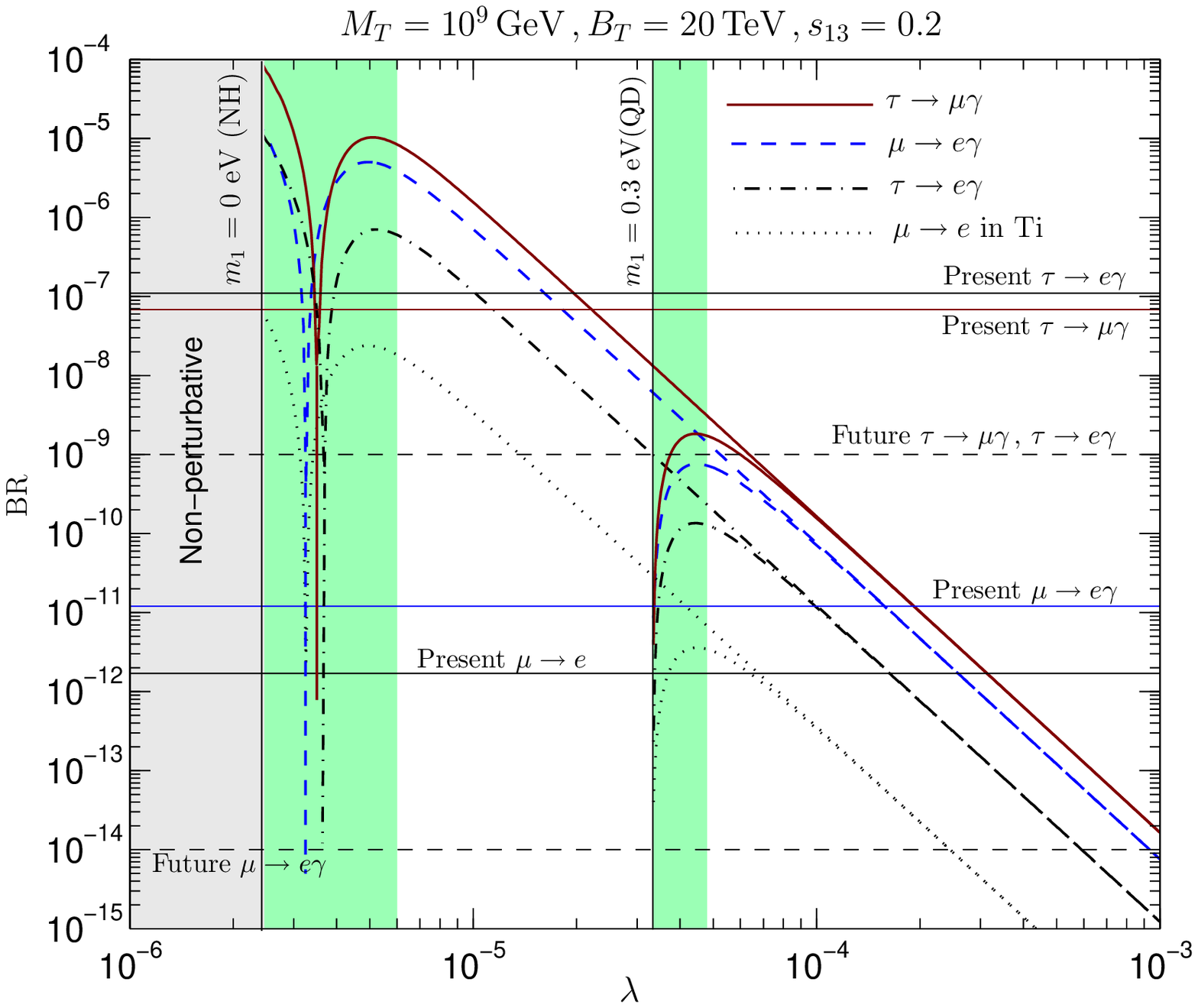} &
\includegraphics[width=7.8cm]{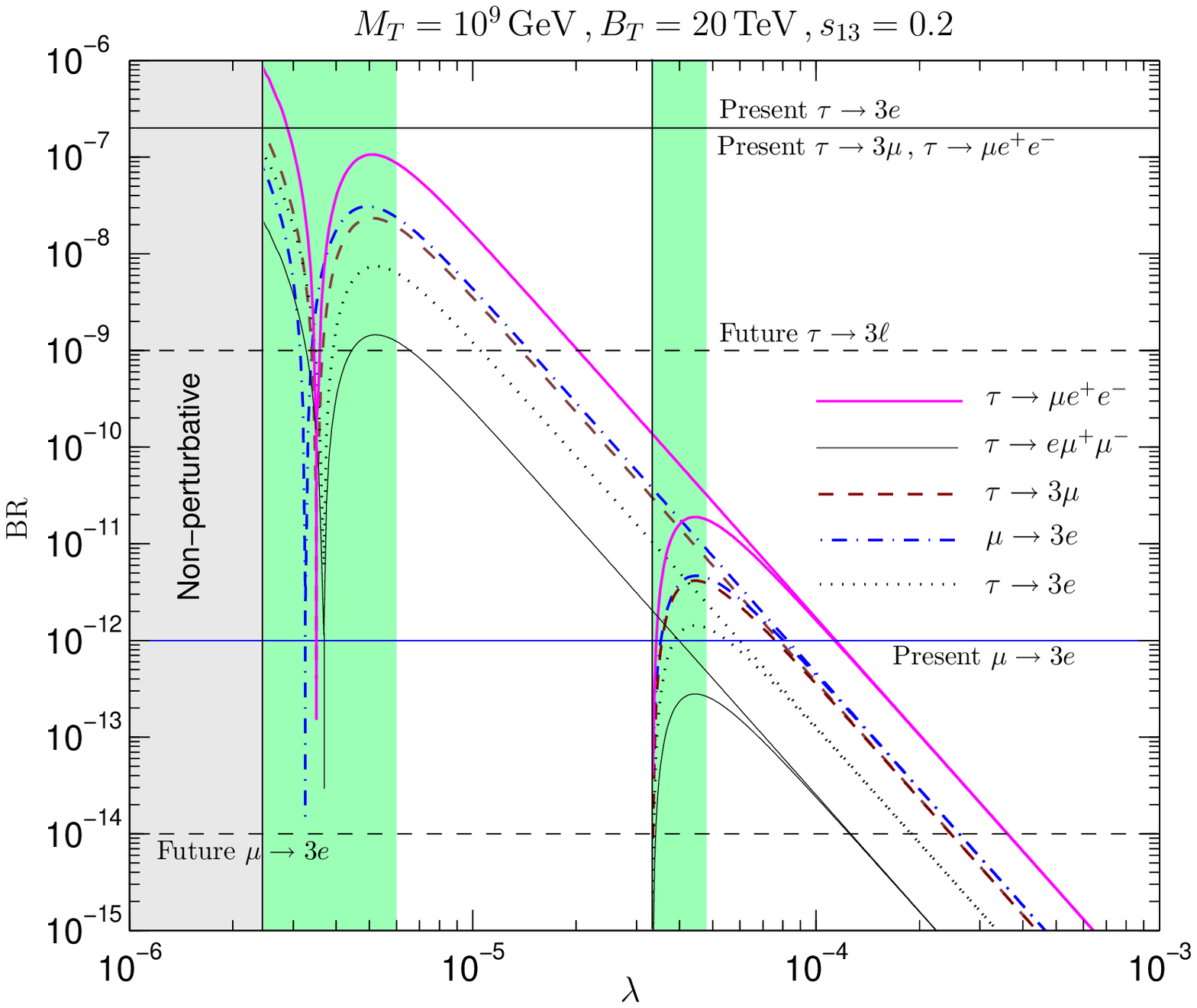}
\end{tabular}
\vspace*{-0.3cm} \captions{\small Left panels: branching ratios of
the radiative decays $\ell_j\rightarrow \ell_i \gamma$ and
$\mu\rightarrow e$ conversion in Ti, as a function of $\la$. Right
panels: branching ratios of the three-body decays $\ell_j \to
3\ell_i$ and $\tau\rightarrow \mu e e\,,\,e\mu\mu$. All the results
are shown as a function of $\la$ for $M_T = 10^{9}~{\rm GeV}$ and
$B_T=20$~TeV taking $s_{13}=0$ (upper plots) and $s_{13}=0.2$ (lower
plots). The left (right) vertical line indicates the lower bound on
$\lambda$ imposed by requiring perturbativity of the Yukawa
couplings $\bY_{T,S,Z}$ when $m_1=0\,(0.3)$~eV [NH (QD) neutrino
mass spectrum]. The regions in green (grey) are excluded by the
$m_{\tilde{\ell}_1} > 100$~GeV constraint (perturbativity
requirement when $m_1=0$\,eV).} \label{f11}
\end{center}
\end{figure}

Finally, we come to a comparative analysis of radiative decays,
3-body decays and $\mu \to e$ conversion in Ti considering the three
types of neutrino spectrum. Fig.~\ref{f11} shows the dependence of
the $\BR$s on $\la$ for $M_T=10^9~{\rm GeV}$ and $s_{13}=0\, (0.2)$
in the upper (lower) panels, assuming a NH spectrum ($m_1 =0~{\rm
eV}$).

For $s_{13}=0$ there is a cancelation dip only in $\BR(\tau \to \mu
\ga)$ (left panel), $\BR(\tau \to \mu ee)$ and  $\BR(\tau \to 3
\mu)$ (right panel) at $\la\sim 3.5\times 10^{-6}$, which lies in
the (green) region excluded by the negative sparticle search, $\la
\leq 6\times 10^{-6}$ (cf. also Fig.~\ref{f5}: upper panel). This
dip originates from the cancelations of the quadratic and quartic
terms in $\bmm^2_\tl$ discussed in Section~\ref{Flavour Structure}.
In $\BR (\mu \to e\ga)$ this would take place at a value of $\la$
smaller by a factor of $\sim (6/78)^{1/2} \sim 0.3$ [see
Eq.~(\ref{NH1})] and so, would fall into the (grey) non-perturbative
range. For $s_{13}=0.2$ the dips are instead present for all the
BRs, as we expect on the basis of Eq.~(\ref{NH2}).
Outside the cancelation regions, the relative ratios of $\BR(\ell_j
\to \ell_i \ga)$ are those announced in Eq.~(\ref{brs1}) and the
$\BR(\ell_j \to \ell_i\ell_k\ell_k)$ are correlated with $\BR(\ell_j
\to \ell_i \ga)$ according to Eq.~(\ref{br3}). For comparison, we
have also plotted all these $\BR$s for the case of the QD spectrum
with $m_1=0.3~{\rm eV}$ which can be obtained by `continously'
rising the mass $m_1$ in the NH case beyond $(\Delta m^2_A)^{1/2}$
[cf. Eq.~(\ref{nh})]. In the QD case the non-perturbative range
extends much above the one relative to the NH (grey) so that the
perturbativity lower bound on $\la$ (indicated by the vertical solid
line) is larger by about one order of magnitude, $\la \geq 3 \times
10^{-5}$. Notice that a more restrictive lower bound is imposed by
sparticle searches (green), $\la \geq 5 \times 10^{-5}$. The
cancelations occur at approximately the same $\la$ (inside the
excluded regions) for all the BRs. For $\la \geq 5 \times 10^{-5}$,
the curves corresponding to the NH and QD are superimposed and so
the two scenarios are not distinguishable [cf. Eqs.~(\ref{yy1}) and
(\ref{yy2})].

Finally, Fig.~\ref{f11} reveals that, for $M_T=10^{9}~{\rm GeV}$ and
assuming a NH spectrum with $s_{13} \approx 0$, only $\BR(\mu \to e
\ga), ~\BR(\tau \to \mu \ga)$, $\BR(\mu \to 3 e)$ and $\CR(\mu\to
e;{\rm Ti})$ are within the future experimental sensitivities for
$5\times 10^{-5} \lsim \la \lsim 7\times 10^{-5}$, while if $\la
> 7\times 10^{-5}$ only $\mu \to e \ga$ and $\CR(\mu\to e;{\rm Ti})$
could be accessible. All the other LFV processes would be
undetectable in the allowed $\la$ range. Then, if for example the
MEG experiment~\cite{MEG} detects $\BR(\mu \to e \ga)$ at the level
of $8\times 10^{-12}$ then $\BR(\tau \to \mu \ga)$, $\BR(\mu \to 3
e)$ and $\CR(\mu\to e;{\rm Ti})$ are expected to be $\sim 2\times
10^{-9}$, $\sim 5\times 10^{-14}$ and $\sim 4\times 10^{-14}$,
respectively. The case of the QD spectrum is similar: in the range
$\la \sim (6 - 7) \times 10^{-5}$ the $\mu e$ LFV decays are
observable but $\BR(\tau \to \mu \ga)$ is predicted to be $\lsim
10^{-9}$, and for larger $\la$ only $\mu \to e \ga$ and $\mu\to e$
conversion would be visible.
\begin{figure}
\begin{center}
\begin{tabular}{cc}
\includegraphics[width=7.8cm]{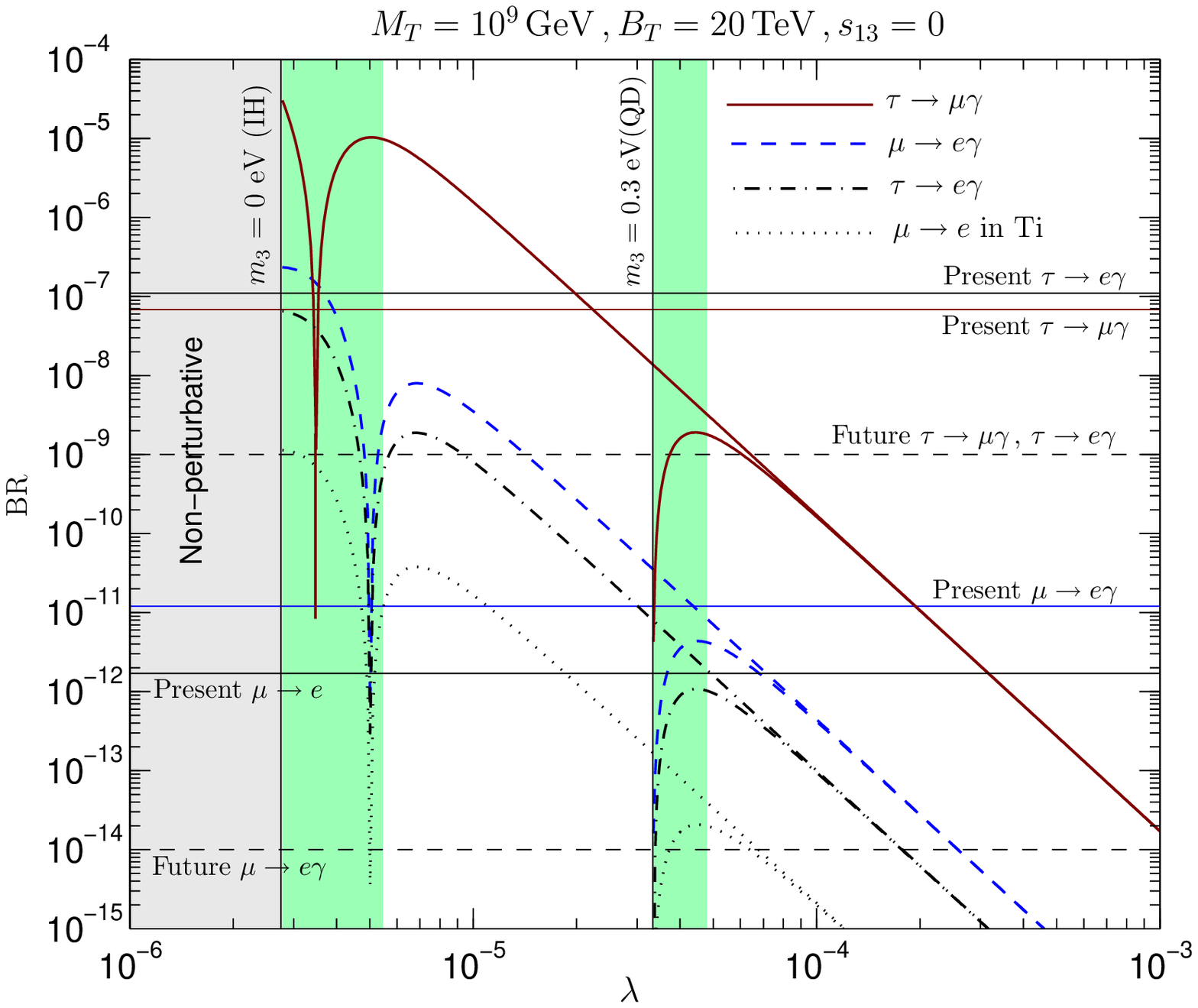} &
\includegraphics[width=7.8cm]{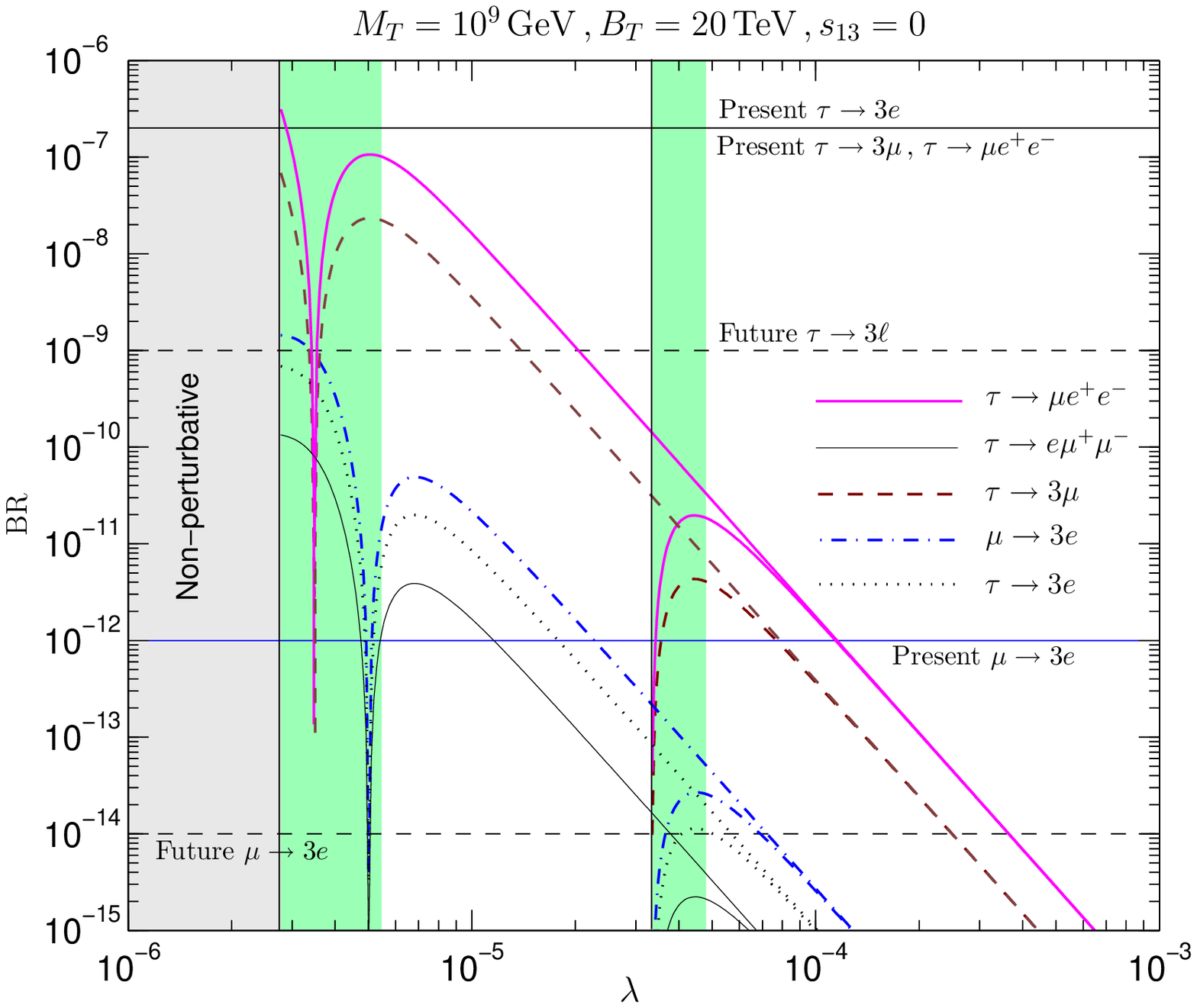}\\
\includegraphics[width=7.8cm]{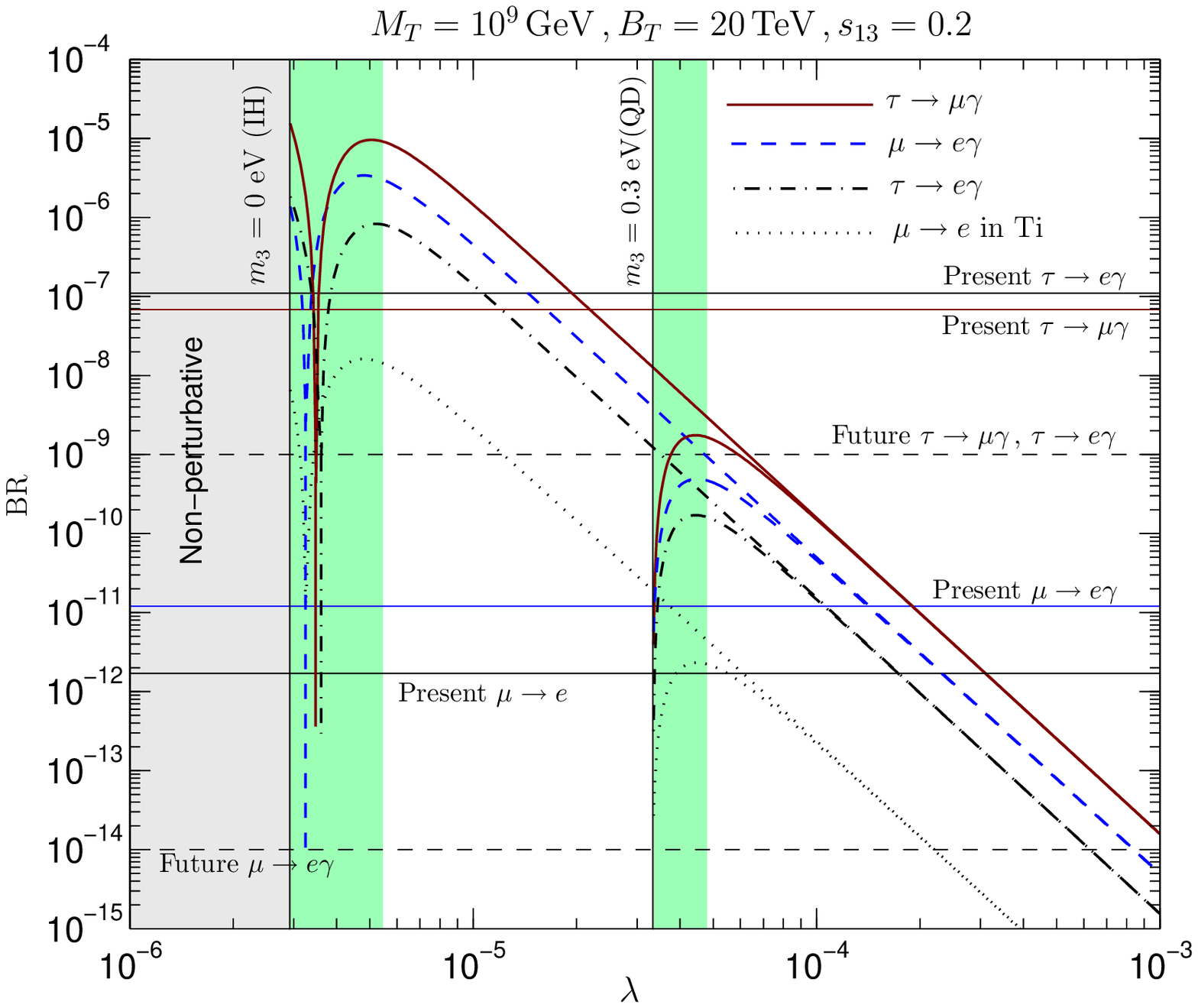} &
\includegraphics[width=7.8cm]{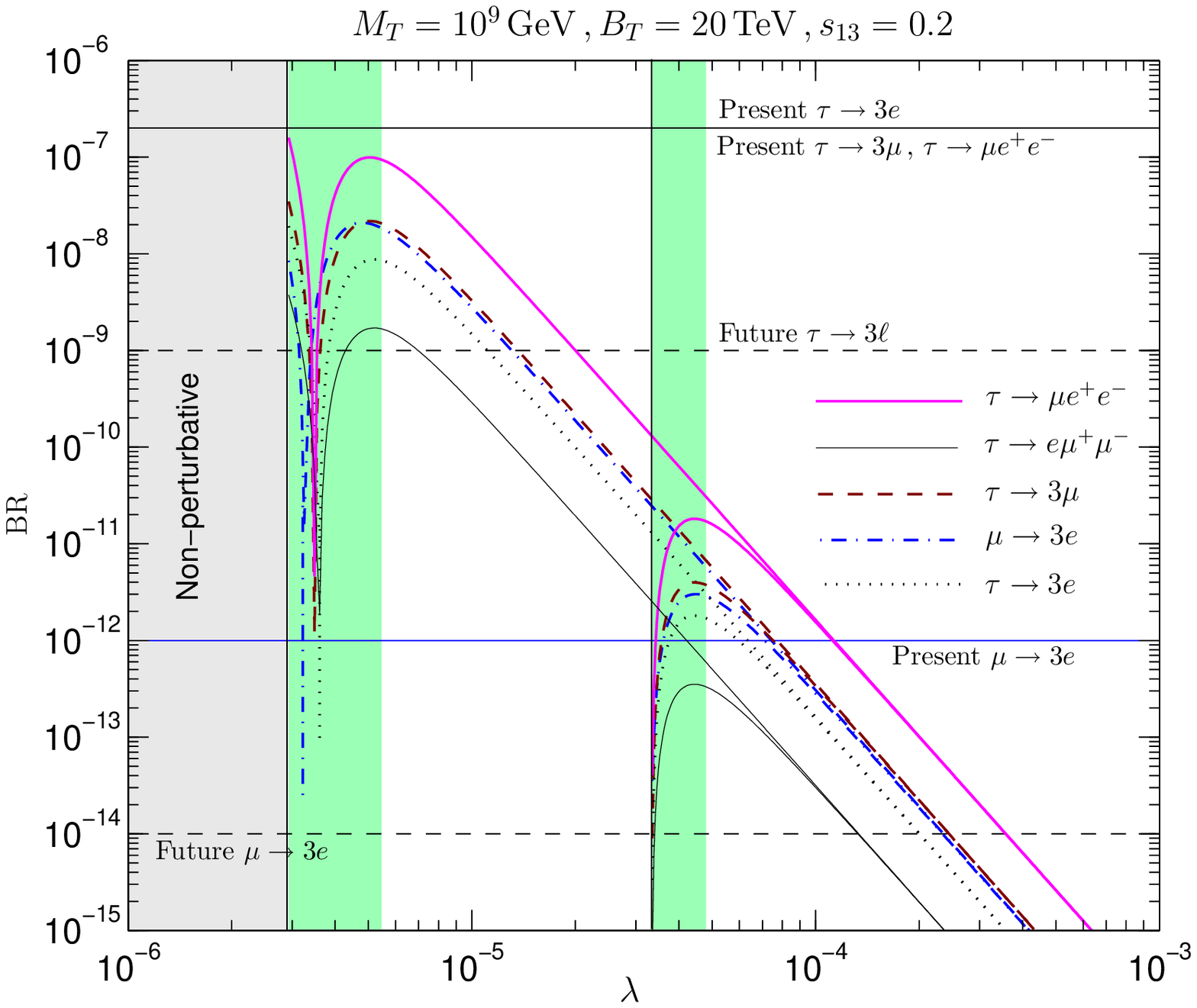}
\end{tabular}
\captions{\small  The same as in Fig.~\ref{f11} for the case of an
IH neutrino mass spectrum ($m_3=0$~eV). For comparison, the results
with $m_3=0.3$~eV are also shown.}
 \label{f12}
\end{center}
\end{figure}
\begin{figure}
\begin{center}
\begin{tabular}{cc}
\includegraphics[width=7.8cm]{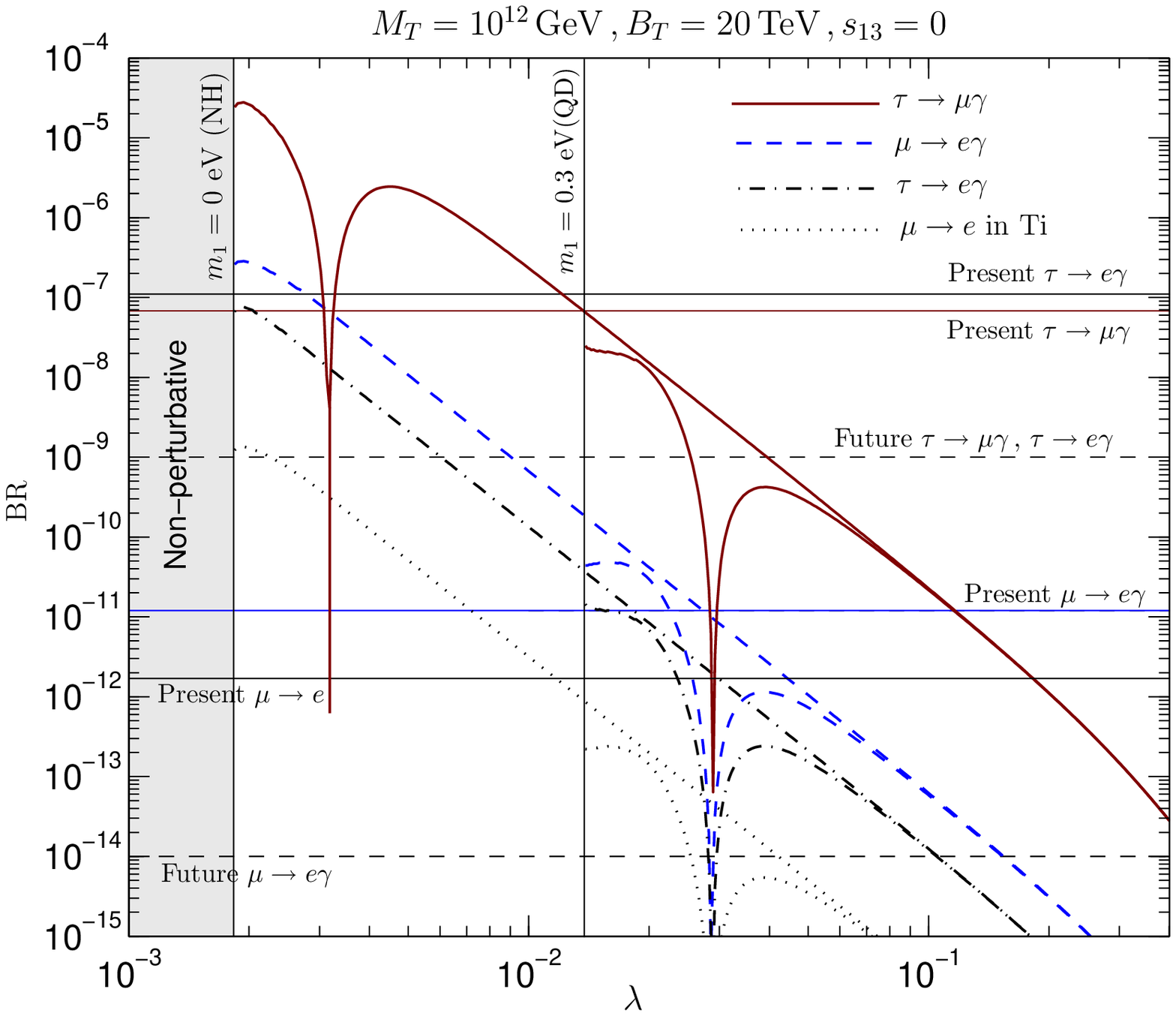} &
\includegraphics[width=7.8cm]{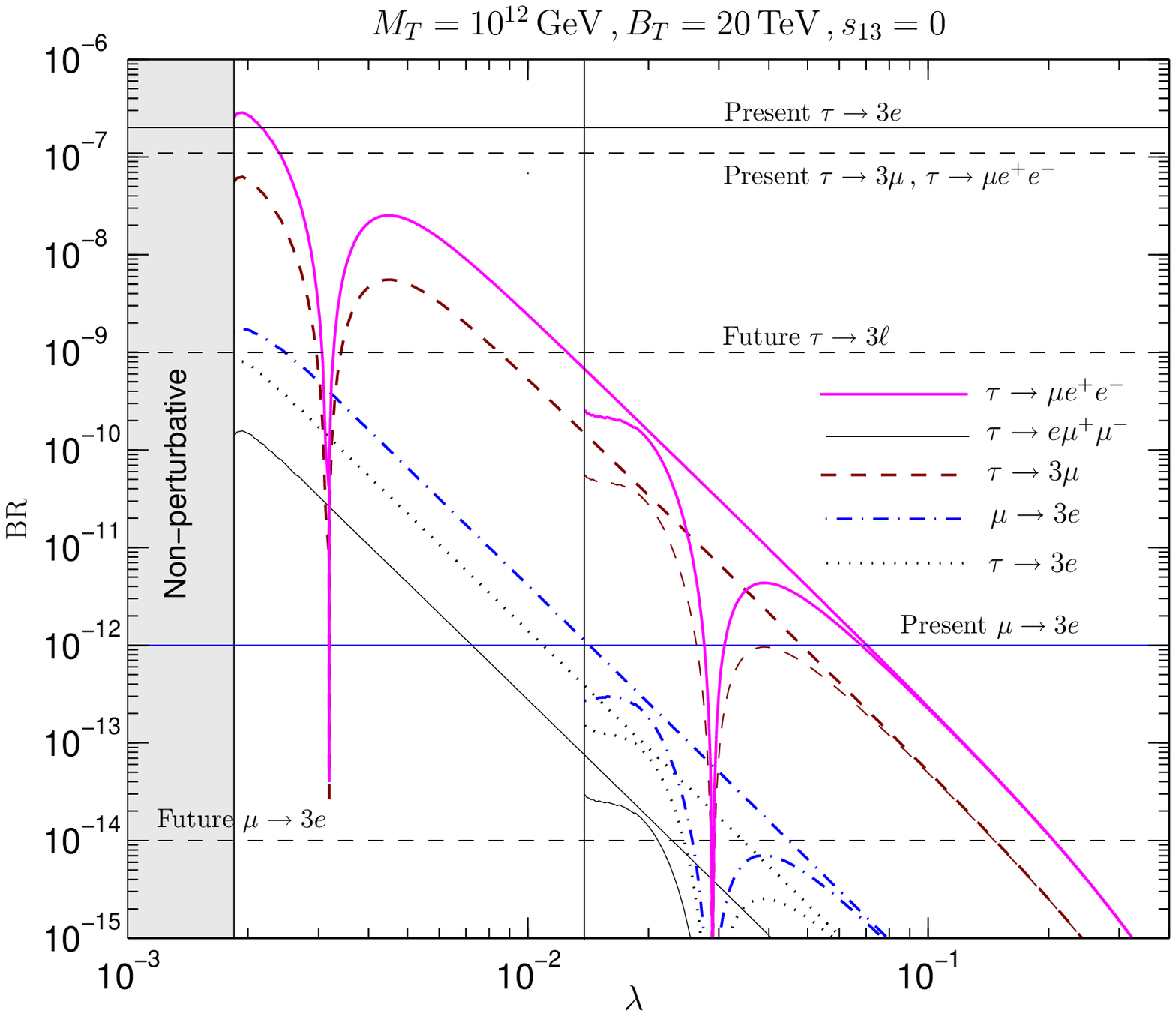}\\
\includegraphics[width=7.8cm]{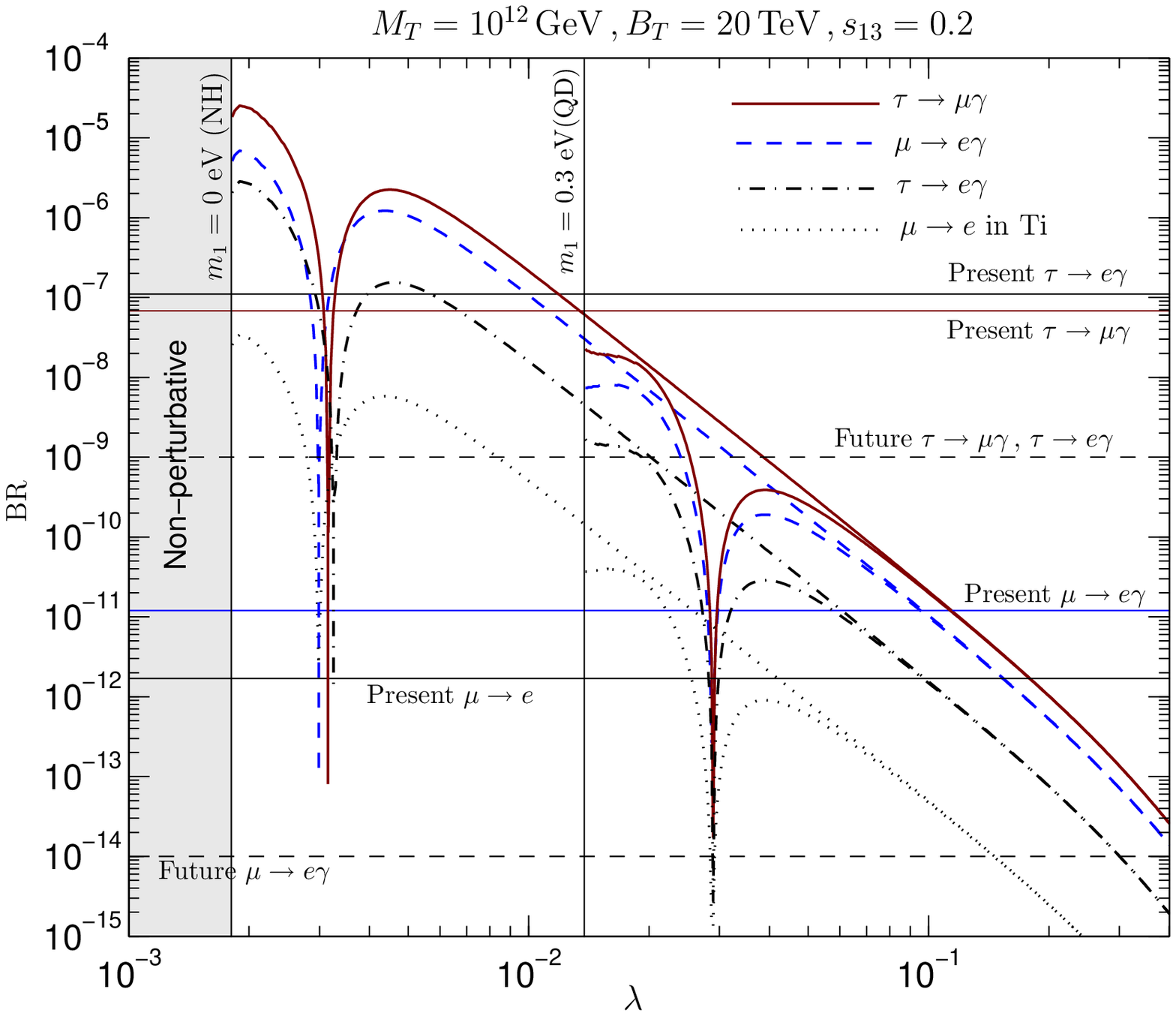} &
\includegraphics[width=7.8cm]{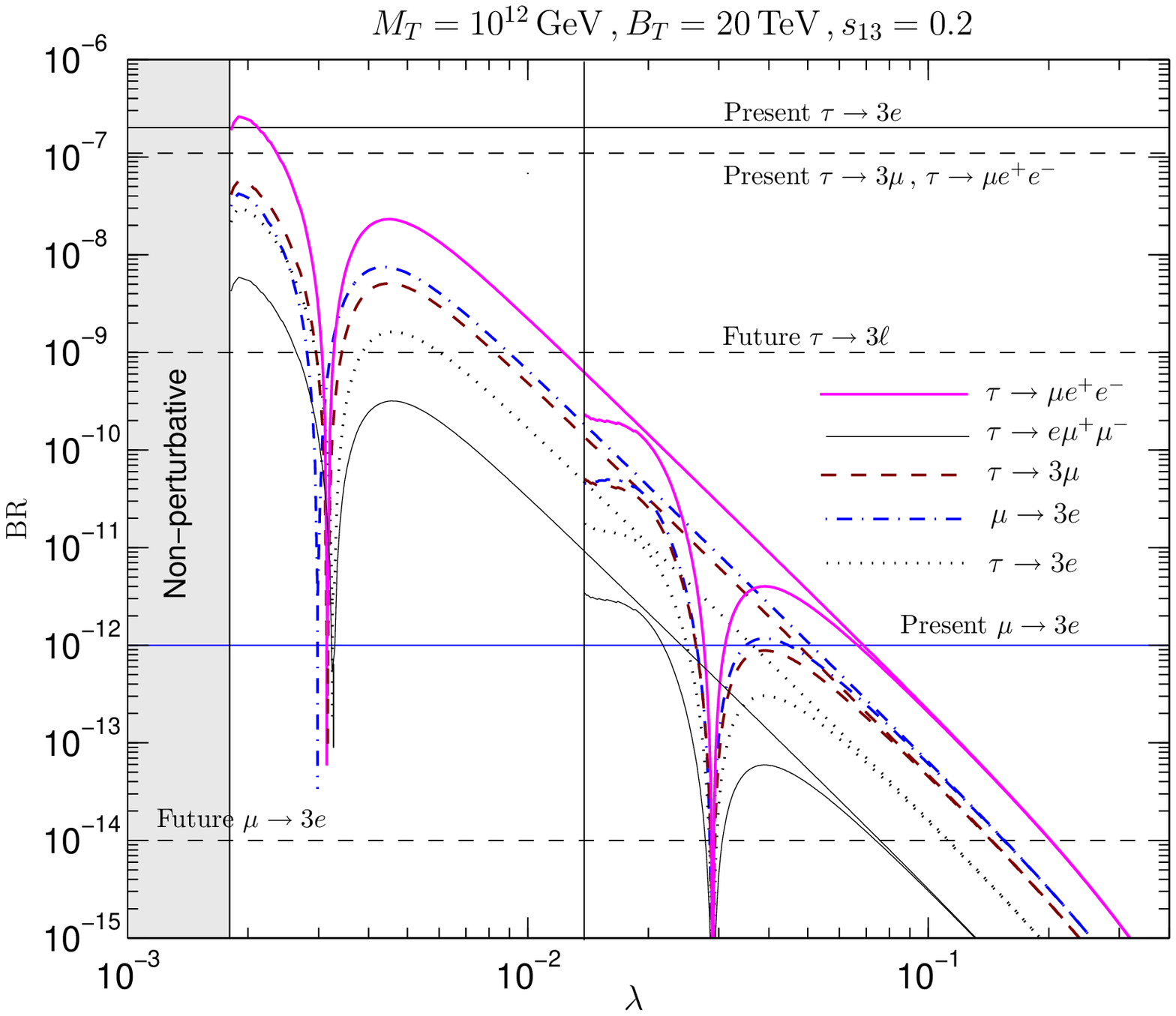}
\end{tabular}
\captions{\small As in Fig.~\ref{f11} but for $M_T=10^{12}~{\rm
GeV}$. }
 \label{f13}
\end{center}
\end{figure}

By switching on $s_{13}$, LFV is enhanced in the $\mu e$ and $\tau
e$ sectors, but it is not essentially altered in the $\tau \mu$
sector. As a consequence, the lower bound on $\la$ imposed by the
present limit on $\BR(\mu \to e  \ga)$ is more restrictive (cf.
Fig.~\ref{f5}) and consequently, the $\BR$s of the $\tau\mu$ sector
are penalised. For instance, detecting $\mu \to e \ga$ with a BR
around $8\times 10^{-12}$ would imply the possibility to measure
also $\BR(\mu \to 3 e)$ and $\CR(\mu\to e;{\rm Ti})$ at the level of
$5\times 10^{-14}$ and $4\times 10^{-14}$, but $\BR(\tau \to \mu
\ga)$ would be $\sim 10^{-11}$ (well below the planned future
capability). This conclusion holds for both the NH and QD spectrum.

Fig.~\ref{f12} presents a similar analysis for the IH spectrum with
$m_3=0~{\rm eV}$. For comparison, we also show the QD case which can
be obtained by pushing the mass $m_3$ to values larger than $(\Delta
m^2_A)^{1/2}$ thus recovering a case very similar to the QD one
reported in Fig.~\ref{f11}. When $s_{13}=0$ (upper panels) the main
difference with the NH case is the fact that all BRs exhibit the
suppression dip in the perturbative range (in the $\tau \mu$ sector
this takes place for a smaller $\la$ with respect to the $\mu e$
one). However, due to the choice of $B_T$ these dips fall into the
(green) range excluded by sparticle searches. Outside these
cancelation regions the values of all these $\BR$s  (and the related
correlations) are essentially the same as those obtained for the NH
case leading to the same phenomenological implications.
\begin{figure}
\begin{center}
\begin{tabular}{cc}
\includegraphics[width=7.8cm]{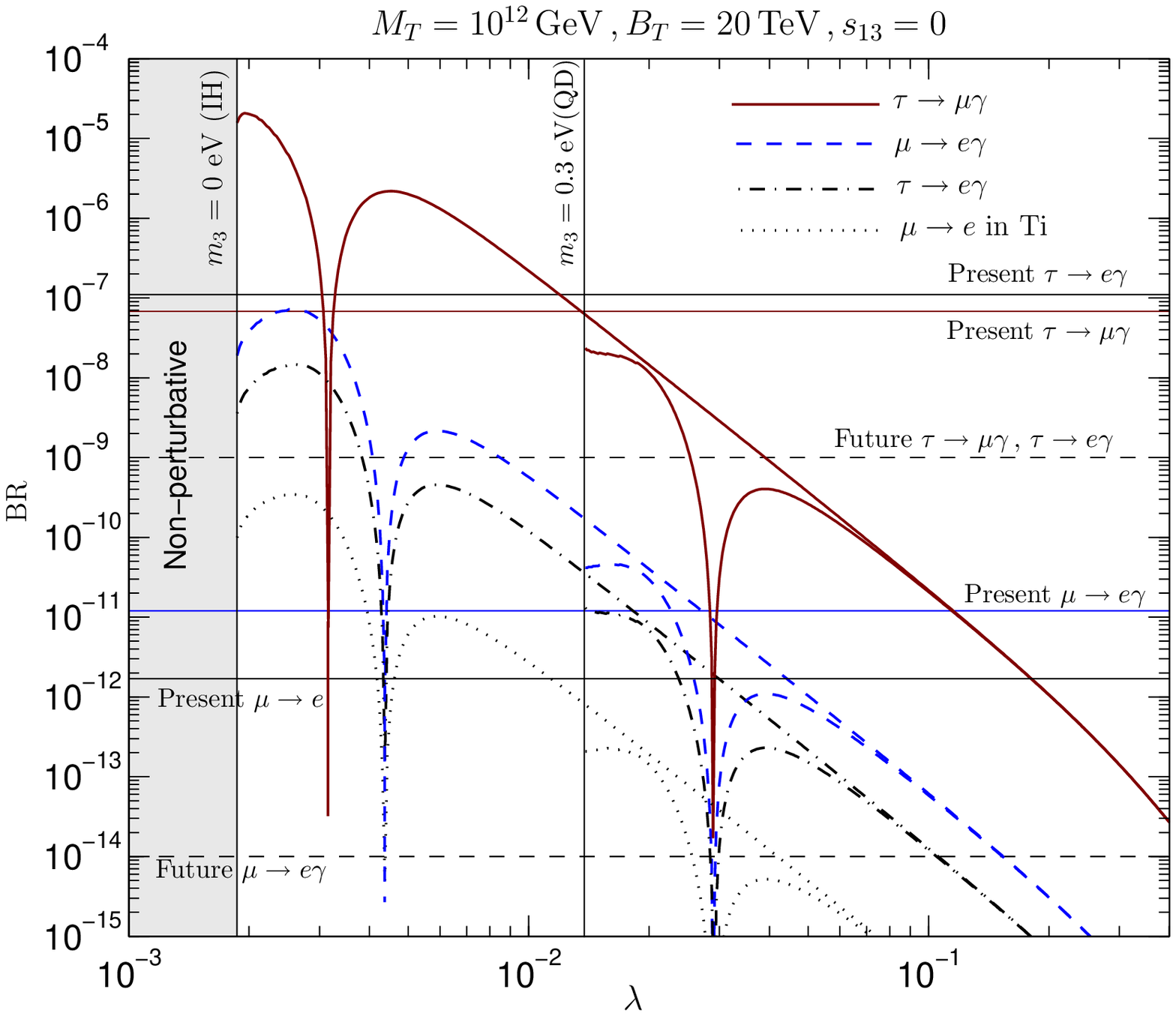} &
\includegraphics[width=7.8cm]{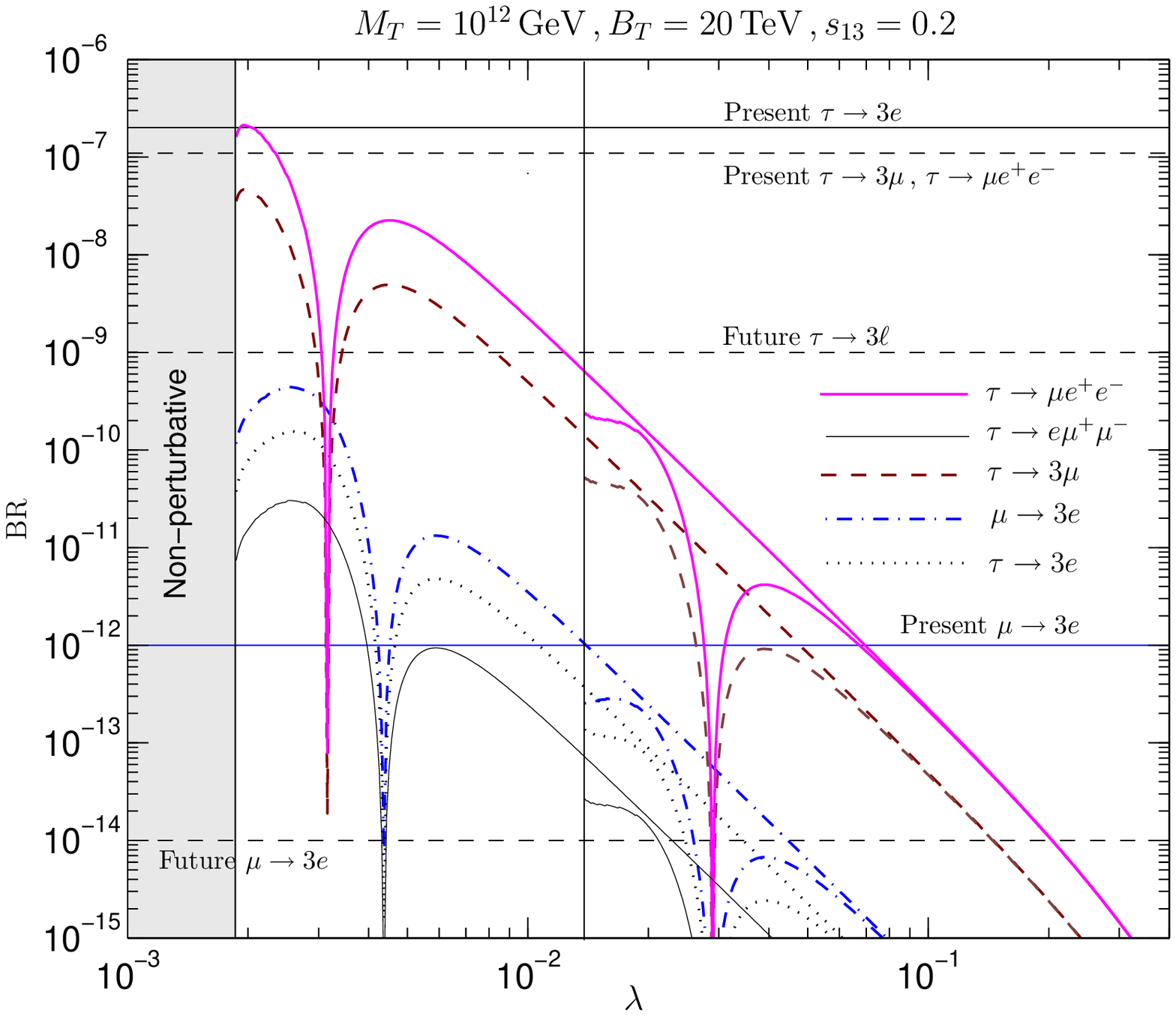}\\
\includegraphics[width=7.8cm]{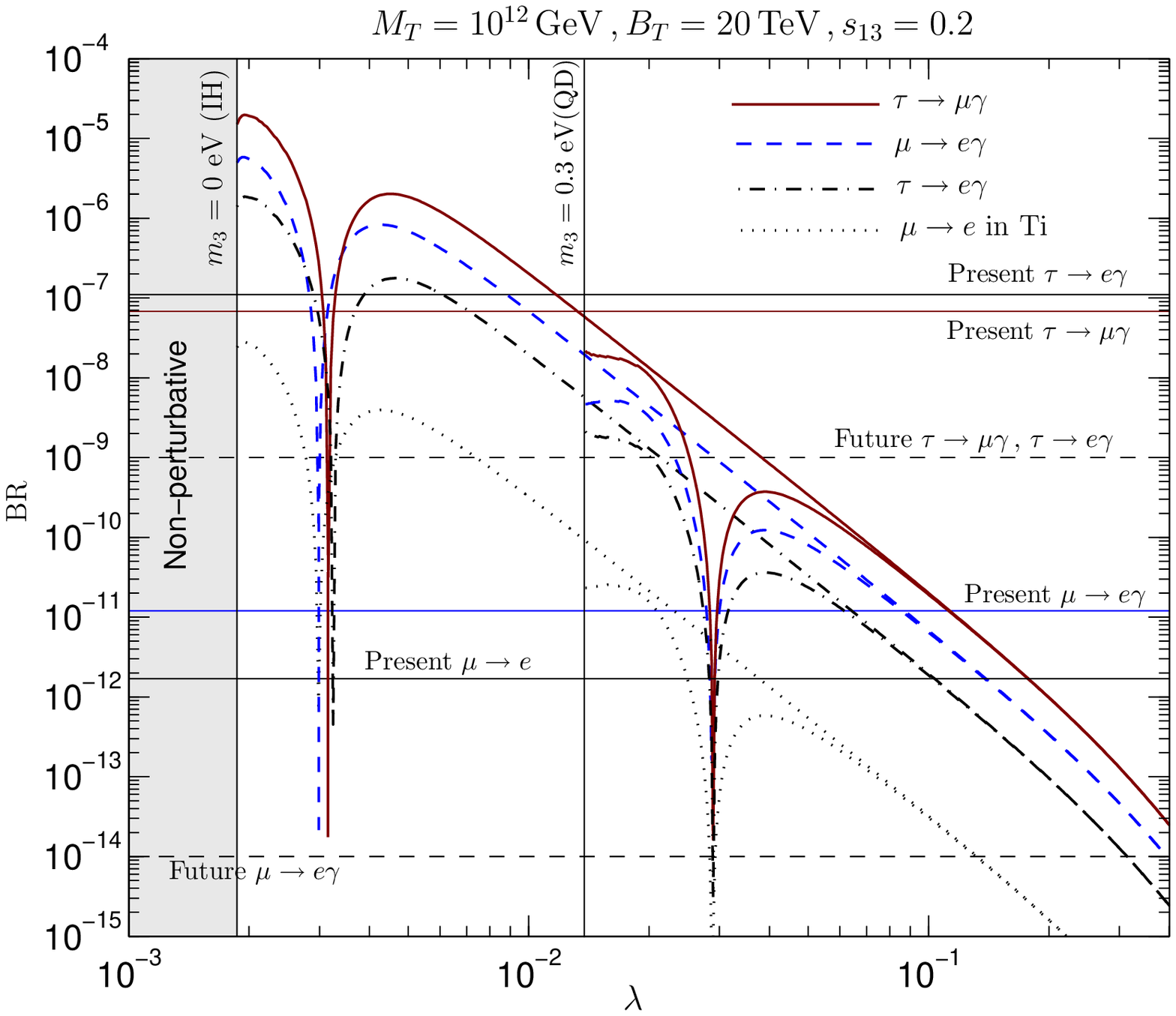} &
\includegraphics[width=7.8cm]{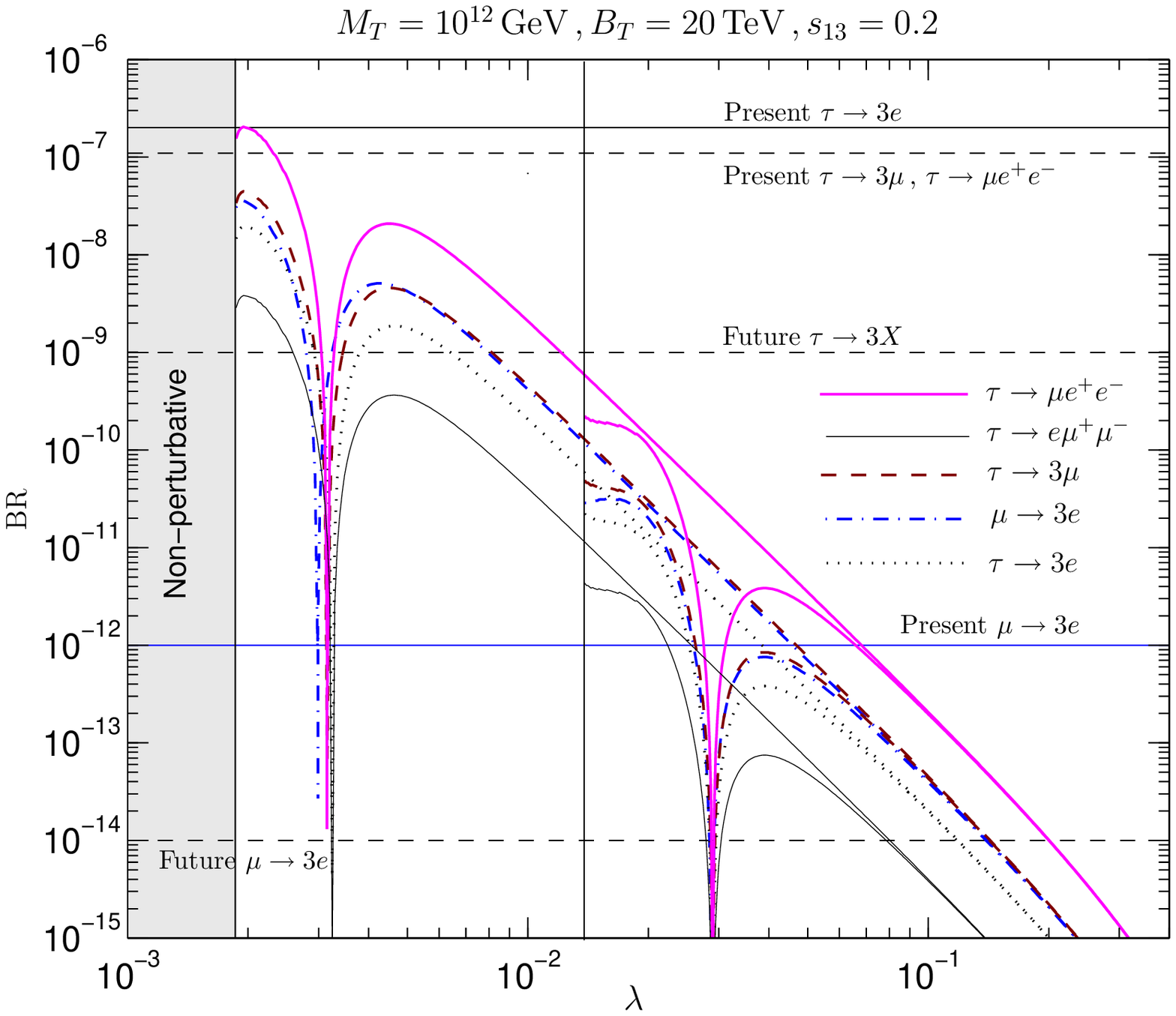}
\end{tabular}
\captions{\small As in Fig.~\ref{f12} but for $M_T=10^{12}~{\rm
GeV}$.}
 \label{f14}
\end{center}
\end{figure}

The IH scenario offers the possibility to reverse the dominance of
the $\mu e$ LFV over the $\tau \mu$ LFV. Suppose that the effective
SUSY breaking is increased up to $B_T \gsim 70~{\rm TeV}$ so that
all the sparticle masses are enhanced by a factor of 3.5. Then the
$\BR(\mu \to e \ga)$ dip would lie in the allowed $\la$
range\footnote{ The effect of increasing $B_T$ does not modify the
perturbativity bounds, but for $B_T \gsim 50~{\rm TeV}$ the
constraint from the negative sparticle searches disappears, as the
comparison between Fig.~\ref{f5} and \ref{f6} has already
demonstrated.} and $\BR(\tau \to \mu \ga)$ would be brought close to
the present bound. As a result, $\mu \to e \ga$, $\mu \to 3 e$ and
$\mu \to e$ conversion might be invisible, whereas $\BR(\tau \to \mu
\ga) \sim 5\times 10^{-8}$ could be measured [perhaps together with
$\BR(\tau \to \mu e e)\sim 5\times 10^{-10}$].

By switching on $s_{13}=0.2$ (lower panels) we find, as for the NH
case, that all the LFV processes undergo a dramatic suppression at
the same $\la$ which, however, occurs in the excluded range. Outside
this interval, the relative ratios of the $\BR$s closely follow the
predictions\footnote{The fact that in Fig.~\ref{f12} $\BR(\tau \to 3
\mu)$ is slightly larger than $\BR(\mu \to 3 e)$, in  contradiction
with the predicitons reported for example in Table~\ref{tb2}, is due
the $Y_\tau$-driven RG running of the neutrino mass $\bmm_\nu$ which
concerns mainly the entries $\tau i\, (i=e, \mu, \tau)$, and is more
sizeable for $s_{13}\neq 0$ with IH spectrum.} (\ref{brs1}),
(\ref{br3}) and (\ref{br4}). In such a case, the planned
experimental sensitivities would allow to test only $\mu \to e \ga$,
$\mu \to 3 e$ and $\mu\to e$ conversion.

Figs.~\ref{f13} and \ref{f14} contain the results of a similar
analysis performed by taking $M_T = 10^{12}~{\rm GeV}$ for the NH
and IH spectrum, respectively. The increasing of $M_T$ implies an
overall decreasing of all the BRs due to a heavier sparticle
spectrum (cf. Fig.~\ref{f8}) with respect to the case with $M_T=
10^{9}~{\rm GeV}$. We again find that for any $\la$ (barring the
cancelation ranges) the several BRs are correlated according to the
model-independent predictions contained in Eqs.~(\ref{brs1}),
(\ref{br3}) and (\ref{br4}). The features of the cancelation dips
are those discussed in Section~\ref{Flavour Structure} and already
encountered for smaller $M_T$. Consider the NH case
(Fig.~\ref{f13}).
At $s_{13}=0$ and $\la$ within $(3 - 4) \times 10^{-2}$, $\mu \to e
\ga, \, \tau \to \mu  \ga$, $\mu \to 3 e$ and $\mu\to e$ conversion
are in the reach of the future experiments in both the NH and QD
cases. From $\la \sim 4 \times 10^{-2}$  up to $1.5 \times 10^{-1}$
only $\mu \to e\ga$ and $\mu \to e$ conversion could be detected.
For $s_{13} = 0.2$ and $\la \geq 0.1$ only $\mu \to e\ga$, $\mu \to
3 e$ and  $\mu \to e$ conversion are accessible.
We arrive at  similar conclusions observing the case of IH spectrum
displayed in Fig.~\ref{f14}: the $\mu e$ sector is more favoured
with respect to the $\tau \mu$ sector, where only the $\tau \to \mu
\ga$ decay could be detected as far as  $s_{13} \approx 0$. However,
this conclusion could be contradicted for larger $B_T$ and
$s_{13}=0$. Consider $B_T\gsim 50~{\rm TeV}$ so that the sparticle
spectrum increases by a factor of 2.5. Then, in the suppression dips
of the $\mu e$ LVF processes (see upper panels in Fig.~\ref{f14}),
$\BR(\tau \to \mu \ga)$ would be close to the present bound and
$\BR(\tau\to \mu  e e ) \sim 10^{-9}$.
Therefore, the future detection of both $\tau \to \mu \ga$ and $\tau
\to \mu e e $, together with a slepton and squark spectrum in the
range $400 - 650 ~{\rm GeV}$ and $1.8 - 2.5 ~{\rm TeV}$,
respectively, would point to a IH neutrino masses and $s_{13}\approx
0$.

\section{Summary and Conclusions}
\label{conclusions}

The future perspectives to detect signals of {\it new physics}
mostly rely on the observation of sparticles at the LHC or LFV
decays at {\it e g.} the B-factories~\cite{taumu:fut}, the incoming
MEG experiment~\cite{MEG}, the Super Flavour factory~\cite{roadmap}
or the PRISM/PRIME experiment at J-PARC~\cite{prime}. It is,
therefore, extremely important to motivate and suggest theoretical
scenarios which can be tested in more than one direction. In this
paper we have presented and discussed in detail a supersymmetric
$SU(5)$ version of the triplet seesaw mechanism in which the triplet
are messengers of both $L$ and SUSY breaking. The key-points of our
model can be outlined as follows:

\begin{itemize}
\item The tree-level exchange of the triplets $T$ generates neutrino
masses, so flavour violation is induced also in the lepton sector of
the SM, as required by the observation of neutrino oscillations. All
the LFV effects are parameterized by a \emph{single} flavour
structure $\bY_T$;
\item The quantum-level exchange of the 15-states $T, \,S$ and $Z$
generates all the SSB mass parameters of the MSSM via gauge and
Yukawa interactions. Their mass scale is determined by the effective
SUSY breaking scale $B_T$. Flavour violation is induced in the mass
matrices $\bmm^2_\tl$ and $\bmm^2_\td$, and in the trilinear terms
$\bA_e$ and $\bA_d$, by the Yukawa couplings $\bY_T$ (and, according
to the $SU(5)$ relation (\ref{y15}), also by $\bY_S$ and $\bY_Z$).
Therefore, the flavour structure of the SSB parameters is provided
by $\bY_T$ or, according to Eq.~(\ref{match}), by the low-energy
neutrino parameters.
\item The number of free parameters is three: the triplet mass
$M_T$, the effective SUSY breaking scale $B_T$ and the unflavoured
coupling constant $\la$.
\end{itemize}

These aspects make the present scenario highly predictive since it
relates neutrino masses and mixing, sparticle and Higgs spectra,
lepton and quark flavour violation in the sfermion masses and
electroweak symmetry breaking. We have performed a complete analysis
of the parameter space spanned by $M_T$, $B_T$ and $\la$ taking into
account the present experimental constraints on the above physical
observables (see Figs.~\ref{f5} and \ref{f6}). This has demonstrated
that there is a region in the parameter space where our framework is
compatible with experiment. In particular, despite the very
constrained structure of our SSB mass parameters, EWSB can be
radiatively realized analogously to the conventional MSSM case. Our
predictions allow us to further test the allowed parameter space as
follows:

\begin{itemize}
\item Regarding the MSSM sparticle spectrum, we predict that the gluino
is the heaviest sparticle while, in most of the parameter space,
$\tilde{\ell}_1$ is the lightest. However, the gravitino is lighter
than the MSSM sparticles. For instance, if $B_T=20$~TeV, the squark
and slepton masses lie in the ranges $800-950$~GeV and
$100-300$~GeV, respectively. The chargino masses are
$m_{\tilde{\chi}^\pm_1} \sim 320$~GeV and $m_{\tilde{\chi}^\pm_2}
\sim 450-550$~GeV. Moreover, $m_{\tilde{\chi}^0_1 }\sim 190$~GeV,
$m_{\tilde{\chi}^0_2} \approx m_{\tilde{\chi}^\pm_1}$ and
$m_{\tilde{\chi}^0_{3,4}} \approx m_{\tilde{\chi}^\pm_2}$. These
mass ranges are within the discovery reach of the LHC. Increasing
the parameter $B_T$ implies a linearly heavier spectrum. The
measurement of only a few sparticle masses will provide a hint on
the value of the effective SUSY breaking scale $B_T$ and a test of
the correlation pattern shown in Fig.~\ref{f8}.

\indent The mass range of the electroweak sparticles implies that
the SUSY contribution to the muon anomalous magnetic moment $\delta
a_\mu$ never exceeds the maximum value of the interval (\ref{gmu}).
Moreover, since $\delta a_\mu>0$, the discrepancy between the
experimental determination and SM prediction is alleviated in our
model.
\item The Higgs sector is characterized by a decoupling regime with
a light SM-like Higgs boson ($h$) and the three heavy states ($H,A$
and $H^\pm$) with mass $m_{H,A,H^\pm}\approx 450-550$~GeV (again,
for $B_T=20$~TeV). These masses increase almost linearly with $B_T$.
\item We have considered several LFV processes: $\mu \to e X$,
$\mu\to e$ conversion in nuclei, $\tau \to e Y$  and $\tau \to \mu
Y$ $(X = \ga, ee$, $Y= \ga, e e, \mu \mu)$. Our framework is
characterized by peculiar LFV and QFV patterns [intimately related
to each other; see Eq.~(\ref{LFratio2})], which are mostly
determined by low-energy neutrino masses and mixing. The size of
QFV, when confronted with the coloured sparticle spectrum, is well
below the present phenomenological bounds extracted from $b\to s
\ga$ and $K^0-\bar{K}^0$ mixing, etc. Therefore, QFV processes do
not constrain our parameter space.
\end{itemize}

Concerning LFV, we stress that strict predictions have been obtained
for the relative branching ratios of the radiative $\ell_j\to\ell_i
\ga$ [see Eq.~(\ref{brs1})] and 3-body $\ell_j\to
\ell_i\ell_k\ell_k$ decays [see Eq.~(\ref{br4})]. The latter, as
well as $\mu \to e$ conversion in Ti, are also correlated with the
radiative decays as shown in Eq.~(\ref{br3}). All these results are
\emph{model-independent} in the sense that they do not depend on
$M_T$, $B_T$ or $\la$ (they are given only in terms of the
low-energy neutrino observables). If the present bound on
$\BR(\mu\to e \ga)$ is saturated, the branching ratios of the
remaining LFV processes are predicted as shown in Table~{\ref{tb2},
where the three types of neutrino spectrum have been considered. The
experimental signatures of our scenario crucially depend on the
value of the lepton mixing
angle $\theta_{13}$:\\

\noindent {\bf Tiny $\bm{\theta_{13}}$}: The analysis has shown
that, in the allowed parameter space, the future experimental
sensitivity will allow to measure at most $\BR(\mu\to e \ga)$,
$\BR(\mu\to 3e)$, $\BR(\tau\to \mu\ga)$ and CR($\mu\to e\,$~Ti)
according to the relations (\ref{brs1}) and (\ref{br3}). In
particular, being $\BR(\tau\to \mu\ga)/\BR(\mu\to e\ga)\sim 300$,
$\BR(\tau\to \mu\ga)$ is expected not to exceed $3\times 10^{-9}$,
irrespective of the type of neutrino spectrum. All the decays
$\tau\to \ell_i\ell_k\ell_k$ would have $\BR <
\mathcal{O}(10^{-11})$.\\

\noindent {\bf Sizeable $\bm{\theta_{13}}$}: If $s_{13}$ is close to
the upper bound (\ref{13}), the $\tau\mu$ sector is hardly
accessible and only the decays $\mu\to e \ga$, $\mu\to 3e$ and $\mu
\to e$ conversion in Ti can be observed in the future. This
conclusion holds for the NH, IH and QD neutrino spectra. For
instance, if the MEG experiment measures $\BR(\mu\to e \ga)\sim
8\times 10^{-12}$, then $\BR(\mu\to 3e)$ and CR($\mu\to e\,$~Ti) are
expected to be $\sim 5\times 10^{-14}$ and $\sim 4\times 10^{-14}$.
The latter is in the reach of the PRISM/PRIME sensitivity. Values of
$s_{13} \sim \mathcal{O}(10^{-1})$ will be explored soon by several
neutrino experiments like MINOS, OPERA and Double
Chooz~\cite{minos}.\\

Deviations to this very specific \emph{model-independent} pattern
occur when $({\bmm^2_\tl})_{ij}$ cancel (see discussion in
Section~\ref{Flavour Structure}). This can be the case if
$s_{13}\approx 0.02(-0.02)$ and the neutrino spectrum is IH (NH).
Then, all the $\mu e$ LFV processes might be invisible and $\tau\to
\mu \ga$ detected with a BR close to the present bound (taking
$B_T=20$~TeV). Moreover, $\tau\to \mu e e$ would be
$\mathcal{O}(10^{-9}-10^{-10})$ and the remaining $\tau$ decays
below $10^{-10}$. Notice that, increasing $B_T$ the BRs are
suppressed since they scale as $1/B_T^4$. This specific value of
$s_{13}$ is in the sensitivity range of Neutrino
Factories~\cite{NF}.

Alternatively, $({\bmm^2_\tl})_{\mu e }$ could be vanishing because
of a cancelation between the quadratic and quartic Yukawa
contributions [see Eq.~(\ref{IH1})]. For the IH case, if $B_T \gsim
50\,$TeV and $s_{13}$ is very small, all the $\mu e$ LFV processes
would be strongly suppressed whereas the $\tau \mu$ sector would be
favoured with $\BR(\tau\to \mu\ga)\sim 5\times 10^{-8}$ and
$\BR(\tau\to \mu e e)\sim \times 10^{-9}$. This scenario is
correlated with a slepton and squark spectrum in the ranges 400-650
GeV (in the limit of the LHC detection capabilities) and above $\sim
2$~TeV (within the LHC reach), respectively.\\

Given the increasing interest on the problem of finding a possible
relation between leptogenesis~\cite{Fukugita:1986hr} and low-energy
neutrino physics~\cite{alllepto}, we would like to comment on this
issue in the framework of our work. Within the present version of
the supersymmetric triplet seesaw mechanism, leptogenesis can be
realized by considering that the soft bilinear term $B_T$ produces a
mass splitting between $T$ and $\bar{T}$, leading to resonant
leptogenesis~\cite{lepto}. For this to work, $B_T$ must be around
the electroweak scale. In our case, since $B_T \gsim 20$~TeV, the
BAU turns out to be too small. Still, leptogenesis could be made
effective either by adding an additional pair of triplets~\cite{hms}
or by including heavy singlet neutrinos~\cite{TN}. The former case
would imply the appearance of one more flavour source
$\bY^{\prime}_T$ which, depending on its size, could have some
impact in the predictions of our framework. Instead, we would like
to comment on the second possibility which requires heavy neutrino
singlets (with mass $M_N$) coupled to the lepton doublets through
the Yukawa couplings $\bY_N$. To maintain the predictions made along
this work we must require that the singlet contribution to neutrino
masses is much smaller then the one generated by the triplet \ie,
$\bY_N^2 \ll \la \bY_T M_N/M_T$. Moreover, $\bY_N \ll \bY_T$ is also
required to suppress LFV arising in the SSB parameters from the
singlet exchange at the quantum level\footnote{For a discussion
related to this in the context of EWSB see~\cite{farza}.} (notice
that the neutrino singlets could couple to the spurion field $X$).
For the purpose of leptogenesis, two different situations can be
envisaged: $M_N > M_T$ or $M_N < M_T$. In the former case, the
$CP$-asymmetry generated through the decay of the triplets into two
leptons is directly proportional to $\bY_N^2/M_N$~\cite{TN}, so it
turns out to be very tiny ($\bY_N^2 \ll \la \bY_T M_N/M_T$). On the
contrary, if $M_N < M_T$ the $CP$-asymmetry is weakly sensitive to
$Y_N$~\cite{TN}, therefore a viable value for the BAU can be
achieved.\\

We conclude our discussion by remarking that our scenario is not
only extremely predictive but it can also be tested in view of the
present and future programmes of LFV and neutrino oscillation
experiments.

\vspace*{1.0cm}

{\bf \large Acknowledgments:} We thank A.~Brignole for valuable
comments and suggestions and T.~Hambye for useful discussions. The
work of F.R.J. is supported by {\em Funda\c{c}{\~a}o para a
Ci{\^e}ncia e a Tecnologia} (FCT, Portugal) under the grant
\mbox{SFRH/BPD/14473/2003},  INFN and PRIN Fisica Astroparticellare
(MIUR). The work of A.~R.~ is partially supported by the project EU
MRTN-CT-2004-503369.

\renewcommand{\theequation}{A-\arabic{equation}}
\setcounter{equation}{0}  
\section*{Appendix A-Extracting the SSB terms from wave function renormalization}

\label{wave}

In this Appendix we derive the general expressions for the soft
supersymmetry breaking scalar masses, the bilinear and trilinear
couplings at the messenger scale $M$. We employ a generalization of
the method suggested in Ref.~\cite{GR} and subsequently presented in
Ref.~\cite{CP}. Consider the case in which the scales of SUSY
breaking and its mediation to the observable sector are determined
by the {\it vev} of the auxiliary and scalar components of a chiral
singlet superfield $X$, $\langle X\rangle = M +\theta^2 F$ (in the
following $M$ is taken to be real and $|F|< M^2$, for consistency of
the method). The leading contributions (at lowest order in $F$) to
the SSB terms arise from $X$-dependent wave-function
renormalizations of the chiral superfields $\bZ_\cq$. The effective
Lagrangian reads\footnote{Here the index $a$ labels either the
superfield and its associated `charges' or only the `charges'; the
context should make clear the case.}
\be
\label{zq} {\cal L}= \int {\rm d}^4\theta \cq^\dagger_a \bZ_\cq(X,
X^\dagger)^a_b \cq^b + \left[\int {\rm d}^2\theta (\mu_{a b} \cq^a
\cq^b + f_{a b c} \cq^a \cq^b \cq^c ) +{\rm h.c.}\right]\,,
\ee
where $\mu_{ab}$ and $f_{abc}$ are superpotential mass parameters
and dimensionless coupling constants, respectively. The
wave-function renormalization $\bZ_\cq$ is a hermitian matrix which
depends on $|X|=\sqrt{X X^\dag}$. Its $\theta$-expansion at
$X=\langle X \rangle$ is given by
\be \left.\bZ_\cq(|X|)\right|_{X=\langle X \rangle} = \bZ_\cq(M) +\frac12 \frac{\partial
\bZ_\cq}{\partial \ln M}\left(\theta^2 \frac{F}{M} + \bar{\theta}^2
\frac{F^\dagger}{M}\right) + \frac14 \frac{\partial^2
\bZ_\cq}{\partial \ln M^2}\theta^2
 \bar{\theta}^2 \frac{F F^\dagger}{M^2}\,.
\ee
After expressing Eq.~({\ref{zq}) in terms of the canonically
normalized superfields $\cq^{'}$ as
\be
\cq = \left(1 + \frac{\bZ^{-1}_\cq}{2} \frac{\partial
\bZ_\cq}{\partial M} \theta^2 F\right)\bZ^{-1/2}_\cq \cq^{'}\,,
\ee
we can  extract the SSB  masses for the scalar component of $\cq$
from the quartic terms $\theta^2\bar{\theta}^2$ in the first
integral and the SSB bilinear and trilinear couplings from the
quadratic terms $\theta^2$ in the second and third integral,
respectively
\bea
\bmm^2_\tcq &=& -\frac14 \bZ^{-1/2}_\cq \left(\frac{\partial^2
\bZ_\cq}{\partial \ln M^2} -  \frac{\partial \bZ_\cq}{\partial \ln
M} \bZ^{-1}_\cq \frac{\partial \bZ_\cq}{\partial \ln M}\right)
\bZ^{-1/2}_\cq \frac{F F^\dagger}{M^2} , \no \\
\bA_{abc}& =& \frac12 \left[ f_{a^{'} bc} \left(\bZ^{-1/2}_\cq
\frac{\partial \bZ_\cq}{\partial \ln M} \bZ^{-1/2}_\cq
\right)^{a^{'}}_{a} + f_{a b^{'}c} \left(\bZ^{-1/2}_\cq
\frac{\partial \bZ_\cq}{\partial \ln M} \bZ^{-1/2}_\cq
\right)^{b^{'}}_{b}
\right. \no \\
&& \left. +f_{a b c^{'}} \left(\bZ^{-1/2}_\cq \frac{\partial
\bZ_\cq}{\partial \ln M} \bZ^{-1/2}_\cq \right)^{c^{'}}_{c}
\right] \frac{F}{M} ,  \no \\
B_{ab}& =& \frac12 \left[ \mu_{a^{'} b} \left(\bZ^{-1/2}_\cq
\frac{\partial \bZ_\cq}{\partial \ln M} \bZ^{-1/2}_\cq
\right)^{a^{'}}_{a} + \mu_{a b^{'}}\left(\bZ^{-1/2}_\cq
\frac{\partial \bZ_\cq}{\partial \ln M} \bZ^{-1/2}_\cq
\right)^{b^{'}}_{b} \right] . \label{ssb2}
\eea
In order to find the explicit expressions for the SSB parameters at
an energy scale $\mu$ we recall that the  $\mu$-dependence of
$\bZ_\cq$ is expressed by the RG equation:
\be\label{rgz}
\left\{\bZ^{-1/2}_\cq, \frac{{\rm d} \bZ^{1/2}_\cq}{{\rm d}t}
\right\}
= \bg_\cq \,,
\ee
where $t=\ln \mu$ and $\bg_\cq$  is the matrix of anomalous
dimension. By defining $\bZ^{1/2}_\cq = 1 + \delta \bZ^{1/2}_\cq$,
where $\delta \bZ^{1/2}_\cq$ encodes the quantum corrections,
Eq.~(\ref{rgz}) reads
\be\label{rgz1}
 \frac{{\rm d} \delta \bZ^{1/2}_\cq}{{\rm dt}} = \frac{\bg_\cq}{2} +
\left\{\delta \bZ^{1/2}_\cq ,\frac{\bg_\cq}{4}\right\}\,,
\ee
at the lowest order. By following the procedure outlined in
Ref.~\cite{CP}, we have to formally integrate the above equation to
obtain $\delta \bZ^{1/2}_\cq$ in terms of the anomalous dimension.
Afterwards, the solution can be plugged into the expressions
(\ref{ssb2}) to extract the SSB terms. For the sake of brevity, we
do not report all the intermediate steps (which can be easily
performed) and, instead, give the final expressions at $\mu= M$:
\bea \label{ssb3}
\bmm^2_\tcq &=& - \frac14 \left[\frac{ {\rm d} \Delta \bg_\cq (M)}
{{\rm d\ln M}} - \left(\Delta \frac{{\rm d}}{{\rm d} \ln M}\right)
\bg_{ \cq}^<(M) \right]
\frac{F F^\dagger}{M^2} ,\no \\
A_{abc} & = & \frac12 \left[f_{a^{'} bc}
\Delta\bg_\cq(M)^{a^{'}}_{a} +f_{a b^{'} c}
\Delta\bg_\cq(M)^{b^{'}}_{b} +
f_{ab c^{'} } \Delta\bg_\cq(M)^{c^{'}}_{c} \right]\frac{F}{M} , \no \\
B_{ab}& = & \frac12 \left[\mu_{a^{'} b} \Delta\bg_\cq(M)^{a^{'}}_{a}
+ \mu_{a b^{'} } \Delta\bg_\cq(M)^{b^{'}}_{b}\right] \frac{F}{M}
\eea
\\
\noindent where $\bg_{\cq}^>(M)$ [$\bg_{\cq}^<(M)$] is the anomalous
dimension above (below) the mass scale $M$, $\Delta\bg_\cq(M) \equiv
\bg_{\cq}^>(M) - \bg_{\cq}^<(M)$  and $(\Delta \frac{{\rm d}}{{\rm
d} \ln M}) \bg_{ \cq}^<(M)$ means considering the difference of the
beta-functions of the couplings contained in $\bg_{\cq}^<(M)$ above
and below $M$. Notice that our result (\ref{ssb3}) for $\bmm^2_\tcq$
differs from the one obtained in Ref.~\cite{CP} because an extra
term proportional to the commutator $[\bg_{\cq}^> , \bg_{\cq}^<]$
appears in that work, which is manifestly inconsistent for a
hermitian quantity such as $\bmm^2_\tcq$. We believe that the
appearance of such a commutator is due to the improper definition of
the RG equation for $\bZ_\cq$ given in Eq.~(3.5) of Ref.~\cite{CP}.

In the following, we provide the explicit expressions for the
relevant anomalous dimensions needed to extract the SSB parameters
from Eq.~(\ref{ssb3}) in our specific framework. The anomalous
dimensions below the scale $M$ are:
\bea   \label{amd}
16 \pi^2 \bg_{ L}^< & =& - \left[2 \bY^\dagger_e \bY_e - 4
\left(\frac{3}{20}
g^2_1 +\frac34 g^2_2\right)\right] , \no \\
16 \pi^2 \bg_{ e^c}^< & =& - \left[4 \bY_e \bY^\dagger_e -
\frac{12}{5} g^2_1 \right] , \no \\
16 \pi^2 \bg_{ Q}^< & =& - \left[2 \bY^\dagger_d \bY_d +2
\bY^\dagger_u \bY_u
 - 4 \left(\frac{1}{60}
g^2_1 +\frac34 g^2_2 +\frac43 g^2_3\right)\right] , \no \\
16 \pi^2 \bg_{ d^c}^< & =& - \left[4 \bY_d \bY^\dagger_d
 - 4 \left(\frac{1}{15}
g^2_1  +\frac43 g^2_3\right)\right] , \no \\
16 \pi^2 \bg_{ u^c}^< & =& - \left[4 \bY_u \bY^\dagger_u
 - 4 \left(\frac{4}{15}
g^2_1 +\frac43 g^2_3\right)\right] , \no \\
16 \pi^2 \bg_{ H_1}^< & =& - \left[2{\rm Tr}(\bY^\dagger_e \bY_e +3
\bY^\dagger_d \bY_d) - 4 \left(\frac{3}{20}
g^2_1 +\frac34 g^2_2\right)\right] , \no \\
16 \pi^2 \bg_{ H_2}^< & =& - \left[6{\rm Tr}(\bY^\dagger_u \bY_u)
 - 4 \left(\frac{3}{20}
g^2_1 +\frac34 g^2_2\right)\right] .
\eea
The differences $\Delta\bg_{\cq} (M)$ are instead the following:
\bea\label{deltaga}
16 \pi^2 \Delta \bg_{ L} & =& - 6 \left( \bY^\dagger_T \bY_T +
 \bY^\dagger_Z \bY_Z \right) ,  \no \\
16 \pi^2 \Delta\bg_{ d^c} & =& - 4 \left( \bY_Z \bY^\dagger_Z
+ 2   \bY_S \bY^\dagger_S\right), \no \\
16 \pi^2 \Delta \bg_{ H_2} & =& - 6|\la|^2 , \no \\
16 \pi^2 \Delta \bg_{ {\cal F} } & =& 0 , ~~~~~~~~ {\cal F} = e^c,
Q, u^c, H_1 .
\eea
Regarding the expressions for the RG equations at one loop in the
MSSM framework with the $(15 +\ov{15})$  $SU(5)$ representation we
refer to Ref.~\cite{ar}. Finally, we obtain the explicit formulas
given in Eqs.~(\ref{soft1}) and (\ref{soft2}).


\renewcommand{\theequation}{B-\arabic{equation}}
\setcounter{equation}{0}  
\section*{Appendix B-Coefficients of the $\bm{\ell_j\ell_i Z}$ operators}  

In this Appendix we compute the coefficients $A^{L, R}$ of the
monopole operators
\be \label{az}
g_Z (A^L_{ji} \bar{\ell}_i \bar{\sigma}^\mu \ell_j + A^R_{ ji}
{\ell}^c_i {\sigma}^\mu \bar{\ell}^c_j +{\rm h.c.})Z_\mu \;\;, \;\;
A^{L (R)}_{ji}= A^{L (R), c}_{ji} +A^{L (R), n}_{ji}
\ee
where $g_Z = g_2/c_W$ ($c_{W} = \cos{\theta_W}$, $\theta_W$ is the
weak mixing angle) and $ A^{L (R), c}$ [$  A^{L (R), n}$] stand for
the contributions from the chargino/sneutrino
[neutralino/charged-slepton] loop diagrams. The two-component spinor
notation is used such that, for example, $\ell_i (\bar{\ell}^c_i)$
is the left-handed (right-handed) component of the lepton $i$ field
$(i=e, \mu, \tau)$. Different one-loop results for such coefficients
have been presented in the literature. For instance, the authors of
Ref.~\cite{HMTY} provided an all-order calculation in the
electroweak breaking effects, while the authors of Ref.~\cite{br}
performed a lowest-order calculation. We found dramatic numerical
discrepancies between the two aforementioned results, which cannot
be ascribed to the approximation used in Ref.~\cite{br}. Moreover,
the authors of Ref.~\cite{AH} have recently re-evaluated the
contributions to $A^{L,R}$ and claimed to have found additional
contributions disregarded in Ref.~\cite{HMTY}. We have independently
performed the all-order computation to compare with the previous
results and to clarify this issue. The notation of Ref.~\cite{HMTY}
has been adopted to define the mass eigenstates of the charged
sleptons $\tilde{\ell}_X, (X=1,\ldots 6)$ , sneutrinos
$\tilde{\nu}_X, (X=1,\ldots 3)$, charginos $\tilde{\chi}^-_A,
(A=1,2)$ and neutralinos $\tilde{\chi}^0_A, (A=1,\ldots 4)$. The
corresponding (unitary) mixing matrices are denoted by $U^\ell_{X
i}, U^\nu_{X i}, (O_L)_{A \alpha}, (O_R)_{A \alpha}$ and $(O_N)_{A
\beta}$ (where $\alpha (\beta)$ is the current-basis index for
charged or neutral gauginos/higgsinos). The relevant interactions of
sleptons/leptons with charginos and neutralinos are:
\be
\label{lg}%
{\cal L} = \bar{\ell}_i C^R_{iA X} \tilde{\nu}_X
\bar{\tilde{\chi}}^+_A +{e}^c_i C^L_{iA X} \tilde{\nu}_X
\tilde{\chi}^-_A + \bar{\ell}_i N^R_{iA X} \tilde{\ell}_X
\bar{\tilde{\chi}}^0_A + {e}^c_i N^L_{iA X} \tilde{\ell}^c_X
\tilde{\chi}^0_A + {\rm h.c.} ,
\ee
where
\bea
\label{vertex}%
& & C^R_{iAX}  =  -g_2(O_R)_{A 1} U^\nu_{X i} \quad,\quad C^L_{iAX}
= g_2 \frac{m_{\ell_i}}{\sqrt{2} m_W \cos\beta}(O_L)_{A 2}
U^\nu_{X i} , \no \\
& & N^R_{iAX} = -\frac{g_2}{\sqrt{2}}\left\{ -[(O_N)_{A 2}+ (O_N)_{A
1} \tan\theta_W] U^\ell_{X i} +   \frac{m_{\ell_i}}{ m_W \cos\beta }
(O_N)_{A 3}U^\ell_{X (i+3)}\right\} , \no \\
& & N^L_{iAX} = -\frac{g_2}{\sqrt{2}}\left[ 2(O_N)_{A 1}
\tan\theta_W U^\ell_{X (i+3)} +   \frac{m_{\ell_i}}{ m_W \cos\beta }
(O_N)_{A 3} U^\ell_{X (i+3)}\right] \,.
\eea
The interactions of charginos and neutralinos with the $Z$ boson are
the following:
\be
\label{l2}%
{\cal L} = - g_Z \left( \ov{\tilde{\chi}^+}_A \bar{\sigma}^\mu
\tilde{\chi}^+_B R_{A B} - \ov{\tilde{\chi}^-}_A \bar{\sigma}^\mu
\tilde{\chi}^-_B L_{A B} + \ov{\tilde{\chi}^0}_A \bar{\sigma}^\mu
\tilde{\chi}^0_B N_{A B} \right)Z_\mu ,
\ee
where
\bea
\label{v2}%
& & R_{A B} = \left[c^2_{\theta_W} (O_R^\ast)_{A 1} (O_R)_{B 1} +
\left(\frac12 -s^2_{\theta_W}\right)  (O_R^\ast)_{A 2} (O_R)_{B
2}\right]\quad , \quad L_{A B} = R^\ast_{A B} |_{R\rightarrow L} , \no\\
& & N_{A B} = \frac12 \left[ (O_N^\ast)_{A 3} (O_N)_{B 3} -
(O_N^\ast)_{A 4} (O_N)_{B 4} \right]
\eea
The chargino contributions are:
\bea
\label{char1L}%
A^{L,c}_{ji} &\!\! = \!\!&\frac{C^R_{iBX} C^{R*}_{jAX}}{16\pi^2}
\left[ R^\ast_{AB} \frac{F(x_{\tilde{\chi}_A \tilde{\nu}_X},
x_{\tilde{\chi}_B \tilde{\nu}_X})+\ln\frac{\mu^2}{M_A^2}}{2} -
\frac{M_{\tilde{\chi}_A}
M_{\tilde{\chi}_B}}{m^2_{\tilde{\nu}_X}}L_{A B} G(x_{\tilde{\chi}_A
\tilde{\nu}_X},
x_{\tilde{\chi}_B \tilde{\nu}_X})  \right. \nn \\
&& \left. -\delta_{A B}  \left(\frac12 - Z^{\ell}_L\right)
\frac{H(x_{\tilde{\chi}_A \tilde{\nu}_X})+\ln\frac{\mu^2}{M_A^2}}{2}\right] , \\
A^{R,c}_{ji} &\!\! =\!\! & A^{L,c}_{ji}|_{L\leftrightarrow R} ,
\label{char1R}
\eea
here   $A, B=1,2$,  $\,X=1,2,3$ and $Z^\ell_{L(R)} = T^3_{\ell_{L
(R)}} - Q_\ell s^2_{\theta_W}$ ($T^3_{\ell_L}=-\frac12,
T^3_{\ell_R}=0$ and $Q_\ell = -1$). (A summation over repeated
indices is understood.) In the above equations the first and second
terms come from the diagrams in which the $Z$ boson line is attached
to the chargino line, the third one where it is attached to the
sneutrino and the fourth term comes from the wave function
renormalization. For completeness, we have also displayed the terms
proportional to $\ln(\mu^2/M_A^2)$ in Eq.~(\ref{char1L}), coming
from the divergent diagrams, where $\mu$ is the renormalization
scale. Obviously, such terms cancel out. As for the argument of the
loop functions we have adopted the convention $x_{ ab}=
m^2_a/m^2_b$, then {\it e.g.} $x_{\tilde{\chi}_A \tilde{\nu}_X} =
M^2_{\tilde{\chi}_A}/m^2_{\tilde{\nu}_X}$. The loop functions are
defined as follows:
\bea
\label{loop}%
F(x,y)&  = &\frac12 + \ln x + \frac{1}{x - y} \left( \frac{x^2\, \ln
x}{1- x} - \frac{y^2\, \ln y}{1- y}\right) ,\no \\
G(x,y) & = &  \frac{1}{x - y}\left(\frac{x\,\ln x}{1- x} - \frac{y\,
\ln y}{1- y}\right) ,
\no \\
H(x) &=& \frac32 + \left[ \frac{ (1-2x) \,\ln x}{(1-x)^2} -
\frac{1}{1-x}\right].
\eea
Using the relation $H(x)=F(x,x)-2\,x\,G(x,x)$, one can easily verify
the validity of the Ward-Takahashi (WT) identity in the
$SU(2)_W\times U(1)_Y$ unbroken phase, which entails the vanishing
of the coefficient, $A^{L(R),c}|_{v_1=v_2=0} = 0$. By exploiting
this identity in Eq.~(\ref{char1L}),
the above expressions (\ref{char1L}) simplify to:
\bea \label{char2L}
A^{L, c}_{ji} &\!\! = \!\!&-\frac{C^R_{iBX} C^{R*}_{jAX}}{16\pi^2}
\left[ (O_R)_{A 2} (O_R^\ast)_{B 2} \frac{F(x_{\tilde{\chi}_A
\tilde{\nu}_X}, x_{\tilde{\chi}_B \tilde{\nu}_X})}{4}\right. \no \\
&&\left.-(O_L^\ast)_{A 2} (O_L)_{B 2} \frac{M_{\tilde{\chi}_A}
M_{\tilde{\chi}_B}}{m^2_{\tilde{\nu}_X}}G(x_{\tilde{\chi}_A
\tilde{\nu}_X}, x_{\tilde{\chi}_B \tilde{\nu}_X})
\right] ,  \\
A^{R, c}_{ji} &\!\! =\!\!& - A^Z_L|_{L\leftrightarrow R} .
\label{char2R}
\eea
The result obtained for $A^L$ in (\ref{char2L}) coincides with that
of Ref.~\cite{HMTY} and is consistent with the one in
Ref.~\cite{br}. The formulas (\ref{char1L}, \ref{char1R}) are also
in agreement with those reported\footnote{ In fact, the agreement
between Eqs.~(\ref{char1L}, \ref{char1R}) and the corresponding
formulas in Ref.~\cite{AH} does not regard the constant terms in the
loop-functions $F(x,y)$ and $H(x)$. Nevertheless, such terms do not
contribute because of the unitarity relations.\label{Hfoot}} in
Ref.~\cite{AH}, which, however, have not been reduced to the form
(\ref{char2L}, \ref{char2R}). We observe that $A^{Z,c}_R\ll
A^{Z,c}_L $ because of the  Yukawa coupling suppression (this
coefficient has been set directly to zero in~\cite{HMTY}).

The coefficients $A^{L(R), n}$ from the neutralino-exchange
contributions are given by:
\bea \label{neut1L}
A^{L, n}_{ji} & \!\!\! =\!\!\! & \frac{N^R_{iBX}
N^{R*}_{jAY}}{16\pi^2} \left\{ \delta_{XY}
\left[N_{AB}^\ast\frac{F(x_{\tilde{\chi}_A \tilde{\ell}_X},
x_{\tilde{\chi}_B \tilde{\ell}_X})+\ln\frac{\mu^2}{M_A^2}}{2} +
N_{AB}\frac{M_{\tilde{\chi}_A}
M_{\tilde{\chi}_B}}{m^2_{\tilde{\nu}_X}}\, G(x_{\tilde{\chi}_A
\tilde{\ell}_X}, x_{\tilde{\chi}_B \tilde{\ell}_X}) \right]\right.\no \\
& \!\! & - \frac{\delta_{A B}}{2}
\left.\left[s^2_{W}\delta_{XY}\left({H(x_{\tilde{\chi}_A
\tilde{\nu}_X})+\ln\frac{\mu^2}{M_A^2}}\right)
 - U^{\ell \ast}_{ X k} U^\ell_{ Y k} \frac{I(
x_{\tilde{\chi}_A \tilde{\ell}_Y}, x_{\tilde{\ell}_X \tilde{\ell}_Y}
)+\ln\frac{\mu^2}{M_A^2}}{2}\right] \right. \no\\
&& \left.+\delta_{A B} \delta_{XY}
 Z^{\ell}_L \frac{H(x_{\tilde{\chi}_A \tilde{\ell}_X}) +\ln\frac{\mu^2}{M_A^2}}{2}\right\}
\no \\ && \no \\ && \no \\
A^{R, n}_{ji} &\!\!\! = \!\!\! & -\frac{N^L_{iBX}
N^{L*}_{jAX}}{16\pi^2} \left\{\delta_{XY}
\left[N_{AB}\frac{F(x_{\tilde{\chi}_A \tilde{\ell}_X},
x_{\tilde{\chi}_B \tilde{\ell}_X})+\ln\frac{\mu^2}{M_A^2})}{2} +
N_{AB}^\ast\frac{M_{\tilde{\chi}_A}
M_{\tilde{\chi}_B}}{m^2_{\tilde{\nu}_X}}\,
 G(x_{\tilde{\chi}_A \tilde{\ell}_X}, x_{\tilde{\chi}_B
\tilde{\ell}_X}) \right\}
\right. \no \\
&& \left.+ \frac{\delta_{A B}}{2} \left[s^2_{W}
\delta_{XY}\left({H(x_{\tilde{\chi}_A
\tilde{\nu}_X})+\ln\frac{\mu^2}{M_A^2}}\right) - U^{\ell \ast}_{Xk}
U^\ell_{ Y k} \frac{I(x_{\tilde{\chi}_A \tilde{\ell}_Y},
x_{\tilde{\ell}_X
\tilde{\ell}_Y})+\ln\frac{\mu^2}{M_A^2}}{2}\right] \right.\no \\
&&\left.-\delta_{A B} \delta_{XY}Z^{\ell}_R \frac{H(
x_{\tilde{\chi}_A
\tilde{\ell}_X})+\ln\frac{\mu^2}{M_A^2}}{2}\right\} , \label{neut1R}
\eea
where $A,B=1, \ldots 4, X=1, \ldots 6$, $k=e,\mu,\tau$ and the loop
function $I(x,y) = 1+ F(x,y)$. The first and second terms derive
from the contributions with  the $Z$ attached to the neutralino
line, the third and fourth from those with the  $Z$ attached to the
slepton line, and the fifth one from the wave-function
renormalization diagram. By using again the WT the above expressions
simplify as\footnote{Using the simplified formulas (\ref{char2L},
\ref{char2R}) and (\ref{neut2L}, \ref{neut2R}) is more convenient
also because cancelations are already accounted for. Needless to say
that the constant numerical addenda appearing in the loop functions
(\ref{loop}) do not contribute to the final amplitudes because of
unitarity of the mixing matrices $U^\nu$ and $U^\ell$. Still, they
are essential to prove the WT identities and then to yield the
simplified formulas (\ref{char2L}, \ref{char2R}) and (\ref{neut2L},
\ref{neut2R}).}:
\bea \label{neut2L} A^{L, n}_{ji}
& \!\!\! = \!\!\! & \frac{N^R_{iBX} N^{R*}_{jAY}}{16\pi^2} \left\{
\delta_{XY} \left[N_{AB}^\ast\frac{F(x_{\tilde{\chi}_A
\tilde{\ell}_X}, x_{\tilde{\chi}_B \tilde{\ell}_X})}{2}
+N_{AB}\frac{M_{\tilde{\chi}_A}
M_{\tilde{\chi}_B}}{m^2_{\tilde{\ell}_X}} G(x_{\tilde{\chi}_A
\tilde{\ell}_X},
x_{\tilde{\chi}_B \tilde{\ell}_X}) \right] \right. \no \\
&& \left. + \delta_{A B}  \left[ U^{\ell\ast}_{ X k} U^\ell_{Y k}
\frac{I( x_{\tilde{\chi}_A \tilde{\ell}_Y}, x_{\tilde{\ell}_X
\tilde{\ell}_Y} )}{4}  - \delta_{XY} \frac{H( x_{\tilde{\chi}_A
\tilde{\ell}_X})}{4}\right] \right\}
\\
A^{R, n}_{ji} &\!\!\! =\!\!\! & -\frac{N^L_{iBX}
N^{L*}_{jAY}}{16\pi^2} \left\{ \delta_{XY}
\left[N_{AB}\frac{F(x_{\tilde{\chi}_A \tilde{\ell}_X},
x_{\tilde{\chi}_B \tilde{\ell}_X})}{2} + N_{AB}^\ast
\frac{M_{\tilde{\chi}_A} M_{\tilde{\chi}_B}}{m^2_{\tilde{\ell}_X}}G(
x_{\tilde{\chi}_A
\tilde{\ell}_X}, x_{\tilde{\chi}_B \tilde{\ell}_X} ) \right] \right. \no \\
&& \left. -\delta_{A B}  U^{\ell\ast}_{X k} U^\ell_{Y k}
\frac{I(x_{\tilde{\chi}_A \tilde{\ell}_Y}, x_{\tilde{\ell}_X
\tilde{\ell}_Y} )}{4}
\right\} . \no\\
&& \label{neut2R}
\eea
Our results (\ref{neut1L}) are compatible with those of
Ref.~\cite{AH} (see, however, the comment in Footnote~\ref{Hfoot}),
and are also compatible with Ref.~\cite{br}. Instead, our final
expressions (\ref{neut2L}) differ from those of Ref.~\cite{HMTY},
because of the third and fourth (third) terms in $A^{L,n}
(A^{R,n})$.

%
\renewcommand{\theequation}{C-\arabic{equation}}
\setcounter{equation}{0}  
\section*{Appendix C-Box coefficients for the $\bm{\ell_j\to \ell_i \ell_k
\ell_k\,(i\neq k)}
$ amplitudes}  
We consider the four-fermion operators which are relevant for the
amplitude of the LFV decays $\ell_j\to \ell_i \ell_k \ell_k$:
\be\label{4f}
(\ov{\ell}_i \smub \ell_j) \left( B^{LL}_{ji;k} \,\bar{\ell}_k \smdb
\ell_k + B^{LR}_{ji;k} \, \ell^c_k \smd  \bar{\ell}^c_k \right) + (
\ell_i^c \smu \ov{\ell}^c_j ) \left( B^{RL}_{ji;k} \,\bar{\ell}_k
\smdb \ell_k + B^{RR}_{ji;k} \, \ell^c_k \smd \bar{\ell}^c_k \right)
+ {\rm h.c.} .
\ee
Each coefficient $B^{M N}\, (M,N=L, R)$ receive contributions from
box diagrams with either charginos/sneutrinos or
neutralinos/charged-sleptons exchange (see Fig.~\ref{f15}):
\be \label{box-c}
B^{M N}_{ji;k} = B^{ M N (c)}_{ji;k} + B^{ M N (n)}_{ji;k}
 \ee
We have obtained the following results valid for $i\neq k$:
\bea
\label{eachbox}%
B^{LL (c)}_{ji;k}& = & \frac{1}{4} J_4(M^2_{\tilde{\chi}_A},
M^2_{\tilde{\chi}_B}, m^2_{\tilde{\nu}_X}, m^2_{\tilde{\nu}_Y})
\left[ C^{R}_{i A Y} C^{R*}_{j A X} C^{R}_{k B X} C^{R*}_{k B Y}
 + C^{R *}_{j A X} C^{R}_{i B X} C^{R}_{k A Y} C^{R*}_{k B Y} \right]
\no \\
B^{L R (c)}_{ji;k}& = & \frac{1}{4} J_4(M^2_{\tilde{\chi}_A},
M^2_{\tilde{\chi}_B}, m^2_{\tilde{\nu}_X}, m^2_{\tilde{\nu}_Y})
C^{R}_{i A Y} C^{R *}_{j A X} C^{L}_{k B X} C^{L *}_{k B Y}
\no \\
& & -\frac12 M_{\tilde{\chi}_A} M_{\tilde{\chi}_B}
I_4(M^2_{\tilde{\chi}_A}, M^2_{\tilde{\chi}_B}, m^2_{\tilde{\nu}_X},
m^2_{\tilde{\nu}_Y})
  C^{R}_{i B X} C^{R *}_{j A X} C^{L}_{k A Y} C^{L *}_{k B Y}
\no \\
B^{LL (n)}_{ji;k}& = & \frac{1}{4} J_4(M^2_{\tilde{\chi}_A},
M^2_{\tilde{\chi}_B}, m^2_{\tilde{\ell}_X}, m^2_{\tilde{\ell}_Y})
\left[ N^{R}_{i B X} N^{R *}_{j A X} N^{R}_{k A Y} N^{R *}_{k B Y}
 + N^{R}_{i A Y} N^{R *}_{j A X} N^{R}_{k B X} N^{R *}_{k B Y}
\right] \no \\
& & + \frac12 M_{\tilde{\chi}_A} M_{\tilde{\chi}_B}
I_4(M^2_{\tilde{\chi}_A}, M^2_{\tilde{\chi}_B},
m^2_{\tilde{\ell}_X}, m^2_{\tilde{\ell}_Y}) \left[ N^{R}_{i B X}
N^{R *}_{j A X} N^{R}_{k B Y} N^{R *}_{k A Y}
 \right.\no \\
 &&\left.+ N^{R}_{i B Y} N^{R *}_{j A X} N^{R}_{k B X} N^{R *}_{k A Y}
\right] \no \\
B^{L R (n)}_{ji;k}& = & \frac{1}{4} J_4(M^2_{\tilde{\chi}_A},
M^2_{\tilde{\chi}_B}, m^2_{\tilde{\ell}_X}, m^2_{\tilde{\ell}_Y})
\left[ N^{R}_{i A Y} N^{R *}_{j A X} N^{L*}_{k B Y} N^{L}_{k B X}
 - N^{R}_{i B X} N^{R *}_{j A X} N^{L*}_{k A Y} N^{L}_{k B Y}
\right. \no \\
&& \left. + N^{R}_{i B Y} N^{R *}_{j A X} N^{L*}_{k A Y} N^{L}_{k B
X} \right] \no \\
&&-\frac14 M_{\tilde{\chi}_A} M_{\tilde{\chi}_B}
I_4(M^2_{\tilde{\chi}_A}, M^2_{\tilde{\chi}_B},
m^2_{\tilde{\ell}_X}, m^2_{\tilde{\ell}_Y}) N^{R}_{i B X} N^{R *}_{j
A X} N^{L *}_{k B Y} N^{L}_{k A Y} \no \\ && \no\\
B^{RR (c)}_{ji;k} & =& B^{ L L(c)}_{ji;k}|_{L\leftrightarrow R} ,
~~~~~~
B^{RL (c)}_{ji;k} = B^{ L R(c)}_{ji;k}|_{L\leftrightarrow R} , ~~~~~~\no \\
B^{RR (n)}_{ji;k} & =&  B^{ L L(n)}_{ji;k}|_{L\leftrightarrow R} ,
~~~~~~ B^{RL (n)}_{ji;k} = B^{ L R(n)}_{ji;k}|_{L\leftrightarrow R}
,
\eea
where the coefficients $N^{L (R)}_{iAX}, C^{L (R)}_{iAX}$ have been
defined in Eq.~(\ref{lg}) and the loop integrals are given as:
\bea\label{intg}
I_4(m_1^2,\ldots,m_4^2) & \equiv & \! {-i \over 16\pi^4}  \!
\int  \! {{\rm d}^4 k \over (k^2 -m_1^2) \ldots  (k^2 -m_4^2)} \no \\
J_4(m_1^2,\ldots,m_4^2) & \equiv & \! {-i \over 16\pi^4}  \! \int
\! {k^2 \,{\rm d}^4 k \over (k^2 -m_1^2) \ldots  (k^2 -m_4^2)}  .
\eea

The box coefficients for the decays $\tau\to \mu e e$ and $\tau \to
e \mu \mu$ correspond to the replacements $(j,i,k)\to(\tau, \mu,e)$
and $(j,i,k)\to(\tau, e,\mu)$, respectively. Our results for these
coefficients are in numerical agreement with those of
Ref.~\cite{br}.
\begin{figure}
\begin{center}
\includegraphics[width=14.0cm]{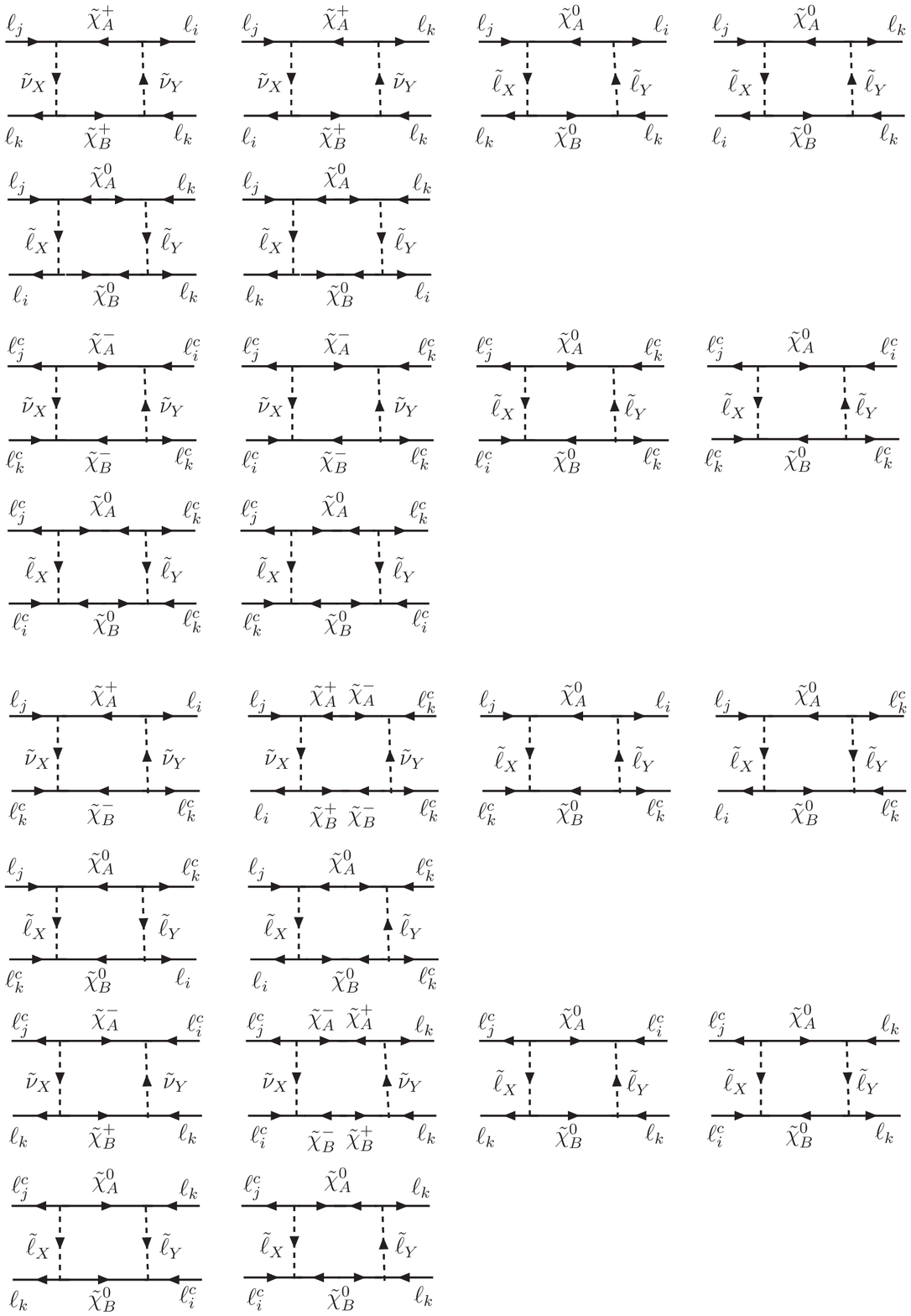}
\end{center}
\captions{\small Box diagrams relevant for the computation of the
coefficients $B^{LL}_{ji;k}$ (first and second rows),
$B^{RR}_{ji;k}$ (third and fourth rows), $B^{LR}_{ji;k}$ (fifth and
sixth rows) and $B^{RL}_{ji;k}$ (seventh and eighth rows) [see
Eqs.~(\ref{4f})-(\ref{eachbox})].} \label{f15}
\end{figure}
%


\newpage
\begin{thebibliography}{99}
\bibitem{W}
S.~Weinberg,
Phys.\ Rev.\ Lett.\  {\bf 43} (1979) 1566.


\bibitem{seesaw}
P.~Minkowski,
Phys.\ Lett.\ B {\bf 67} 421 (1977); M. Gell-Mann, P. Ramond and R.
Slansky, in {\it  Supergravity}, eds.\ P. Van Nieuwenhuizen and D.
Freedman (North-Holland, Amsterdam, 1979),p.~315; T. Yanagida, in
{\it Proceedings of the Workshop on the Unified Theory and the
Baryon Number in the Universe}, eds.\ O. Sawada and A. Sugamoto
(KEK, Tsukuba, 1979), p.~95; S.L. Glashow, in {\it Quarks and
Leptons}, eds.\ M. L\'evy et al., (Plenum, 1980, New-York), p. 707;
R.N.~Mohapatra and G. Senjanovi\'{c}, Phys.\ Rev.\ Lett.\ {\bf 44},
912 (1980).

\bibitem{t0}
R.~Foot, H.~Lew, X.~G.~He and G.~C.~Joshi,
Z.\ Phys.\ C {\bf 44} (1989) 441;
E.~Ma,
Phys.\ Rev.\ Lett.\  {\bf 81} (1998) 1171;
B.~Brahmachari, E.~Ma and U.~Sarkar,
Phys.\ Lett.\ B {\bf 520} (2001) 152.


\bibitem{ss2}
R. Barbieri, D.V. Nanopolous, G. Morchio and F. Strocchi, Phys.\
Lett.\ B {\bf 90}, 91 (1980);
R.~E.~Marshak and R.~N.~Mohapatra,
VPI-HEP-80/02
{\it Invited talk given at Orbis Scientiae, Coral Gables, Fla., Jan
14-17, 1980}; T.~P.~Cheng and L.~F.~Li,  Phys.\ Rev.\ D {\bf 22},
2860 (1980);
M. Magg and Ch.\ Wetterich, Phys.\ Lett.\ B {\bf 94}, 61 (1980);
J.~Schechter and J.~W.~F.~Valle,
Phys.\ Rev.\ D {\bf 22}, 2227 (1980);
G.~Lazarides, Q.~Shafi and C.~Wetterich,
Nucl.\ Phys.\ B {\bf 181}, 287 (1981);
R.N. Mohapatra and G. Senjanovic, Phys.\ Rev.\ D {\bf 23}, 165
(1981);
E. Ma and U. Sarkar, Phys.~Rev.~Lett.~{\bf 80}, 5716 (1998).


\bibitem{casas}
J.~A.~Casas, V.~Di Clemente, A.~Ibarra and M.~Quiros,
Phys.\ Rev.\ D {\bf 62} (2000) 053005.


\bibitem{ellisnan}
J.~R.~Ellis and D.~V.~Nanopoulos,
Phys.\ Lett.\ B {\bf 110}, 44 (1982);
R.~Barbieri and R.~Gatto,
Phys.\ Lett.\ B {\bf 110}, 211 (1982);
M.~J.~Duncan,
Nucl.\ Phys.\ B {\bf 221}, 285 (1983);
J.~F.~Donoghue, H.~P.~Nilles and D.~Wyler,
Phys.\ Lett.\ B {\bf 128}, 55 (1983).



\bibitem{lfv1}
F.~Borzumati and A.~Masiero, Phys.\ Rev.\ Lett. {\bf 57}, 961
(1986).

\bibitem{lfv2}
F.~Gabbiani and A.~Masiero,
Nucl.\ Phys.\ B {\bf 322}, 235 (1989).

\bibitem{ar}
A.~Rossi, Phys.\ Rev.\ D {\bf 66}, 075003 (2002).

\bibitem{DI}
See A.~Casas and A.~Ibarra in Ref.~\cite{rgess} and S.~Davidson and
A.~Ibarra,
JHEP {\bf 0109} (2001) 013.


\bibitem{HMTY}
J.~Hisano, T.~Moroi, K.~Tobe and M.~Yamaguchi,
Phys.\ Rev.\ D {\bf 53},  2442 (1996).

\bibitem{rgess}
J.~Hisano, D.~Nomura and T.~Yanagida,
Phys.\ Lett.\ B {\bf 437},  351 (1998);
J.~Hisano and D.~Nomura,
Phys.\ Rev.\ D {\bf 59} (1999) 116005;
J.~R.~Ellis, M.~E.~Gomez, G.~K.~Leontaris, S.~Lola and
D.~V.~Nanopoulos, Eur.\ Phys.\ J.\ C {\bf 14} (2000) 319;
J.~A.~Casas and A.~Ibarra,
Nucl.\ Phys.\ B {\bf 618},  171 (2001);
D.~F.~Carvalho, J.~R.~Ellis, M.~E.~Gomez and S.~Lola,
Phys.\ Lett.\ B {\bf 515} (2001) 323;
S.~Lavignac, I.~Masina and C.~A.~Savoy,
Phys.\ Lett.\ B {\bf 520}, 269 (2001);
~A.~Kageyama, S.~Kaneko, N.~Shimoyama and M.~Tanimoto, Phys.\ Rev.\
D {\bf 65} (2002) 096010;
~A.~Masiero, S.~K.~Vempati and O.~Vives,
Nucl.\ Phys.\ B {\bf 649},  189 (2003);
%
K.~S.~Babu, B.~Dutta and R.~N.~Mohapatra, Phys.\ Rev.\ D {\bf 67}
(2003) 076006;
T.~Blazek and S.~F.~King, Nucl.\ Phys.\ B {\bf 662} (2003) 359;
S.~T.~Petcov, S.~Profumo, Y.~Takanishi and C.~E.~Yaguna, Nucl.\
Phys.\ B {\bf 676} (2004) 453;
M.~Hirsch, J.~C.~Romao, S.~Skadhauge, J.~W.~F.~Valle and
A.~Villanova del Moral, Phys.\ Rev.\ D {\bf 69} (2004) 093006;
E.~Arganda and M.~J.~Herrero,
Phys.\ Rev.\ D {\bf 73}, 055003 (2006);
L.~Calibbi, A.~Faccia, A.~Masiero and S.~K.~Vempati,
arXiv:hep-ph/0605139;
P.~Hosteins, S.~Lavignac and C.~A.~Savoy, arXiv:hep-ph/0606078;
S.~Antusch, E.~Arganda, M.~J.~Herrero and A.~M.~Teixeira,
arXiv:hep-ph/0607263.


\bibitem{af1}
F.~R.~Joaquim  and A.~Rossi,
Phys.\ Rev.\ Lett.\  {\bf 97} (2006) 181801.

\bibitem{hms}
T.~Hambye, E.~Ma and U.~Sarkar,
Nucl.\ Phys.\ B {\bf 602} (2001) 23.

\bibitem{PDG}
S.~Eidelman {\it et al.}  [Particle Data Group],
Phys.\ Lett.\ B {\bf 592} (2004) 1.

\bibitem{MFV1}
L.~J.~Hall and L.~Randall,
Phys.\ Rev.\ Lett.\  {\bf 65} (1990) 2939.

\bibitem{MFV2}
V.~Cirigliano, B.~Grinstein, G.~Isidori and M.~B.~Wise,
Nucl.\ Phys.\ B {\bf 728} (2005) 121; 
B.~Grinstein, V.~Cirigliano, G.~Isidori and M.~B.~Wise,
arXiv:hep-ph/0608123;
S.~Davidson and F.~Palorini,
arXiv:hep-ph/0607329;
G.~C.~Branco, A.~J.~Buras, S.~Jager, S.~Uhlig and A.~Weiler,
arXiv:hep-ph/0609067.


\bibitem{bim}
V.~S.~Kaplunovsky and J.~Louis, Phys.\ Lett.\ B {\bf 306} (1993)
269;
A.~Brignole, L.~E.~Ibanez and C.~Munoz,
Nucl.\ Phys.\ B {\bf 422} (1994) 125 [Erratum-ibid.\ B {\bf 436}
(1995) 747].

\bibitem{univers}
L.~J.~Hall, V.~A.~Kostelecky and S.~Raby,
Nucl.\ Phys.\ B {\bf 267} (1986) 415.

\bibitem{univers2}
R.~Barbieri and L.~J.~Hall,
Phys.\ Lett.\ B {\bf 338} (1994) 212.

\bibitem{cmrv}
E.~J.~Chun, A.~Masiero, A.~Rossi and S.~K.~Vempati,
Phys.\ Lett.\ B {\bf 622}, 112 (2005).


\bibitem{gcu}
C.~Bachas, C.~Fabre and T.~Yanagida,
Phys.\ Lett.\ B {\bf 370}, 49 (1996);
S.~Chaudhuri, G.~Hockney and J.~D.~Lykken,
Nucl.\ Phys.\ B {\bf 469}, 357 (1996);
T.~Han, T.~Yanagida and R.~J.~Zhang,
Phys.\ Rev.\ D {\bf 58}, 095011 (1998);
~B.~Brahmachari,
Phys.\ Rev.\ D {\bf 65}, 067502 (2002);
I.~Dorsner and P.~F.~Perez, Nucl.\ Phys.\ B {\bf 723}, 53 (2005);
I.~Dorsner, P.~F.~Perez and R.~Gonzalez Felipe, Nucl.\ Phys.\ B
{\bf 747}, 312 (2006).

\bibitem{HS1}
P.~Langacker and B.~D.~Nelson, Phys.\ Rev.\ D {\bf 72} (2005) 053013
[arXiv:hep-ph/0507063].

\bibitem{HS2}
M.~Cvetic, I.~Papadimitriou and G.~Shiu,
Nucl.\ Phys.\ B {\bf 659} (2003) 193 [Erratum-ibid.\ B {\bf 696}
(2004) 298]; 
M.~Cvetic and P.~Langacker,
arXiv:hep-th/0607238.

\bibitem{yeyd}
H.~Georgi and C.~Jarlskog,
Phys.\ Lett.\ B {\bf 86}, 297 (1979);
J.~A.~Harvey, P.~Ramond and D.~B.~Reiss,
Phys.\ Lett.\ B {\bf 92}, 309 (1980); For more recent works see for
example:
K.~S.~Babu and S.~M.~Barr,
Phys.\ Rev.\ D {\bf 56}, 2614 (1997);
Z.~Berezhiani,
Phys.\ Lett.\ B {\bf 417}, 287 (1998);
R.~Barbieri, L.~J.~Hall, S.~Raby and A.~Romanino,
Nucl.\ Phys.\ B {\bf 493}, 3 (1997);
Z.~Berezhiani and A.~Rossi,
JHEP {\bf 9903}, 002 (1999);
Z.~Berezhiani and A.~Rossi,
Nucl.\ Phys.\ B {\bf 594}, 113 (2001);
G.~Altarelli, F.~Feruglio and I.~Masina,
JHEP {\bf 0011} (2000) 040.



\bibitem{doubtrip}
S.~Dimopoulos and H.~Georgi,
Nucl.\ Phys.\ B {\bf 193} (1981) 150;
N.~Sakai,
Z.\ Phys.\ C {\bf 11}, 153 (1981);
E.~Witten,
Nucl.\ Phys.\ B {\bf 188}, 513 (1981).


\bibitem{gmall}
M.~Dine, W.~Fischler and M.~Srednicki,
Nucl.\ Phys.\ B {\bf 189}, 575 (1981);
S.~Dimopoulos and S.~Raby, Nucl.\ Phys.\ B {\bf 192}, 353 (1981);
L.~Alvarez-Gaume, M.~Claudson and M.~B.~Wise, Nucl.\ Phys.\ B {\bf
207}, 96 (1982);
C.~R.~Nappi and B.~A.~Ovrut, Phys.\ Lett.\ B {\bf 113}, 175 (1982);
S.~Dimopoulos and S.~Raby, Nucl.\ Phys.\ B {\bf 219}, 479 (1983);
M.~Dine and A.~E.~Nelson, Phys.\ Rev.\ D {\bf 48}, 1277 (1993);
M.~Dine, A.~E.~Nelson and Y.~Shirman, Phys.\ Rev.\ D {\bf 51}, 1362
(1995).

\bibitem{GR-PR}
For a recent review see G.~F.~Giudice and R.~Rattazzi,
Phys.\ Rept.\  {\bf 322}, 419 (1999).

\bibitem{dns}
M.~Dine, Y.~Nir and Y.~Shirman,
Phys.\ Rev.\ D {\bf 55} (1997) 1501
[arXiv:hep-ph/9607397].



\bibitem{GR}
L.~V.~Avdeev, D.~I.~Kazakov and I.~N.~Kondrashuk,
Nucl.\ Phys.\ B {\bf 510} (1998) 289;
G.~F.~Giudice and R.~Rattazzi,
Nucl.\ Phys.\ B {\bf 511}, 25 (1998);
N.~Arkani-Hamed, G.~F.~Giudice, M.~A.~Luty and R.~Rattazzi,
Phys.\ Rev.\ D {\bf 58}, 115005 (1998);
C.~E.~M.~Wagner, Nucl.\ Phys.\ B {\bf 528}, 3 (1998).

\bibitem{CP}
Z.~Chacko and E.~Ponton,
Phys.\ Rev.\ D {\bf 66} (2002) 095004. 


\bibitem{bm}
For a string-inspired realization of GMSB see \emph{e.g.}
E.~Floratos and C.~Kokorelis,
arXiv:hep-th/0607217.

\bibitem{DS}
G.~R.~Dvali and M.~A.~Shifman,
Phys.\ Lett.\ B {\bf 399} (1997) 60;
%
Z.~Chacko, E.~Katz and E.~Perazzi,
Phys.\ Rev.\ D {\bf 66} (2002) 095012.

\bibitem{fit}
See {\em e.g. }, M.~C.~Gonzalez-Garcia, Phys.\ Scripta {\bf T121}
(2005) 72;
G.~L.~Fogli, E.~Lisi, A.~Marrone and A.~Palazzo, Prog.\ Part.\
Nucl.\ Phys.\  {\bf 57}, 742 (2006);
A.~Strumia and F.~Vissani,
Nucl.\ Phys.\ B {\bf 726}, 294 (2005);
R.~N.~Mohapatra and A.~Y.~Smirnov,
arXiv:hep-ph/0603118;
J.~W.~F.~Valle,
arXiv:hep-ph/0603223.



\bibitem{exp}
M.~L.~Brooks {\it et al.}  [MEGA Collaboration],
Phys.\ Rev.\ Lett.\  {\bf 83}, 1521 (1999).

\bibitem{MEG}
M.~Grassi  [MEG Collaboration],
Nucl.\ Phys.\ Proc.\ Suppl.\  {\bf 149} (2005) 369.

\bibitem{exp-rad}
B.~Aubert {\it et al.}  [BABAR Collaboration],
Phys.\ Rev.\ Lett.\  {\bf 95}, 041802 (2005).

\bibitem{taumu:fut}
A.~G.~Akeroyd {\it et al.}  [SuperKEKB Physics Working Group],
arXiv:hep-ex/0406071.

\bibitem{exp-rade}
B.~Aubert {\it et al.}  [BABAR Collaboration], Phys.\ Rev.\ Lett.\
{\bf 96}, 041801 (2006).

\bibitem{sindrum}
U.~Bellgardt {\it et al.}  [SINDRUM Collaboration],
Nucl.\ Phys.\ B {\bf 299} (1988) 1.

\bibitem{mu3e}
J.~Aysto {\it et al.},
arXiv:hep-ph/0109217;
B.~L.~Roberts, M.~Grassi and A.~Sato,
Nucl.\ Phys.\ Proc.\ Suppl.\  {\bf 155} (2006) 123.


\bibitem{babar}
B.~Aubert {\it et al.}  [BABAR Collaboration],
arXiv:hep-ex/0312027.

\bibitem{belle}
Y.~Yusa, H.~Hayashii, T.~Nagamine and A.~Yamaguchi  [Belle
Collaboration],
eConf {\bf C0209101} (2002) TU13 [Nucl.\ Phys.\ Proc.\ Suppl.\  {\bf
123} (2003) 95] [arXiv:hep-ex/0211017].

\bibitem{list}
M.~Hodgkinson  [BaBar Collaboration],
Nucl.\ Phys.\ Proc.\ Suppl.\  {\bf 144} (2005) 167.

\bibitem{sindrum2}
J.~Kaulard {\it et al.}  [SINDRUM II Collaboration],  Phys.\ Lett.\
B {\bf 422} (1998) 334.


\bibitem{prime}
Y.~Mori {\emph et al.} [PRISM/PRIME Working group], LOI at J-PARC
50-GeV PS, LOI-25 [{\tt http://
psux1.kek.jp/$\sim$jhf-np/LOIlist/LOIlist.html}].

\bibitem{lep}
R.~Barate {\it et al.}  [LEP Working Group for Higgs boson
searches], Phys.\ Lett.\ B {\bf 565} (2003) 61.

\bibitem{FH}
S.~Heinemeyer, W.~Hollik and G.~Weiglein,
Comput.\ Phys.\ Commun.\  {\bf 124}, 76 (2000).

\bibitem{MP}
See for a review see {\it e.g.}, M.~Passera,
Nucl.\ Phys.\ Proc.\ Suppl.\  {\bf 155} (2006) 365.

\bibitem{cdfd0}
[Tevatron Electroweak Working Group],
arXiv:hep-ex/0603039.

\bibitem{rgeeq}
See \eg, N.~K.~Falck, Z.\ Phys.\ C {\bf 30}, 247 (1986);
S.~P.~Martin and M.~T.~Vaughn, Phys.\ Rev.\ D {\bf 50}, 2282 (1994).


\bibitem{gsmspec}
For studies on the sparticle spectrum in GMSB models see \eg,
S.~Dimopoulos, S.~D.~Thomas and J.~D.~Wells, Nucl.\ Phys.\ B {\bf
488}, 39 (1997);
A.~Strumia, Phys.\ Lett.\ B {\bf 409}, 213 (1997).

\bibitem{BPMZ}
D.~M.~Pierce, J.~A.~Bagger, K.~T.~Matchev and R.~j.~Zhang,
Nucl.\ Phys.\ B {\bf 491} (1997) 3.

\bibitem{LHC}
For a review see N.~V.~Krasnikov and V.~A.~Matveev, Phys.\ Usp.\
{\bf 47}, 643 (2004) [Usp.\ Fiz.\ Nauk {\bf 174}, 697 (2004)] and
{\tt http://CMSinfo.cern.ch/Welcome.html/CMSdocuments/CMSplots}.

\bibitem{higgs}
M.~Carena and H.~E.~Haber,
Prog.\ Part.\ Nucl.\ Phys.\  {\bf 50} (2003) 63, and references
therein.

\bibitem{br}
A.~Brignole and A.~Rossi,
Nucl.\ Phys.\ B {\bf 701}, 3 (2004).

\bibitem{Hexc}
K.~S.~Babu and C.~Kolda,
Phys.\ Rev.\ Lett.\  {\bf 89} (2002) 241802;
A.~Dedes, J.~R.~Ellis and M.~Raidal, Phys.\ Lett.\ B {\bf 549}
(2002) 159.


\bibitem{brh}
A.~Brignole and A.~Rossi,
Phys.\ Lett.\ B {\bf 566} (2003) 217.


\bibitem{AH}
See E.~Arganda and M.~J.~Herrero in \cite{rgess}.


\bibitem{deltas}
F.~Gabbiani, E.~Gabrielli, A.~Masiero and L.~Silvestrini,
Nucl.\ Phys.\ B {\bf 477}  (1996) 321;
J.~Foster, K.~i.~Okumura and L.~Roszkowski,
JHEP {\bf 0603} (2006) 044.

\bibitem{roadmap}
M.~A.~ Giorgi {\it et al.} [SuperB group], INFN Roadmap Report,
March 2006.


\bibitem{minos}
M.~Komatsu, P.~Migliozzi and F.~Terranova,
J.\ Phys.\ G {\bf 29}, 443 (2003);
P.~Migliozzi and F.~Terranova, Phys.\ Lett.\ B {\bf 563}, 73 (2003);
P.~Huber, J.~Kopp, M.~Lindner, M.~Rolinec and W.~Winter, JHEP {\bf
0605}, 072 (2006).

\bibitem{NF}
For a recent rewiew, see A.~Blondel, A.~Cervera-Villanueva,
A.~Donini, P.~Huber, M.~Mezzetto and P.~Strolin,
arXiv:hep-ph/0606111.


\bibitem{Fukugita:1986hr}
M.~Fukugita and T.~Yanagida,
Phys.\ Lett.\ B {\bf 174} (1986) 45.


\bibitem{alllepto}
G.~C.~Branco, R.~Gonzalez Felipe, F.~R.~Joaquim and M.~N.~Rebelo,
Nucl.\ Phys.\ B {\bf 640} (2002) 202;
%
W.~Buchmuller, P.~Di Bari and M.~Plumacher,
Nucl.\ Phys.\ B {\bf 643} (2002) 367;
%
G.~C.~Branco, R.~Gonzalez Felipe, F.~R.~Joaquim, I.~Masina,
M.~N.~Rebelo and C.~A.~Savoy,
Phys.\ Rev.\ D {\bf 67} (2003) 073025;
%
S.~Pascoli, S.~T.~Petcov and W.~Rodejohann,
Phys.\ Rev.\ D {\bf 68} (2003) 093007;
%
E.~K.~Akhmedov, M.~Frigerio and A.~Y.~Smirnov,
JHEP {\bf 0309} (2003) 021;
%
G.~F.~Giudice, A.~Notari, M.~Raidal, A.~Riotto and A.~Strumia,
Nucl.\ Phys.\ B {\bf 685} (2004) 89;
%
T.~Hambye, Y.~Lin, A.~Notari, M.~Papucci and A.~Strumia,
Nucl.\ Phys.\ B {\bf 695} (2004) 169;
%
%
R.~Gonzalez Felipe, F.~R.~Joaquim and B.~M.~Nobre,
Phys.\ Rev.\ D {\bf 70} (2004) 085009;
%
W.~Buchmuller, P.~Di Bari and M.~Plumacher,
Annals Phys.\  {\bf 315} (2005) 305;
%
G.~C.~Branco, R.~Gonzalez Felipe, F.~R.~Joaquim and B.~M.~Nobre,
Phys.\ Lett.\ B {\bf 633} (2006) 336;
%
A.~Abada, S.~Davidson, F.~X.~Josse-Michaux, M.~Losada and A.~Riotto,
JCAP {\bf 0604} (2006) 004;
%
E.~Nardi, Y.~Nir, E.~Roulet and J.~Racker,
JHEP {\bf 0601} (2006) 164;
%
%
A.~Abada, S.~Davidson, A.~Ibarra, F.~X.~Josse-Michaux, M.~Losada and
A.~Riotto,
arXiv:hep-ph/0605281;
%
V.~Cirigliano, G.~Isidori and V.~Porretti,
arXiv:hep-ph/0607068;
K.~Bhattacharya, N.~Sahu, U.~Sarkar and S.~K.~Singh,
arXiv:hep-ph/0607272.


\bibitem{lepto}
G.~D'Ambrosio, T.~Hambye, A.~Hektor, M.~Raidal and A.~Rossi,
Phys.\ Lett.\ B {\bf 604}, 199 (2004);
%
For previous works on resonant leptogenesis in different contexts
see \eg,
M.~Flanz, E.~A.~Paschos and U.~Sarkar,
Phys.\ Lett.\ B {\bf 345} (1995) 248 [Erratum-ibid.\ B {\bf 382}
(1996) 447];
%
L.~Covi, E.~Roulet and F.~Vissani,
Phys.\ Lett.\ B {\bf 384} (1996) 169;
%
M.~Flanz, E.~A.~Paschos, U.~Sarkar and J.~Weiss,
Phys.\ Lett.\ B {\bf 389} (1996) 693;
%
G.~D'Ambrosio, G.~F.~Giudice and M.~Raidal,
Phys.\ Lett.\ B {\bf 575} (2003) 75;
%
Y.~Grossman, T.~Kashti, Y.~Nir and E.~Roulet,
Phys.\ Rev.\ Lett.\  {\bf 91} (2003) 251801;
%
Nucl.\ Phys.\ B {\bf 692} (2004) 303.

\bibitem{TN}
T.~Hambye and G.~Senjanovic,
Phys.\ Lett.\ B {\bf 582} (2004) 73;
S.~Antusch and S.~F.~King,
Phys.\ Lett.\ B {\bf 597} (2004) 199; 
T.~Hambye,
arXiv:hep-ph/0412053;
T.~Hambye, M.~Raidal and A.~Strumia,
Phys.\ Lett.\ B {\bf 632} (2006) 667.

\bibitem{farza}
Y.~Farzan, JHEP {\bf 0502} (2005) 025.
\end{thebibliography}
\end{document}